\documentclass[12pt]{report} 
\usepackage{thesis,epsfig}
\usepackage{amssymb}
\usepackage{latexsym}
\usepackage{verbatim}
\DeclareGraphicsExtensions{.jpg,.eps}

\title{Purifying Quantum States: Quantum and Classical Algorithms} 
\author{Eric Dennis} 
\date{September 2003}

\newcommand{\mri}[0]{\mathrm{i}}

\begin{document}

\pagenumbering{roman}
\maketitle

\begin{approval}
\noindent 
The dissertation of Eric Dennis is approved.\\

\vspace{1in}

\noindent 
------------------------------------------------------------------------\\
Herschel Rabitz (Princeton University)\\

\bigskip
\bigskip

\noindent 
------------------------------------------------------------------------\\
David Awschalom\\

\bigskip
\bigskip

\noindent 
------------------------------------------------------------------------\\
Matthew Fisher\\

\bigskip
\bigskip

\noindent 
------------------------------------------------------------------------\\
Philip Pincus\\

\bigskip
\bigskip

\noindent 
------------------------------------------------------------------------\\
Atac Imamoglu, Committee chair\\

\vfil

\begin{centering}
July 2003\\
\end{centering}

\end{approval}

\makecopyright

\selectfont

\begin{acknowledgements} 

I would like to thank my advisors Atac Imamoglu and Hersch Rabitz for
their guidance and instruction, David Awschalom for giving me the
opportunity to see what a qubit really looks like, John Preskill for his
methodical lessons in quantum information, and the grad students and
post-docs who have helped and entertained me along the way, in particular
Mike Werner, Ignacio Sola, Jay Kikkawa, Jay Gupta, Dan Fuchs, Andrew
Landahl, Dave Beckman, and Daniel Gottesman. I am also grateful to Nino
Zanghi, Sheldon Goldstein, and Roderich Tumulka for their intransigence,
and to Travis Norsen for his level-headedness. I have been supported by
the Army Research Office, the Princeton Plasma Physics Laboratory, and the
National Science Foundation.

\end{acknowledgements}

\begin{vita}

\selectfont
\begin{centering}
\textbf{Vita}\\ 
Eric Dennis\\
\end{centering}

\smallskip
\noindent \emph{Education}\\

\noindent Bachelor of Science (honors), California Institute of
Technology,
June 1998.\smallskip\\
\noindent Master of Science in Physics, UC Santa
Barbara, Sept. 2000.\smallskip\\
\noindent Doctor of Philosophy in Physics, UC Santa
Barbara, Sept. 2003 (expected).\\

\smallskip
\noindent \emph{Professional Employment}\\

\noindent Summers 1994-98: Research Assitant, California
Institute of Technology.\smallskip\\
\noindent 1998-2000: Research Assistant, Dept. of Physics,
UC Santa Barbara.\smallskip\\
\noindent 2000-03: Research Assistant, Dept. of Chemistry,
Princeton University.\\

\smallskip
\noindent \emph{Publications}\\

\noindent E. Dennis, ``Toward fault-tolerant computation without
concatenation,'' Phys. Rev. A \textbf{63} 052314 (2001); presented as an
invited talk at 1999 SQuINT annual conference.\smallskip\\
\noindent E. Dennis, A. Kitaev, A. Landahl, J. Preskill, ``Topological
quantum memory,'' J. Math. Phys. \textbf{43} 4452 (2002).\smallskip\\
\noindent E. Dennis, H. Rabitz, ``Path integrals, stochastic trajectories,
and hidden variables,'' Phys. Rev. A \textbf{67} 033401 (2003); presented
as an invited talk at CECAM workshop in Lyon, France (Sept. 2002).\\

\smallskip
\noindent \emph{Awards}\\

\noindent Three PPST Fellowships, Princeton Plasma Physics
Laboratory (6/01--6/03).\\

\smallskip
\noindent \emph{Fields of Study}\\

\noindent Major Field: Quantum Information\smallskip\\
\noindent Study in Quantum Error Correction with Prof. Atac
Imamoglu.\smallskip\\
\noindent Study in Experimental Condensed Matter Physics with Prof. David
Awschalom.\smallskip\\
\noindent Study in Quantum Control and Simulation with Prof. Herschel
Rabitz.\\

\end{vita}

\selectfont

\begin{abstract}

The problems of decoherence and destructive measurement make quantum
computers especially unstable to errors. We begin by exploring Kitaev's
framework for quantum error correcting codes that exploit the inherent
stability of certain topological quantum numbers. Analytic estimates and
numerical simulation allow us to quantitatively assess the conditions
under which this stability persists. In order to allow quantum
computations to be performed on data stored in these and other
unconcatenated codes, we present a probabilistic algorithm for executing
an encoded Toffoli gate via preparation of a certain three-qubit entangled
state as a computational resource. The algorithm works to iteratively
purify an initial input state, conditional on favorable measurement
outcomes at each stage in the iteration.

Analogous iterative purification methods are identified in the context of
classical algorithms for simulating quantum dynamics. Here an ensemble of
stochastic trajectories over a classical configuration space is
iteratively driven toward a fixed point that yields information about a
quantum evolution in an associated Hilbert space. We test this procedure
for a restricted subspace of the Heisenberg ferromagnet.

The trajectory ensemble concept is then applied to the problem of
mechanism identification in closed-loop quantum control, where a precise
definition of mechanism in terms of these trajectories leads to methods
for extracting dynamical information directly from the kinds of
optimization procedures already realized in control experiments on
optical-molecular systems. We demonstrate this on simulated data for
controlled population transfer in a seven level molecular system under
shaped-pulse laser excitation.

\end{abstract}

\tableofcontents
\listoffigures

\chapter{Elements of Quantum Information}
\label{chap-qi}
\pagenumbering{arabic}

\section{Universal Quantum Simulators}

It is a bromide among particle physicists that the standard model contains
(or that some still undefined string/M theory may someday contain) 
all the rest of physics at lower
energy scales and that the problem is just the difficulty of actually
computing its implications on these scales. It is a bromide among
condensed matter physicists that chemistry and materials science are
likewise reducible to one large $N$-particle Schrodinger equation. And it
is a bromide among chemists that biology is reducible in some such way to
chemistry.

But why are even the most sophisticated modern computers unable to
seriously penetrate these age-old disciplinary boundaries? It appears that
the main obstacle to exploiting this reductionism is native to quantum
mechanics, and it is not so subtle. Consider how a computer could
represent a purely classical, physical system. Assuming a discretized
state space so that for $N$ distinguishable particles each may occupy one
of $S$ possible states, the total amount of information necessary for a
computer to store one particular configuration of the system is just $N
\log_2 S$ bits. One particular state of the corresponding quantum system,
however, will require for its description a set of $S^N$ complex numbers,
one for every possible classical state. This means the number of bits
necessary for a computer to store this description will grow
exponentially, not linearly, with $N$. As $N$ gets large, not only does it
become computationally intractable to predict the future behavior of the
system---it is intractable even to fully specify its present state.

Something here is not quite right, though. The desired result of a given
quantum simulation may not necessarily include a complete quantum
description of some final state. The heat capacity of a metal for
instance---while requiring a quantum mechanical treatment in certain
circumstances---is still just one number. It is only the intermediate
stages in the calculation which presumably require consideration of these
intractable quantum state vectors. But perhaps there are alternative
representations which obviate such huge memory sinks. Indeed there are,
and they succeed in eliminating the exponential blow-up of storage and
computation costs for certain kinds of systems. However, no such
techniques exist that work on a more general level, and their
effectiveness at particular problems is something of an open question to
be probed one problem at a time.

This leads to a more pessimistic appraisal: if \emph{ab initio}
computation is in general so expensive, why bother with it at all? If one
is interested in the behavior of a generic quantum system in the
laboratory, perhaps it will always be easier to perform actual experiments
on the system itself. Let the system perform the computations for you!
Indeed the physical system may be regarded as a kind of ultra-special
purpose computer, fit for exactly one problem. Perhaps one is more
ambitious, however, and might attempt to use one physical system that is
especially accessible in the laboratory in order to simulate some other
kind of system entirely. For instance there are proposals to study
gravitational systems by resort to analogies in condensed matter
\cite{ether}, whereby the material relations for certain liquids are seen
to fortuitously mimic some aspects of the Einstein field equations.

Even bolder would be to imagine that certain quantum systems, by virtue of
some special types of interactions, are able to simulate large classes of
quantum systems in a way which is less fortuitous and more by design. In
essence, exactly this is what is going on in a classical computer when it
simulates other classical systems. There is no real physical similarity
between a silicon microchip and the turbulent flow of nitrogen gas which
it simulates via a fluid dynamics algorithm. Why can't such a
computational universality exist in the quantum domain as well?

This was the motivation of Feynman when he proposed the idea of a
\emph{quantum computer} as a universal quantum simulator \cite{feynman}.

A sufficiently well controlled quantum mechanical system that can be
initialized in a chosen state, evolved by tunable interactions, and
measured with high fidelity deserves the title quantum computer if these
interactions are sufficiently universal as to mimic a large class of other
quantum systems bearing little physical relationship to that of the
computer. The crux of quantum computation is in this notion of
computational universality. The difference between a liquid Helium system
mimicking the quantum fluctuations of space-time and a quantum computer
mimicking, say, the molecular dynamics of water is that tomorrow the
quantum computer may be used to simulate something entirely different;
whereas, the liquid Helium system will always be stuck on the same
physical problem of quantum gravity.

As opposed to the Helium system, which achieves its mimickery through a
kind of mathematical coincidence, the quantum computer works (rather,
would work, if ever one were built) by detailed independent control of
many many individualized components. The basic unit of the quantum
computer is the quantum bit (or \emph{qubit})\footnote{Sometimes ``qubit''
will be further abbreviated as just ``bit'' when the context is not one
comparing quantum and classical information.}. One qubit is a quantum
system represented by a two-dimensional Hilbert space with basis states
conventionally denoted as $|0\rangle$ and $|1\rangle$. $N$ qubits are
represented by the space spanned by all states $|x\rangle$ where $x$ is an
$N$-bit binary string. The dimension of this space thus grows as $2^N$.

A general quantum system with classical configuration space $Q$ may then
be represented as an $N$-qubit system by discretizing $Q$ into $2^N$
elements. In order to simulate the system evolution in the quantum
computer, this evolution must be discretized in time, and each resulting
unitary step must be translated into operations to be performed on the
qubits. In particular if the classical configuration space $Q =
(0,1)^{\otimes N}$ for $N$ distinguishable particles in 1d is discretized
such that the $i$-th coordinate can be represented by a bit string $x_i$
(containing, say, $S$ bits), we can write the infinitesimal system
evolution operator as
\[
U(t,t+\Delta t) = 
e^{-\mri H\Delta t} \approx e^{-\mri T\Delta t}e^{-\mri V\Delta t}
\]
where the kinetic term matrix elements $\langle x_i|T|x_j \rangle$ vanish
unless $i=j$. Here, we are taking the state of the computer to
comprise a superposition of terms
\[
|x_1\rangle \cdots |x_N\rangle
\]
where the length $S$ quantum register $|x_i\rangle$ encodes the
instantaneous value of the $i$-th coordinate. If we assume only pairwise
interactions, then the potential operator may be written
\[
V = \sum_{ijkl}
	|x_i,x_j\rangle \langle x_i,x_j|V|x_k,x_l\rangle \langle x_k,x_l|
\, .
\]
In order to let $e^{-\mri V\Delta t} \approx 1 - iV\Delta t$ act on our
registers, we thus need to perform a quantum computation in which
arbitrary pairs of registers are interacted and, based on the initial
values of those registers, transformed into a superposition whose
coefficients are determined by the $\langle x_i,x_j|V|x_k,x_l\rangle$
terms above. Moreover, the values of these terms must themselves be
computed through arithmetical operations in the quantum computer, for
instance if they are given analytically as polynomials in
$x_i-x_k$ and $x_j-x_l$.

A given arithmetic operation between two registers will be accomplished by
a sequence of operations carried out on the physical qubits constituting
these registers, either individually or in pairs (or possibly in higher
order combinations). This is the basic idea of a \emph{quantum gate}. The
essential difference between a quantum and a classical gate is simply that
in the quantum case, a register consisting of a single term, e.g.\
$|01\rangle$, may be taken into a register consisting of a superposition
of terms, e.g.\ $|00\rangle + |10\rangle$. More generally, a quantum gate
is just a unitary transformation applied to the qubits it acts on.

In order to apply the gate corresponding to the unitary operator
$e^{-\mri V\Delta t}$ in this case we must simply allow it to act on every
combination of two coordinate registers $|x_i\rangle$ and
$|x_j\rangle$---with each such action reduced to a sequence of bit-wise
gates on the individual qubits constituting the registers. This amounts
to O$(N^2)$ operations. Propagating from $t=0$ out to $t=T$ will thus
require O$(N^2 T/\Delta t)$ operations, clearly scaling polynomially in
the system size $N$. (See \cite{lloyd} for a more detailed treatment of
quantum computers as simulators.)

\section{Steering a Quantum Computer}

A classical computer, because its gates take a single bit-string only into
another single bit-string, will need to perform a separate computation for
every term in a superposition like $\sum_{i_1\cdots i_N}|x_{i_1} \cdots
x_{i_N}\rangle$ if it is to simulate a generic evolution for the above 
quantum system. This 
implies a number of computations scaling exponentially in
$N$. On the other hand the quantum computer is in effect carrying out all 
these operations
\emph{in parallel}. More generally, consider an $N$ bit string $x$, and a
function $f(x)$ outputing a single bit. A single classical computation of
$f$ might yield something like:
\[
x = 0101001\cdots01 \rightarrow f(x) = 1
\]
whereas, if the function $f$ can be realized as a unitary transformation
$U$ so that $U|x\rangle|0\rangle = |x\rangle|f(x)\rangle$, then a quantum
computer could perform the following as single function call:
\[
\sum_x |x\rangle|0\rangle \rightarrow \sum_x |x\rangle|f(x)\rangle 
\]
where the action of $U$ has been distributed over all terms in the sum by
virtue of its linearity. This ``massive parallelism'' is the essence of
the exponential speed-up realized by a quantum computer in the problem of
quantum simulation (and likewise in Shor's factoring algorithm \cite{shor}).

As stated this parallelism seems too good to be true. It seems we can
trivially crack NP-complete problems (e.g.\ optimization problems like the
traveling salesman problem) by simply encoding all possible solutions in
a quantum register (e.g.\ $\sum_x |x\rangle$) and devising some $U$ to
perform computations on all these possible solutions in parallel. Indeed
this is too good to be true, for the necessity of reading out the computer
at the end of the computation takes on a new significance in the quantum
domain. Reading out entails measuring the register, which in this case,
will collapse the register into the term corresponding to a single one of
the candidate solutions $|x\rangle$. This irreversible process has
destroyed almost all of the information encoded in our quantum state.

Indeed in the above sketch of a quantum simulation algorithm we cannot
read out the entire state vector at $t=T$. We would expect to define some
Hermitian operator of interest, and effectively measure that operator
alone. The massive parallelism is only half the battle---this second
aspect, being clever about what to measure at the end, is equally
important. Another way of looking at this is not as some special
measurement to be made at $t=T$, but rather a sequence of additional
operations to be performed on the encoded quantum information that will
allow us to extract the desired properties of our final state through more
standardized (single qubit) measurements. It is the task of quantum
algorithms to devise operations that can make use of this kind of massive
but restricted parallelism.

An essential issue in designing such algorithms is the question of what
primitive operations are assumed available to act on qubits. Without any
constraints here, we could simply define some final state that encodes the
solution to a problem of interest, assert that such a state is reachable
through \emph{some} unitary transformation from the initial state
$|00\cdots 0\rangle$, and declare victory. Rather, what we would like is a
small set of primitive unitary operations involving a limited number of
qubits at one time, such that when applied in combination they can produce
any desired unitary (with arbitrary accuracy) on the space of all the
qubits jointly. Such a set of primitive operations is called a
\emph{universal gate set}, and indeed the first fundamental results in the
field concerned the construction of such a set.

The problem of finding a compact universal gate set and demonstrating its
universality---i.e.\ the ability to generate arbitrary unitaries over a
Hilbert space with arbitrarily many qubits---is closely related to the
problem of the controlability of a continuous quantum system whose
evolution is given by some Hamiltonian
\[
H = H_\mathrm{int} + \sum_i \alpha_i(t) H_i \,.
\]
Here $H_\mathrm{int}$ corresponds to the internal dynamics of the system,
and the $H_i$ correspond to controllable external interactions imposed by
the experimenter, so that the couplings $\alpha_i(t)$ serve as tunable
control knobs. Such a system is referred to as controllable if by suitable
tuning of the $\alpha_i(t)$, any state $|\psi(T)\rangle$ may be reached
from some fixed initial state $|\psi(0)\rangle$, possibly subject to
certain constraints imposed on the magnitude of the $\alpha_i(t)$ and
their derivatives for $t \in (0,T)$.

The essential resource we have in trying to produce motions in the
Hilbert space which are not generated simply by a single
$H_i$ is the possibility of composing such motions as the following:
\[
e^{-\mri H_i\Delta t}e^{-\mri H_j\Delta t}
	e^{\mri H_i\Delta t}e^{\mri H_j\Delta t}
\]
which to lowest order results in a motion generated by the commutator
$[H_i,H_j]$. Moreover, given these $H_i$, we can generate all possible
iterated commutators, and the question becomes whether the algebra of all
these commutators exhausts the space of all Hermitian matrices. If so,
then the set $\{H_i\}$ is universal over all possible Hermitian generators
\cite{hersch}, directly analogous to a universal gate set. The difference
is principally just that our control knobs here may be adjusted
continuously in time, while in the quantum computing context, the control
knobs are simply the choices of which gates to apply in what order and are
therefore discrete in nature.

We will be interested in problems of continuous quantum control later. Now
it will suffice just to recognize the possibility of universal gate
sets, which constitute our most elementary tool box in the design of
quantum algorithms. For example, it is well known \cite{preskill} that one
universal gate set can be built from just two gates: (i) a generic single
qubit rotation, e.g.\ $e^{\mri \theta X/2}$, where $\theta$ is an
irrational multiple of $\pi$, and (ii) the controlled-not (C-NOT) gate,
which acts on two qubits.\footnote{The notation $X$, $Y$, and $Z$ is
commonly used for the Pauli operators $\sigma_x$, $\sigma_y$, and
$\sigma_z$.} We can fully specify this C-NOT gate by its action on the
basis elements in a two-qubit Hilbert space:
\begin{eqnarray*}
&\mathrm{C-NOT}& \\
|00\rangle &\rightarrow& |00\rangle \\
|01\rangle &\rightarrow& |01\rangle \\
|10\rangle &\rightarrow& |11\rangle \\
|11\rangle &\rightarrow& |10\rangle \\
\end{eqnarray*}
where we say that the C-NOT is taken \emph{from} the first qubit (the
``control'' qubit) \emph{to} the second one (the ``target'' qubit). The
action of this gate on a general two-qubit state is then implied by the
fact that the gate is a linear operation. 

It should be noted that the one-qubit rotation combined with the C-NOT may
be used, first, to generate another one-qubit rotation that does not
commute with the original one. These two can then be used to generate all
possible one-qubit rotations \cite{preskill}. The significance of the
requirement that the original one-qubit rotation involve an angle $\theta$
that is an irrational multiple of $\pi$ is that, otherwise, the commutator
algebra associated with these two rotations would close on itself before
exploring the entire SU$(2)$ of qubit rotations---yielding only a finite
number of reachable rotations.

Some such gate as the C-NOT is crucial to exploit the massive parallelism
of a quantum computer because it is an \emph{entangling} operation, i.e.\
it has the power to take two initially unentangled qubits and produce a
final state in which they are entangled, for instance:
\[
|00\rangle + |10\rangle \rightarrow |00\rangle + |11\rangle \, .
\]
Whereas the initial state may be factored, hence has no entanglement, the
final state cannot. Obviously massive parallelism in a quantum computer
would require more than just two-qubit entanglement. The idea in using the
universal gate set defined above is that more complicated entanglements
among many qubits can be built up by successive applications of the C-NOT
gate to different pair of qubits.

\section{Unitary and Decoherent Errors}

At this point we have laid out one very significant application of a
quantum computer (simulating quantum systems) and the basic capabilities
that must be realized in any physical implementation of such a device. In
principle, then it seems like quantum engineers would now be at a place
analogous to that of Mauchly and Eckert on the eve of the ENIAC project.
Obviously formidable technical problems remain, but no basic theoretical
obstacles seem to remain.

Except for one thing, which was probably more of a nuisance
than a serious obstacle for Mauchly and Eckert: the problem of error
correction. Due to the imperfection of components and action of gates,
etc., there will always be errors creeping into any computation. For a
classical computer, there will always be the occasional 0 that is
accidentally flipped to a 1 and vice versa. Uncontrolled, these errors
will tend to build up and totally corrupt the computation after an amount
of time depending on the computer's error rates.

It was von Neumann who first exhibited a straightforward method to handle
this problem. If we want a particular bit to store a 1, for example, we
should actually use not one physical bit for the job but a couple, say
$N$. We will simply set all of these bits to 1. After a little time has
elapsed, some of these bits may have accidentally flipped to 0. We can
combat this tendency simply by checking the bit values and majority voting
in order to flip the erroneous bits back. (Note: here we cannot just flip
all the bits back to 1 irrespective of the how many flipped, because that
would require us to have stored a \emph{separate} record of what the
correct value of the bit was, namely 1.)

If the probability of any individual bit having flipped over this time is
no more than some bound $p$, then as $N$ gets large the chances that one
half or more of these bits have accidentally flipped---hence the
probability that our majority vote will fail to restore the proper bit
state---goes down exponentially with $N$. This is the essential property
of an error correcting code: exponential security, or, in other words,
arbitrarily good security with only a logarithmic overhead of additional
bits.

Constructing this particular error correcting code, that is the assignment
of $N$ physical bits to encode one ``logical'' bit by pure repitition, was
quite simple. In fact much more sophisticated codes exist for classical
computers; however, they do not differ in the general nature of this
scaling between the overhead required to implement the code (number of
additional bits) and the security provided by the code (probability of
failure).

Unfortunately, the problem of error correction is qualitatively harder in
the quantum domain. In fact, from the advent of quantum algorithms in the
1980's to the first proposal for a viable quantum error correcting code in
1995 there were serious doubts about even the theoretical
possibility of such a code \cite{landauer}.

There are two essential obstacles unique to maintaining the integrity of
quantum information: decoherence and state reduction. Decoherence refers
to the tendency of quantum systems to become increasingly entangled with
their environments so that the interference effects necessary for quantum
algorithms become gradually less pronounced, until they are finally
reduced to unmeasurability. Loosely, decoherence is what turns a quantum
system into a classical one, hence one incapable of allowing us to cash in
on massive parallelism.

Suppose, for instance, we are interested in measuring a physical effect
associated with some quantum mechanical phase (e.g.\ an Aharanov-Bohm
effect) in a two-level system. Let us first describe the process in the
absence of any decoherence. Thus we might start with our system in the
unnormalized state $|0\rangle +|1\rangle$, and then enact the
phase-generating process, which may be described by some unitary evolution
\[
|0\rangle +|1\rangle \rightarrow |0\rangle +e^{2\mri \phi}|1\rangle \, .
\]
In order to measure $\phi$, we might perform a simple quantum gate on this
single qubit, which is to apply a Hadamard rotation defined by
the basis state transformations (neglecting normalization):
\begin{eqnarray*}
&\mathrm{Hadamard}& \\
|0\rangle &\rightarrow& |0\rangle + |1\rangle\\
|1\rangle &\rightarrow& |0\rangle - |1\rangle \, .\\
\end{eqnarray*}
This gives
\[
|0\rangle +e^{2\mri \phi}|1\rangle \rightarrow \cos\phi|0\rangle -
\mri\sin\phi|1\rangle
\]
which allows us to transform the phase information into
amplitude information through a process of quantum interference. We may
now simply measure $Z$, i.e. the bit-value of this qubit. (Such a
measurement, as opposed to a direct measurement of phase, is what we
generally assume is available to us.) Repeated trials will be able to
determine $\phi$ with high accuracy.

We can now consider the effect of decoherence by explicitly accounting for
the quantum state of the environment surrounding our system. Suppose the
environment starts in some state $|E\rangle$, unentangled with one qubit,
so that we have the total initial state
\[
(|0\rangle +|1\rangle)|E\rangle \, .
\]
Suppose as well that during the phase-generating process, some interaction
between our qubit and the environment causes the latter to either respond
($|E\rangle \rightarrow |E^\prime\rangle$) or not respond ($|E\rangle
\rightarrow |E\rangle$) depending on the state of the qubit, i.e.\
\[
(|0\rangle + |1\rangle)|E\rangle \rightarrow 
|0\rangle|E\rangle + e^{2\mri \phi}|1\rangle|E^\prime\rangle \, .
\]
After applying a Hadamard rotation, this becomes
\[
|0\rangle(|E\rangle + e^{2\mri \phi}|E^\prime\rangle) \, + \,
|1\rangle(|E\rangle - e^{2\mri \phi}|E^\prime\rangle) \, .
\]
Now, assuming the effect on the environment is eventually amplified enough
that $\langle E|E^\prime\rangle=0$, measuring the qubit will have equal
probability of yielding $|0\rangle$ or $|1\rangle$ independent of $\phi$,
hence it will reveal exactly nothing about $\phi$.

In other words, the simple quantum computation that in the absence of
decoherence allowed us to extract phase information from our quantum
state, has been rendered useless in the presence of decoherence---more
precisely, in the presence of total decoherence, since we assumed $\langle
E|E^\prime\rangle=0$. In this case we say that the qubit has totally
decohered in the basis $\{|0\rangle,|1\rangle\}$. If $\langle
E|E^\prime\rangle \ne 0$, the qubit will have decohered only partially in
this basis, and as $\langle E|E^\prime\rangle$ approaches zero, the number
of measurements necessary to extract $\phi$ with a given accuracy will
approach infinity.

The crux of the decoherence problem in regard to storing and using quantum
information lies in the ubiquity of what we classify as ``environment,''
that is: everything not part of the finely controlled system which
constitutes the quantum computer itself. Everything from ambient
electromagnetic field modes (whether occupied or unoccupied), to air
molecules capable of scattering off components of the computer, to the
atoms constituting a substrate for these components, to other neglected
degrees of freedom within the components themselves---all are environment.
Because of this environmental ubiquity, each individual qubit will face an
independent array of possibilities for it to decohere. Loosely, each qubit
will have a constant (or bounded from below by a constant) probability of
decohering within a given time interval, independent of the other qubits.
Therefore, the probability that the unaided computer will maintain any
fixed degree of coherence goes down exponentially with the number of
qubits for sufficiently complicated (i.e. time-intensive) computations.

\section{Quantum Error Correcting Codes}

So far, the exact same comments may be made for a classical computer whose
bits each independently face a fixed probability of being accidentally
flipped. The problem for a quantum computer is that the basic method of
error correction, the repitition code, fails trivially for a system of
qubits.

The quantum analog of the repetition code is easy to construct.
Encode one logical qubit in $N$ physical qubits by the assignment
\begin{eqnarray*}
\mathrm{logical} & & \mathrm{physical}\\
|0\rangle &\rightarrow& |000\cdots0\rangle\\
|1\rangle &\rightarrow& |111\cdots1\rangle
\end{eqnarray*}
Suppose now we have encoded the logical state $\alpha|0\rangle +
\beta|1\rangle$ in $n$ qubits. After some time has past, some of these $n$
qubits will have suffered errors, for example having their bit values
flipped:
\[
\alpha|00000\rangle + \beta|11111\rangle \rightarrow \alpha|00101\rangle +
\beta|11010\rangle
\]
(we have taken $n=5$ to illustrate bit flips on the third and fifth bit.)
This is a unitary error---errors associated with decoherence processes
might also occur. But neglecting decoherence for the moment, we
would then like to measure all the qubits and somehow majority vote to
determine how we should restore our state back to its original (pre-error)
form. The problem is that even measuring only a single qubit collapses the
state to a single computational basis element, here either $|00101\rangle$
or $|11010\rangle$. We may majority vote and correctly
reconstruct, e.g., $|11010\rangle$ into $|11111\rangle$; however, we have
eliminated all the quantum information comprising $\alpha$ and $\beta$ in
our original state. Indeed we have learned something about these parameters
by the probabilistic nature of this state reduction, but we have
transformed some of the quantum information into classical information
and simply destroyed the rest.

A simple strategy to overcome this unfortunate state reduction is to
measure not each individual bit value---i.e.\ measure $Z_i$ on each qubit
$i$---but to measure only the bit values of each bit relative to its
neighbors, which means measuring the product $Z_i Z_{i+1}$ where $i+1$ is
taken modulo $n$. Since both terms $|11010\rangle$ and $|11010\rangle$ are
eigenstates with the same eigenvalue for any one of these measurements, we
will not have collapsed our state. And a majority vote will allow us to
restore the original error-free state. But how exactly are we to measure
this product operator $Z_i Z_{i+1}$? 

Assuming we are able to measure single qubit operators like $Z_i$ itself,
such product operators may be measured by simple procedures involving
\textit{ancilla} qubits---qubits that are used just as a temporary scratch
pad and whose final state is not important for the overall computation. To
measure $Z_i Z_{i+1}$, we initialize one ancilla qubit $a$ in the state
$|0\rangle$, perform a C-NOT from qubit $i$ to $a$, another C-NOT from
$i+1$ to $a$, and then measure $Z_a$. These operations ensure that any
eigenstate of $Z_i Z_{i+1}$ will, with the addition of the ancilla $a$,
also be an eigenstate of $Z_a$, with the same eigenvalue. So measuring
$Z_a$ is equivalent, in terms of the information revealed and the effect
on the total state, to measuring $Z_i Z_{i+1}$ itself.

Thus we can overcome what seems the basic obstacle of revealing too much
information when we make measurements necessary to correct errors in our
state. Still, a uniquely quantum problem remains. We have only discussed 
the kind
of error (whether decoherent or unitary) that applies to the bit-value of
the state, as opposed to its phase. (In the decoherent case this relates
to the basis in which we assume decoherence.) Suppose in the above error
process, along with bit flip errors, our qubits can undergo phase errors:
$|0\rangle \rightarrow |0\rangle$ but $|1\rangle \rightarrow -|1\rangle$.
In particular suppose the first qubit $i=1$ alone suffers such a phase
error. The above parity measurements will not reveal any information about
this error, and we will end up not with the original error-free state, but
with the logical state $\alpha|0\rangle - \beta|1\rangle$. Thus a phase
error occurring in even a single qubit is enough to undermine this kind of
bit-parity code.

A dual kind of code can be constructed that does exactly the opposite: it
corrects phase errors but not bit errors. This phase-parity code is
identical to the bit-parity code if we just prepare our state by applying
a Hadamard rotation to each qubit, hence the encoding is
\begin{eqnarray*}
|0\rangle &\rightarrow& (|0\rangle + |1\rangle)
	(|0\rangle + |1\rangle)\cdots
	(|0\rangle + |1\rangle)\\
|1\rangle &\rightarrow& (|0\rangle - |1\rangle)
	(|0\rangle - |1\rangle)\cdots
	(|0\rangle - |1\rangle) \, .
\end{eqnarray*}
The effect of this is to change a physical bit error into a physical phase
error and vice versa, which is how we can use the repetition idea above to
fix phase errors. With this encoding the measurement procedure prior to
our majority vote involves measuring $X_iX_{i+1}$ operators, not
$Z_iZ_{i+1}$ operators.

Realizing the simple kind of duality between bit and phase errors
illustrated by the construction of this code, Peter Shor first proposed a
fully quantum error correcting code in that it corrects both bit and
phase errors \cite{9qubit}. Here, one logical qubit is encoded in nine
physical qubits as follows
\begin{eqnarray*}
|0\rangle &\rightarrow& (|000\rangle + |111\rangle)
	(|000\rangle + |111\rangle)
	(|000\rangle + |111\rangle)\\
|1\rangle &\rightarrow& (|000\rangle - |111\rangle)
	(|000\rangle - |111\rangle)
	(|000\rangle - |111\rangle) \, .
\end{eqnarray*}
The measurement procedure is to measure $Z_i Z_{i+1}$ for neighboring bits
among the first three qubits taken as a set, and likewise for the second
three, and then for the final three---the corresponding results are part
of what is called the \emph{error syndrome}. If a single bit among the
nine had flipped, this syndrome will reveal which one and allow us to
correct that error. To complete the syndrome we measure the operators $X_1
X_2 X_3$ and $X_4 X_5 X_6$ and $X_7 X_8 X_9$, which is analogous to the
$X_i$ measurements in the pure phase code above. If a single phase error
occurs, say to qubit 6, this will show up as a $-1$ result when we measure
$X_4 X_5 X_6$, indicating that we must choose either $X_4$, $X_5$, or
$X_6$ and apply it to our state. This corrects the phase error and returns
us to our original error-free state.

In the above, we have been somewhat cavalier about the kinds of errors
that might occur in our computer. Even if errors are unitary in nature and
not decoherent, it is unlikely that they will be either pure phase or pure
bit errors. However any arbitrary unitary error for $n$ qubits may be
expressed as a direct sum of terms consisting of the identity and products
of bit and phase errors. After such an error occurs, one sees the
measurements specified above have the affect of collapsing the error
itself into a set of pure phase and pure bit errors.

A similar phenomenon occurs in regard to errors resulting from
decoherence. The act of measuring our qubit system has the fortunate
effect of actually unentangling it from the environment and transforming
decoherence errors into a set of unitary (and in fact pure bit/phase)
errors \cite{reliable}. If sufficiently many errors occur (e.g.\ more than
one bit and one phase error in the 9-qubit code above), however, the
result may be that our recovery operations end up transforming our logical
state $\alpha|0\rangle + \beta|1\rangle$ into either $\alpha|1\rangle +
\beta|0\rangle$ or $\alpha|0\rangle - \beta|1\rangle$. We thus snowball
errors to the physical qubits into a full-blown error to the encoded
quantum information. Recovery has failed.

This possibility is dealt with by constructing more sophisticated quantum
codes, which permit more and more physical errors before a logical error
is precipitated. In fact the pattern of Shor's 9-qubit code plainly
suggests a recursive generalization. This code can be thought of as
possessing not just two levels of qubits---the physical and the
logical---but three levels. We have the physical qubits, then we have
intermediate qubits encoded under $|0\rangle \rightarrow |000\rangle$
and $|1\rangle \rightarrow |111\rangle$, and finally we have the logical
qubits encoded under $|0\rangle \rightarrow (|0\rangle + |1\rangle)^3$ and
$|1\rangle \rightarrow (|0\rangle - |1\rangle)^3$, where the qubits used
for this last encoding are themselves not physical bits but the
intermediate bits. We might just as well iterate this and employ
$4,5,\ldots,L$ levels of encoding. Here we have suggested alternating
between bit and phase codes in this iteration; however, we can also regard
this as iterating a single code, in this case the 9-qubit code. In fact
other codes, e.g.\ involving five or seven qubits, that protect against
bit and phase errors are likewise amenable to this kind of hierarchical
scheme.

Codes generated in such a way are called \textit{concatenated codes}, and
constitute the first systematic, scalable means of quantum
error correction \cite{7}.

The motivation for constructing such large codes is simply to leverage the
trade-off between computational overhead (number of physical bits per
logical bit) and informational security. Traditionally, security was
measured in terms of how many errors to the physical qubits it would take
in order that error correction procedures would turn out to corrupt the
logical qubits. In the 9-qubit code, it takes in fact three errors. This
code is thus said to have ``distance'' three. The overhead/security
trade-off may be measured for a given code as a ratio of its distance to
its block size---the number physical qubits necessary to form a block that
encodes one logical qubit.

As we increase the block size $n$, we might hope that code distance $d$
has a linear asymptotic scaling with $n$. If through each round of error
correction each qubit will have some error probability $p$ (or even
bounded above by $p$), we would be satisfied if $p < d/n$, for then the
chances of recovery failure would fall exponentially with $n$, making it
relatively painless to achieve very high accuracy.

However it is by no means necessary that $p < d/n$ in order to obtain such
an exponential scaling. In particular it is possible to imagines codes in
which $d/n \rightarrow 0$ as $n \rightarrow \infty$ but that are still
exponentially secure in this limit. It may be that although it is possible
that only $d$ errors cause recovery failure, this becomes highly unlikely
for large $n$, unlikely enough to overwhelm the high probability of
merely realizing $d$ errors in the first place. 

Let us examine this issue in the case of concatenating the 9-qubit code
through $L$ levels, and obtain a rough estimate of the chances of recovery
failure assuming both bit and phase errors have an independent probability
$p$ per round of error correction. Consider the first level of the
code, where we are dealing with physical qubits. The probability of two
bit errors occurring (on two separate qubits) is then about $9p \cdot 8p =
72p^2$, and likewise for two phase errors. For small $p$, the probability
of either of these two cases is then about twice that, $144p^2$. Thus $p
\rightarrow 144p^2$ is the mapping from the error rate at level 1 to that
at level 2. At level 3, our ``physical'' qubits are actually the logical
qubits at level 2, which have error rate $144p^2$, so the logical qubits
at level 3 will fail with probability $144(144p^2)^2$. Iterating this
calculation, gives that the overall failure rate $F$ for the code up to
$L$ levels is
\[
F = 144^{2^{L-1} + 2^{L-2} + \cdots + 1}p^{2^L} = (144p)^{2^L} =
(144p)^{n^{0.315}} \, ,
\]
where we have expressed $2^L$ in terms of the block size $n=9^L$, with
$\log_9 2 \approx 0.315$. $F$ thus has the desired property that it dies
exponentially in $n$ (rather a power of $n$) if the physical error rate
$p<1/144$. This is a very simple kind of threshold result. The crude
estimate $p_c = 1/144$ for the physical error rate threshold, or the
``critical'' error rate, would therefore serve as a benchmark for
evaluating the viability of a given physical implementation of the
concatenated 9-qubit code.

In the above we have glossed over one very important point. By taking the
error rate at level $l$ as simply the failure rate at level $l-1$, we have
assumed that error correction at $l$ will be just as easy as error
correction at $l-1$. This is obviously not true. There are 9 times as many
qubits in a block at $l$ than at $l-1$. We will therefore require many
more recovery operations and measurements at the higher level. And
crucially: the recovery operations and measurements that we employ
to correct errors are themselves liable to cause additional errors in the
computer. Our own operations are faulty, and our codes must be designed
to take this into account. Moreover, because we must make joint
measurements on multiple qubits, e.g.\ measuring $X_1 X_2 X_3$ in the
9-qubit code itself, there is the possibility of an error in one qubit
contaminating other qubits in the block. In designing our recovery
procedures, we must be sure that the tendency for our own actions to
spread errors does not overbalance the error correction achieved through
those procedures.

Mathematically, this means that the error rate mapping between levels
$l-1$ and $l$, which we had taken as
\[
p_{l-1} \rightarrow p_l = 144p_{l-1}^2 \, ,
\]
will now explicitly involve $l$ in a more complicated manner. Determining
this $l$ dependence requires an analysis of how errors are both generated
and spread by our recovery operations. Systematic calculations have been
performed in this manner showing that fault-tolerant recovery is possible
with a general class of concatenated codes, and better estimates of the
critical error rate(s) are on the order of $p_c \sim 10^{-4}$ as well as
comparable thresholds for the accuracy of the (physical qubit) gates 
\cite{7}.

What we have addressed so far is only the problem of storing quantum
information with these codes. Additional questions arise when we want to
also perform quantum gates and make measurements of the logical qubits
encoded in such blocks. In other words, we need to determine what sequence
of operations to perform on the physical bits themselves that will result
in the application of a desired operation to the logical qubit(s). For
instance, if we want to perform a bit flip ($X$ gate) on a logical qubit
stored with the 9-qubit code, we need to act on the state with operators
that flip the three relevant phases, e.g.\ with the operators $Z_1$,
$Z_4$, and $Z_7$. The physical operation $Z_1Z_4Z_7$ is therefore
equivalent to the logical $X$ operation. Similar correspondences have to
be found for all members of a desired universal gate set in order to allow
for universal computation on the encoded quantum information. Given a
specification of some gate on the encoded information in terms of a
sequence of gates on physical qubits, we must then analyze the propensity
for errors to build up as a result. This will lead to similar threshold
results corresponding to these physical gates \cite{7}. For example, the
accuracy of a physical C-NOT gate will have to be below a certain critical
value in order that using it in the performance of a specified logical
gate will (with high probability) not spark a cascade of physical qubit
errors that may damage the encoded information.

Both the problem of storing and of performing long computations with
quantum information have been essentially solved on the theoretical level
with these concatenated codes. However, the solution is not unique, and
it seems new solutions will be necessary in order to bridge the gap
between, on the one hand, the assumptions entering into present estimates
of the performance of concatenated codes and, on the other hand, the
forseeable experimental frontier in quantum information.

Although it has not been emphasized above, one important assumption is
that physical qubits separated by large distances in the computer may be
gated together efficiently. For instance, measuring an operator like $Z_1
Z_4 Z_7$ at the highest level of a quadruply concatenated 9-qubit code
(i.e.\ one with $L=4$) will require gating pairs of qubits separated
typically by on order of 100 other qubits---even if we assume qubits
distributed over a two-dimensional lattice. Clearly this scenario poses a
daunting experimental problem. Gating qubits will always require some kind
of well controlled physical interaction to take place between them, and
the experimental possibility of achieving this pair-wise over a very large
array is highly restricted. 

One strategy might seek to confront this experimental difficulty head-on,
by using specially designed physical environments in which quantum
information may be exchanged over large distances, for example by coupling
two quantum dots (qubits) through a very high finesse QED cavity mode
\cite{atac}. Another strategy would seek to confront the problem first at
the quantum software level, that is: to design quantum error correcting
codes which minimize the necessity of long-distance interactions between
qubits. Such is the goal of an alternative paradigm for error correction
invented by Alexei Kitaev, which develops a connection between the idea of
the stability to errors in a quantum code and that of the stability to
deformations in the topology of a 2-dimensional surface.

\chapter{Topological Quantum Memory}
\label{chap-memory}

\section{Lattice Codes}

The basic idea of fault-tolerance is to store information in such a way
that any little errors occurring in the computer's components cannot do
serious damage.  But this is not a new idea; \emph{topological}
information has this same character, being invariant to local deformations
of a specified geometry. To exploit this analogy, one must find a way to
view a block of qubits as a geometrical object with the information
encoded in the block corresponding to some kind of topological invariant.

This is the idea behind Kitaev's framework for quantum error-correcting
codes \cite{8} \cite{11}, which organize qubits as edges in a 2d lattice
on a torus. Kitaev found codes for which certain crucial recovery
operations (syndrome measurements) are all \emph{local} on the lattice,
never involving more than a few neighboring qubits.  Thus errors are
severely limited in their propagation without the necessity of complicated
fault-tolerant gate constructions---fault-tolerance is introduced at a
more fundamental level. Moreover, fatal error processes are seen to arise
only in the aftermath of large scale topological breakdowns in a recovery
algorithm to be specified.

\begin{figure} 
\centering
\scalebox{.2}{\includegraphics{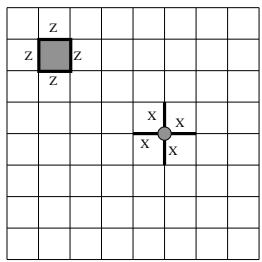}}
\caption{\small 
A plaquette operator $B$, comprising four $Z$'s, and a star operator $A$,
comprising four $X$'s.
}
\label{checks2}
\end{figure}

\begin{figure} 
\centering
\scalebox{.25}{\includegraphics{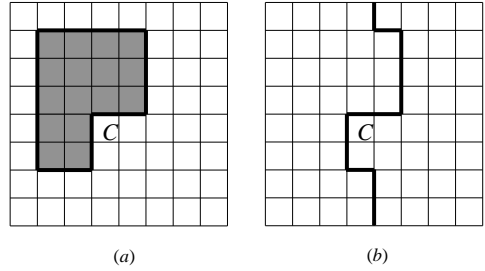}}
\caption{\small
A contractible loop (a), and a non-contractible loop (b) on the lattice.
}
\label{homo2}
\end{figure}

The \emph{toric code} TOR$(k)$ uses $2k^2$ physical bits arranged in a
$k\times k$ lattice (with edges identified) to encode two logical bits.
Its stabilizer---i.e. the group of all transformations that do not affect
the encoded information---is generated by \emph{star} and \emph{plaquette}
operators
\[
A_s = \prod_{+s} X \;\;\;\;\;\;\;\; \mbox{and} \;\;\;\;\;\;\;\; B_p =
\prod_{\Box p} Z
\]
respectively, where ``$+s$'' denotes the four edges emanating from vertex
$s$ and ``$\Box p$'' denotes the four edges enclosing face $p$ (see Fig.\
\ref{checks2}). The code subspace is that fixed by $A_s$ and $B_p$ for all
$s$ and $p$.  Note that any $A_s$ shares either zero or two edges with any
$B_p$, so all the stabilizer operators commute.  Because of the two
operator identities $\prod_{s} A_s = 1$ and $\prod_{p} B_p = 1$, only
$2k^2 - 2$ of the stabilizer generators are independent, giving $2k^2 -
(2k^2 - 2)=2$ encoded qubits, i.e., a 4-dimensional code subspace.
The connection to topology arises from the fact that the plaquette
operators generate exactly the set of \emph{contractible} loops of $Z$'s
on
the lattice (see Fig.\ \ref{homo2}). Likewise, the star operators generate
exactly the the set of contractible loops of $X$'s on the dual lattice
(the
lattice obtained by rotating every edge by $90^\circ$ about its midpoint).

To see how this is reflected in the assignment of logical basis elements
(``codewords''), let us find them explicitly. Consider the (unnormalized)
state
\[
|\bar 0 \bar 0 \rangle \equiv \prod_{s} \left(1+A_s\right) |0 \cdots 0 \rangle = \left(1 + \sum_s A_s + 
\sum_{s < s'} A_s A_{s'} + \cdots \right)|0 \cdots 0\rangle
\]
where $|0 \cdots 0 \rangle$ refers to all the $2k^2$ physical bits and the
barred bit-values indicate logical qubits. Any $B_p$ applied to this state
commutes through all $1+A_s$ factors and leaves $|0 \cdots 0 \rangle$
fixed, so $|\bar 0 \bar 0 \rangle$ is a +1 eigenstate of all the plaquette
operators.  $|\bar 0 \bar 0 \rangle$ is also fixed by each star operator
because, $A_r$ commutes through all the $1+A_s$ factors until it finds
$1+A_r$, and $A_r(1+A_r)=1+A_r$.  Thus $|\bar 0 \bar 0 \rangle$ can be
taken as a codeword.  To see the topological nature of this state expand
the product as above.  Each term in the sum represents a pattern of
contractible co-loops (loops on the dual lattice) and the sum is equally
weighted over all such patterns.  In this sense, all ``geometries'' are
summed over, leaving only topological information as far as error chains
are concerned.  Operating on $|\bar 0 \bar 0 \rangle$ with any
contractible co-loop of $X$'s just permutes terms of the sum, leaving the
state unchanged.

\begin{figure} 
\centering
\scalebox{.25}{\includegraphics{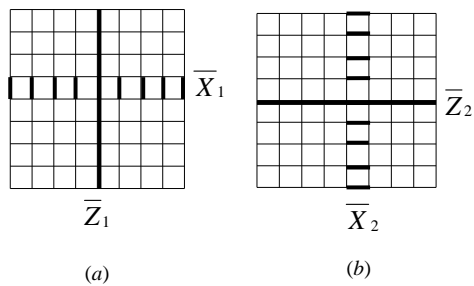}}
\caption{\small
The logical $X$ and $Z$ operators for qubits 1 and 2. Each logical $Z$ is
given by a non-contractible loop of physical $Z$'s, and each logical $X$
by
a non-contractible co-loop of physical $X$'s.
}
\label{logical2}
\end{figure}

More generally, any loop of $Z$'s or co-loop of $X$'s, contractible or
not, commutes with all the stabilizer operators.  If the loop or co-loop
is contractible it fixes all codewords, but if non-contractible it
non-trivially transforms the code subspace.  In fact we can take $X_1|\bar
0 \bar 0 \rangle$, $X_2|\bar 0 \bar 0 \rangle$, and $X_1 X_2|\bar 0 \bar 0
\rangle$ as the three remaining codewords, where $X_i \equiv
\prod_{c_{xi}} X$ is given by a non-contractible co-loop of $X$'s running
across the lattice horizontally along the path $c_{x1}$ or vertically
along $c_{x2}$ (see Fig.\ \ref{logical2}).  (Here the index $i$ refers to
lofical not physical qubits.) Thus $X_1$ and $X_2$ act as the logical
$X$'s for bits 1 and 2.  The logical $Z$'s are given by $Z_i \equiv
\prod_{c_{zi}} Z$, a non-contractible loop of $Z$'s running horizontally
($i=2$) or vertically ($i=1$) across the lattice. Note that $c_{xi}$ and
$c_{zi}$ run in perpendicular directions so that $X_i$ and $Z_i$
anticommute.  Also note that these constructions only depend on topology:  
the paths defining any of these operators may be ``continuously'' deformed
without affecting their action on the code subspace since any such
deformation corresponds to applying a contractible loop operator, which
fixes all codewords.

Suppose we have a state in the code subspace and apply an open co-chain of
$X$'s along some co-path $P$ between faces $q$ and $p$. This changes the
quantum numbers for $B_q$ and $B_p$ from $+1$ to $-1$, generating
``particles'' at $q$ and $p$.  Now the sum-over-geometries is such that
the resulting state would be exactly the same if we had used not $P$ but
some $P'$ which is obtained by ``continuously'' deforming $P$ with its
endpoints fixed. Information about which of the topologically equivalent
co-paths is taken washes away in the superposition because $P$ and $P'$
differ only by a contractible co-loop of $X$'s, which belongs to the
stabilizer.  Likewise, applying a chain of $Z$'s between vertices $r$ and
$s$ generates a dual kind of particle at $r$ and one at $s$, with the same
topological character.  Given a lattice state we can measure all the star
and plaquette operators to obtain a \emph{syndrome} which just lists the
locations of all the particles present on the lattice.  To correct the
errors indicated by the presence of the star (plaquette) particles we must
group all the particles in pairs, connect each pair with a chain
(co-chain) of our own, and apply $Z$ ($X$) operators to the qubits along
these chains (co-chains).  Leaving aside the possibility of measurement
errors, which will be addressed below, this transforms an arbitrary
pattern of errors into a number of closed loops on the (dual) lattice (see
Fig.\ \ref{droplets}).  What we want is that all these closed loops be
contractible so that the logical qubits are left undisturbed. If one of
the loops is non-contractible we will have unwittingly applied one of the
$X_i$ or $Z_i$ operators to our state, causing an error in the encoded
information.

In principle, it only takes $k/2$ errors lying along one non-contractible
(NC) loop to undermine TOR$(k)$ irrespective of our particle pairing
algorithm.  But it would be exponentially improbable as $k$ gets large,
that if just $k/2$ errors occur they would be positioned in just the right
way to do this.  In general, measuring all the star and plaquette
operators will collapse the lattice state into a superposition of
codewords all acted on by a definite set of single qubit phase ($Z$) and
bit flip ($X$) errors. If decoherence/error processes act independently on
separate qubits, and in a relatively uniform way, they will give rise to a
certain probability, $p_z$, for each qubit to undergo a phase error, and
perhaps a different probability, $p_x$, to undergo a bit error.  
Depending on $p_x$ and $p_z$ and on what algorithm we use to pair
particles, there will be some probability that we are tricked into
generating an NC loop when we think we have merely corrected errors.  If
this happens our state is corrupted, but we will see that such a recovery
failure can be made exponentially improbable as $k$ increases, a result
reminiscent of concatenated codes.

\section{Repetition Code as a 1d Lattice Code}

For practice and later reference let us examine the 1d equivalent of
TOR$(k)$, which uses a circle of $k$ qubits instead of a $k \times k$
toric lattice.  The plaquette operators do not exist here, and the star
operator associated with vertex $s$ becomes the product of $X$'s over the
two qubits touching $s$.  In its own right this code, which is dual to
repitition code discussed in Chapter \ref{chap-qi}, is worthless because a
single bit flip error causes a logical bit flip error.  But understanding
the statistics of $z$-error chains will prove useful for analysis of
TOR$(k)$.

Suppose our lattice code state is picked from an ensemble in which each physical qubit suffers a $z$-error with 
probability $p$, independent of all the other qubits. For example, we might have $\mathbf{n}=(0,1,1,0,0,0,1,0)$, which 
describes the 1d lattice (with ends identified)
\[
\mbox{---\,$\leftrightarrow$\,$\leftrightarrow$\,---\,---\,---\,$\leftrightarrow$\,---}
\]
where $\leftrightarrow$ indicates a $z$-error. Measuring the syndrome, we
determine the locations of all $z$-error chain endpoints, in this case
\[
\mbox{---$\cdot$---\,---$\cdot$---\,---\,---$\cdot$---$\cdot$---}
\]
We must now guess which endpoints are connected to which others and apply
our own \emph{recovery chain} of $Z$'s between each pair of ``connected''
endpoints to cancel the errors.  In 1d there are only two possible guesses
corresponding to two complementary patterns of errors on the circle.  So
if we guess wrong the combination of errors and recovery chains will
encircle the lattice, giving $\mathbf{n} = (1,\ldots,1)$ hence a logical
phase error.  Otherwise we will have successfully corrected all the
errors, giving $\mathbf{n}=(0,\ldots,0)$.  Assuming the error probability
$p$ is relatively small, the obvious algorithm for particle pairing would
be to favor the minimum total length of recovery chains. (For a 2d analog
of this minimum distance algorithm, see \cite{11}.) This algorithm,
however, is highly non-local on the lattice; consider instead the
following quasi-local alternative. First pair all particles separated by
only one edge; contested pairings may be resolved randomly.  Then pair any
remaining particles separated by two edges, etc., until all particles are
accounted for.  In 1d this algorithm may produce a number of recovery
chains which overlap hence cancel each other, always resulting in one of
the two basic guesses.

The failure probability $F$, here referring to the probability of causing a logical phase error,
derives from the set of all possible error configurations which can trick
the algorithm into forming an NC loop of
$z$-errors.  In particular we have the bound
\begin{equation}\label{eq:1} 
F  \; = \; \langle n_i \rangle \; \le \;
k\sum_{n=n^{(k)}}^{\infty} h_1(n)p^n
\end{equation}
where $h_1(n)$ is the number of different error chains that the algorithm
can generate with a fixed number $n$ of $z$-errors and starting from a
fixed vertex. In other words, $h_1(n)$ is the number of ways $n$
$z$-errors can trick the algorithm into flipping all the bits inbetween
instead of correcting the erroneous bits themselves.  The lower limit
$n^{(k)}$ in the sum is the fewest number of errors necessary to cause the
algorithm to generate an NC loop on a circle of size $k$. The ensemble
average $\langle n_i \rangle$ refers to an arbitrary component of
$\mathbf{n}$, evaluated \emph{after} recovery chains have been applied.
One might then expect $F$ is a kind of order parameter describing the
topological order of error chains on the lattice. We shall see that for
$p$ below a certain critical error rate $p_c$, our recovery algorithm
maintains the lattice in a highly stable phase where $\langle n_i \rangle
\ll 1$ so NC loops are very unlikely. In the thermodynamic limit
$k\rightarrow \infty$, $\langle n_i \rangle = 0$ in this phase, but what
we want to know is exactly how small $\langle n_i \rangle$ is as a
function of $k$.
  
To study $F$ let us first calculate $n^{(k)}$, or equivalently calculate
the maximum length $l(n)$ of an $[n]$-chain---that is, an error chain
generated by our algorithm and containing $n$ errors.  Clearly $l(2)=3$,
since two lone errors can be separated by at most one edge if they are to
be paired by our algorithm.  This makes the $[2]$-chain
$\leftrightarrow$\,---\,$\leftrightarrow$ where the middle link is a
recovery chain. Now if we take two of these $[2]$-chains and join them
through the longest possible recovery chain (itself 3 edges long), what we
have is the longest possible $[4]$-chain.  We can continue to build up
maximal $[2^L]$-chains in this highly symmetrical, Cantor set pattern, and
we find $l(2^L) = 3^L$.  Generating an NC loop requires an error chain of
length at least $k/2$, so if $k/2$ is a power of 3 we have $n^{(k)} =
(k/2)^\beta$ where $\beta = \log_3 2 \approx 0.6309$.  If $k/2$ is not a
power of 3, the maximum chain will have to involve asymmetric joining
processes, which serve only to decrease its length relative to the Cantor
chain trend. Thus $l(n)=n^{1/\beta}$ serves as an upper bound on chain
length in general, but it will prove useful to have an explicit expression
when $n$ is inbetween powers of 2.

Consider the sub-chain structure of the maximal $[2^L-2^M]$-chain. We may emulate the Cantor pattern by dividing the 
$2^L-2^M$ errors into two identical $[2^{L-1}-2^{M-1}]$-chains and extending the longest possible recovery chain 
between 
them. Iterate the process for each of these two chains, etc., until we have reduced the lot into 
$[2^{L-M}-1]$-chains and can go no 
further. Now it is not hard to determine $l(2^N-1)$. As the $2^N-1$ errors join in successive levels, they look just like 
a 
Cantor chain, except at each level there is always one runt sub-chain shorter than the rest. At the first level, 
$[1]$-chains join in pairs to become $[2]$-chains
($\leftrightarrow$\,---\,$\leftrightarrow$), except one is left unpaired 
resulting 
in the runt $[1]$-chain at the second level. Now the $[2]$-chains join in pairs, except one joins the runt giving the 
runt 
$[3]$-chain
($\leftrightarrow$\,---\,$\leftrightarrow$\,---\,$\leftrightarrow$),
etc. The number of edges lost at each level 
relative to the corresponding Cantor chain are as follows: 2 edges at the first level; another 2 edges at the second; and 
at an arbitrary level, a number of edges equal to the sum of all previous losses. Summing 
the 
series yields a 
total relative loss of exactly $2^N$ edges, so $l(2^N-1) = l(2^N)-2^N = 3^N-2^N$. Taking all 
of 
our 
$[2^{L-M}-1]$-chains as units in one big Cantor pattern, one finds
\begin{equation} \label{l(n)}
l(n) = (3^{L-M} - 2^{L-M})3^M = 3^L - \left({\textstyle \frac{3}{2}}\right)^M 2^L,
\end{equation}
which is the desired expression for maximal chain length when $n=2^L-2^M$ is inbetween powers of 2, giving the correct 
results for the limits $M=1,L-1$. Note $l(n) \le n^{1/\beta}$ with equality when $n$ is a power of 2. 

To bound the chain counting function $h_1(n=2^L)$, consider all ways an $[n]$-chain can be decomposed 
into an $[m]$-chain $S$ and an $[n-m]$-chain $S'$ joined by a recovery chain $R$. Neither $S$ nor $S'$ can contain 
any recovery chains longer than $R$, which means that $S$ cannot contain any recovery chains 
longer than $S'$ itself and vice versa. Thus we can write
\begin{equation}\label{eq:2}
h_1(n) \le \sum_{m=1}^{n-1} h_1(m|m_<) h_1(n-m|m_<) \cdot (m_<^{1/\beta}+1)
\end{equation}
where $m_<$ is the lesser of $m$ and $n-m$, and ``$|m)$'' reads ``given that there are no recovery chains longer than 
$m^{1/\beta}$,'' which is the maximum length of an $[m]$-chain. The factor 
$m_<^{1/\beta}+1$ counts all possible recovery chains $R$, including the ``0-chain.'' To bound the sum, 
let us find the maximum value of $h_1(m|m_<)h_1(n-m|m_<)$ over all possible $m$ or, without loss of generality, over 
$1\le 
m 
\le n/2$. In general we expect the number of different 
chains to increase with increasing chain length, so we should find the $m$ which allows for the maximum possible summed 
length $l_{SS'}$ of $S$ and $S'$, corresponding to the two $h$ factors. For a given $m$ we have
\[
l_{SS'}(m) = l(m) + l(n-m|m).
\]
We know $l(m)$ from (\ref{l(n)}), and we can calculate $l(n-m|m)$ by finding 
the error configuration which saturates the ``$|m)$'' constraint. This is done by dividing the $n-m$ errors into groups 
of 
$m$ errors, arranging errors within each group in the Cantor form, and linking these groups together through recovery 
chains of maximum 
length. 
Together with (\ref{l(n)}), and using $l(m)=m^{1/\beta}$, this yields
\begin{equation}\label{lss}
l_{SS'}(m) = 2\frac{n-m}{m}l(m) = \frac{2}{m}\left[(n-m)n^{1/\beta}-n(n-m)^{1/\beta}\right].
\end{equation}
It is straight-forward to show this function is strictly increasing over $1\le m\le n/2$, so that the maximum is achieved 
at $m = n/2$, which choice should then also maximize $h_1(m|m_<)h_1(n-m|m_<)$. Using (\ref{eq:2}) and the fact that 
$h_1(m|m) = h_1(m)$ we have 
\[
h_1(n) \le \Sigma_1(n)h_1(n/2)^2 \;\;\;\;\; \mbox{where} \;\;\;\;\; \Sigma_1(n) \equiv \sum_{m=1}^{n-1} 
(m_<^{1/\beta}+1).
\]
Iterating the bound yields
\begin{equation}\label{eq:3}
h_1(n) \le \left[h_1(1)\prod_{L=1}^{\infty}\Sigma_1(2^L)^{2^{-L}} \right]^n = (8.872\ldots)^n. 
\end{equation}
with the aid of some numerical evaluation.  We might have put $h_1(1)=2$ but instead use $h_1(1)=1$ because 1d error
chains cannot double-back on themselves.  (At a chain's starting point $h_1(1)=2$ holds, but this has exponentially small
effect for a long chain.)  Now (\ref{eq:1}) implies a concise bound on the (phase error) failure probability for 
this 1d algorithm:
\begin{equation}\label{eq:4}
F \le k \left(\frac{p}{p_c}\right)^{(k/2)^\beta}
\end{equation}
where the actual accuracy threshold $p_c$ is no less than $1/8.872$. 

\section{Recovery with Perfect Measurements}

We must now extend the algorithm to 2d (again assuming no measurement
errors), so that something like (\ref{eq:4}) applies to both $z$-errors
and $x$-errors. To simplify analysis we make no use of correlations
between phase and bit flip errors, so $x$-error correction on the dual
lattice is formally identical to $z$-error correction on the lattice, and
only the latter is addressed below.

\begin{figure} 
\centering
\scalebox{.2}{\includegraphics{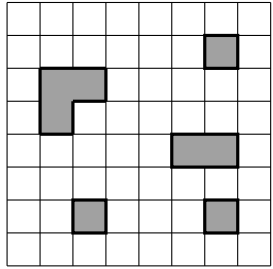}}
\caption{\small
At the end of one recovery round, all open error chains have been
transformed into closed loops on the lattice. Recovery is successfull if,
as above, all these loops are contractible. 
}
\label{droplets}
\end{figure}

``Two particles separated by a distance $l$ on the lattice'' means that
the shortest path between them contains $l$ edges. The locus of vertices
equidistant from a given vertex looks like a diamond.  So, given a
particle $s$ in the algorithm's $t$-th step, we need to search for
partners over all vertices on a diamond of radius $t$ centered on $s$.  
As the algorithm proceeds from $t=1$, error chains close into loops and
join with one another until no open chains are left (see Fig.\
\ref{droplets}).

A bound on the failure probability is obtained as before, but we must calculate a new chain counting function $h_2(n)$
since a given 1d chain may wander across the 2d lattice along many different paths.  Consider an $[m]$-chain with 
endpoints $r$ and $s$, which is to join an $[n-m]$-chain with endpoints $r'$ and $s'$. If the joining occurs through $s$ 
and $s'$, then $s$ must be closer to $s'$ than to $r$. So, taking $s$ 
fixed, $s'$ must be somewhere within the diamond centered on $s$ and passing through $r$. This diamond has radius at most 
$m^{1/\beta}$ hence contains at most $2m^{1/\beta}(m^{1/\beta}+1)+1$ vertices.  Thus in 2d we have
\begin{equation}
h_2(n) \le \Sigma_2(n)h_2(n/2)^2 \;\;\;\; \mbox{where} \;\;\;\; \Sigma_2(n) \equiv \sum_{m=1}^{n-1} 
2m_<^{1/\beta}(m_<^{1/\beta}+1)+1
\end{equation}
and iteration gives
\[
h_2(n) \le \left[h_2(4)^{1/4}\prod_{L=3}^{\infty} \Sigma_2(2^L)^{2^{-L}} \right]^n = (75.38\ldots)^n
\]
with the aid of some numerical evaluation. Note we have halted iteration after reaching $h_2(4)$ in order to improve the 
bound. We have bounded $h_2(4)$ itself by using diagrams to
count all possible $[4]$-chains. For example,
\newcommand{\FR}[2]{{\textstyle \frac{#1}{#2}}}
\[
(\leftrightarrow\,\leftrightarrow)(\;\cdot\; + \;\mbox{---}\; +
\;\mbox{---\,---}\;)(\leftrightarrow\,\leftrightarrow) = 
(4\cdot3)(1+3+7\cdot\FR{4}{3}\cdot\FR{1}{4})(3\cdot 3)
\]
is the contribution to $h_2(4)$ from the joining of two $[2]$-chains, each length 2. The numbers arise as follows: 
we start at some fixed vertex and have 4 choices for positioning the first error, leaving 3 choices for the second 
error. 
Our recovery chain may have length 0, 1, or 2, giving a number of choices equal to 1, 3, or 7 respectively. Then we have 
$3\cdot3$ ways to position the next two errors. If the recovery chain is two edges long, however, there are $4\cdot3$ 
ways 
to position these two errors, hence the factor of $\frac{4}{3}$ above. The factor $\frac{1}{4}$ arises from the fact that 
pairing ambiguities are resolved randomly by the recovery algorithm. If the recovery chain has length 2, there is only a 
1/4 chance that the $[2]$-chains will join as above (for this to happen, one of the two interior vertices must be chosen 
first for pairing, and then it must be paired with the other interior vertex). We may compute three other diagrams 
allowing 
for either of the $[2]$-chains to have length 3, and we obtain the bound $h_2(4) \le 4997$. (In counting arrangements of 
a 
length 3 chain we consider two separate cases, namely when the endpoints are separated by one edge and by three edges.)

The failure probability bound in 2d is thus
\begin{equation}\label{eq:5}
F \; \le \; k^2 \!\! \sum_{n=(\frac{k}{2})^\beta}^{\infty} h_2(n)p^n = k^2
\left(\frac{p}{p_c}\right)^{(k/2)^\beta}
\end{equation}
with critical probability $p_c \ge 1/75.38$. This result applies equally to $x$-error and $z$-error correction.

However, we shall see that this represents an over-estimate in regard to the exponent $(k/2)^\beta$.  The bound
(\ref{eq:5}) was derived by assuming that all chains of $(k/2)^\beta$ or
more errors generate an NC loop, hence result in
failure.  But, for instance, of all the chains with $n=(k/2)^\beta$ errors
only a few can generate an NC loop because 
every
error must be placed in just the right spot along a perfectly straight line for the chain to achieve the necessary 
endpoint
separation.  In general, the fraction of chains starting from a given
vertex and capable of generating an NC loop on our
$k\times k$ lattice will be some function $f(n)$, tending to unity as $n$ gets large.  This function should multiply the
chain counting function $h_2(n)$ in our failure probability bound (\ref{eq:5}).

We know the $n^{(k)}=(k/2)^\beta$ contribution to $F$ should go like $p^{n^{(k)}}$ because once a starting point is 
picked,
the positions of the $n^{(k)}$ errors are essentially all fixed if the
chain is to generate an NC loop.  Thus 
$f(n^{(k)})=(p_c)^{n^{(k)}}$ to get the right term for $n = n^{(k)}$ in (\ref{eq:5}).
Also $f(n) \le 1$ by definition, so we can bound
\[
f(n) \le \left[p_c + \frac{n^{1/\beta} - k/2}{n_0^{1/\beta} - k/2} (1-p_c)\right]^n
\]
for some $n_0 > (k/2)^\beta$.  Using this in (\ref{eq:5}) with $h_2(n)$ multiplied by $f(n)$ and finding the maximum term
in the sum allows us to sharpen (\ref{eq:5}) insofar as $n_0$ exceeds $(k/2)^\beta$.  Physically $n_0$ represents the
saturation point at which adding one more error to a chain stops having so great an effect on the chances of its being 
able
to generate an NC loop.  To get a hold on the value of $n_0$ first
consider
only geodesic error chains---chains of 
extremal 
length for 
fixed endpoints.
Statistically this category will be dominated by nearly diagonal chains. But a diagonal chain must have length at least 
$k$ 
to generate a 
NC loop, so adding one more error will be irrelevant
only if $n \ge k^\beta$.  In general, error chains will be sub-geodesic, so that we expect the saturation point $n_0$ to 
exceed $k^\beta$.  Using this as a bound, we find the maximum term in the sum of (\ref{eq:5}), if it exists, satisfies
\[
\frac{p}{p_c} \le \frac{k}{2l-k} \exp\left[-\frac{l}{\beta(l-k/2)}\right] 
\]
where $l \equiv n^{1/\beta}$ and we have neglected $O(p_c)$ corrections.  If this is satisfied by no $l < k$, the end 
term 
with $l=k$ is the maximum term.  Since the above function is strictly increasing on $0\le l \le k$, this occurs
if $p/p_c$ exceeds the right hand side above evaluated at $l=k$, which is
$e^{-2/\beta} \approx 0.0420$.  So we have the 
estimate
\begin{equation}\label{eq:6}
F \sim k^{2+\beta} \left(\frac{p}{p_c}\right)^{k^\beta}
\end{equation}
where the exponent $k^\beta$ applies rather than $(k/2)^\beta$ if $p\ge
0.0420\,p_c$. 

\begin{figure} 
\centering
\scalebox{.5}{\includegraphics{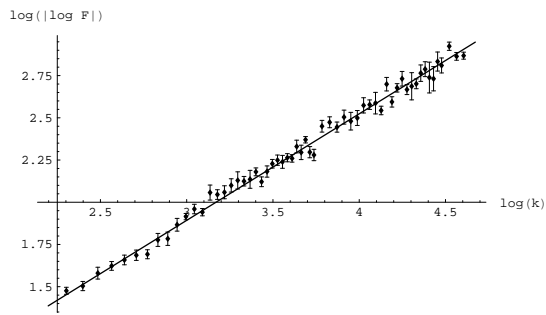}}
\caption{\small
Recovery failure rates $F$ as a function of lattice size $k$, obtained
from Monte Carlo simulation of errors. The vertical axis is taken as the
logarithm of $k^\beta$ values extracted from failure rate data and is
effectively a double logarithm of $F$. The line represents our analytic
result for the exponent $\beta = \log_3 2$.
}
\label{noghosts}
\end{figure}

We have sought to test these results through numerical simulations of the
recovery process. For a given torus size $k$, we perform $\sim 10,000$
individual recovery simulations for eight values of $p$ from 0.01 to 0.07.  
Each Monte Carlo run starts by generating a random pattern of errors, each
edge with error probability $p$, and implements the expanding diamonds
algorithm until all particles are paired.  Recovery success or failure is
determined by checking for NC loops.  For each $p$, $F$ is just given as
the failure frequency, which we fit with (\ref{eq:6}) as a function of
$p$, for fixed $k$, yielding the $k^\beta$ exponent as a fitted paramter
value in a log-log plot. (Here we neglect the prefactor $k^{2+\beta}$.)
The logarithms of these extracted $k^\beta$ values are plotted against $k$
in Fig.\ \ref{noghosts}. According to (\ref{eq:6}), the result should be a
line with slope $\beta = \log_3 2 \approx 0.6309$.  The measured slope is
quite close:  $0.627\pm 0.008$.  The intercept---predicted as the log of
the coefficient of $k^\beta$ in (\ref{eq:6}), hence zero---is measured to
be $0.02\pm 0.03$. Measured values of the accuracy threshold $p_c$ for
each $k$ are all comfortably consistent with the bound $p \ge 1/75.38$
obtained above.

In picking the data to fit, we must select a maximum $p$ (or,
equivalently, $F$) since the actual scaling for which (\ref{eq:5}) is a
bound must break down at some $p$ greater than the bound obtained for
$p_c$.  Here, a cut-off at $F=.05$ was applied.  We also select a minimum
$F$ to limit Poisson scatter, chosen to minimize the standard error for
our measured value of $\beta$. Scatter in the plot arises not only from
Poisson fluctuations but also from the fact that actual values of $\beta$
for finite $k$ differ from the theoretical value $\log_3 2$, which is
really an asymptotic ($k\rightarrow \infty$) prediction. These finite $k$
effects involve the interpolation which must take place between Cantor
chains whose lengths are all powers of 3. 

\section{Recovery with Imperfect Measurements}

Until now we have assumed perfect syndrome measurements.  But suppose we
err in measuring each star operator $A_s$ with some probability $q$ (which
might be the same order as the physical bit error probability $p$).  This
mistake would lead us to think a particle (``ghost particle'') exists at
$s$ when there really is none, or that no particle exists at $s$ (``ghost
hole'') when there really is one.  Because the basic expanding diamonds
algorithm becomes unstable when ghosts are introduced, we must modify it
and apply our failure probability analysis (chain counting, etc.)
to the modified version.

Imagine recovery (with measurement errors) via expanding diamonds.  We
would generate contractible loops, hence correct real errors, but recovery
chains would also connect ghost particles to one another and to real
particles.  Once a chain connects to a ghost particle it can no longer
propagate from that endpoint because there is no pre-existing error chain
to continue it to another particle.  So in addition to all the loops
generated by recovery, the lattice would also be left with open chains
that carry over to the next round as if they arose from spontaneous
errors.  Failure might occur, as before, by the generation of an entire
NC loop in one recovery round or, now, over many rounds.

Unfortunately, the left-over chains quickly begin to dominate the failure
rate.  Suppose $q$ were small enough that in a particular round just two
ghosts occurred.  They would typically be separated by a distance O$(k)$
on the lattice.  Since they are the only ghosts around, and open chains
end only on ghosts, these two will be connected by a recovery chain, which
will give an O$(1)$ chance of failure in the next round, independently of
$k$.  We cannot remedy this situation simply by repeating syndrome
measurements a number of times to increase confidence in their results.
The reason is that no matter how many times we repeat, there will always
be processes involving just a few errors and ghosts (hence occurring with
bounded probability) that corrupt the supposedly verified syndrome. For
instance suppose, exactly half-way through a series of repeated rounds of
syndrome measurement, a real error occurs with endpoints $r$ and $s$, but
we err in measuring $A_r$.  Majority voting after the final round would
trick us into accepting $s$ as a real particle, but not $r$, effectively
generating a ghost in our verified syndrome.

So we need a better algorithm.  The first thing to realize is that we will
inevitably leave open chains behind from one round to the next.  The only
way to prevent ghosts from generating long chains is to be suspicious of
calls to connect widely separated particles.  Suppose we are lead to
consider generating a recovery chain between two particles separated by a
distance $l$ in the $T$th round.  Should we do it?  If $l$ is large, the
hypothetical chain is more likely a pair of ghost particles.  But age is
also important:  the longer the particles have been around (left
unconnected in previous rounds), the less likely they are to be ghosts.  
To keep track of particle age information, imagine a 3d lattice
comprising 2d shelves representing successive rounds of syndrome
measurement.  Particles which are the endpoints of left-over chains will
be registered from their birth to the present, forming vertical ``world
lines."  Pairing particles in the current round should be done by
reference both to their spatial separation on the current 2d shelf and
also their \emph{temporal} separation, i.e., the number of rounds
having elapsed between their respective births.

Ghosts can eclipse particles or join onto chains themselves, either
way causing an age discrepancy between chain endpoints.  However we attempt to correct these types of errors, there is
always the additional possibility that we generate more of them ourselves.  As we shall see, the kinds of processes which
result in chains with large spatial displacements in 2d have analogs in 3d which generate large temporal displacements.
And as before, the more defects (now including ghosts), the greater these separations can be.  This suggests we treat
temporal and spatial separations on the same footing:  the algorithm should connect particles according to some definite
combination of their spatial and temporal separations.  A natural generalization of expanding diamonds is found by
extending the 2d spatial metric on the lattice into a 3d space-time metric:
\begin{equation}\label{lstar}
l_\ast = l + \alpha |\Delta T|
\end{equation}
defines the 3d distance $l_\ast$ in terms of the spatial and temporal displacements $l$ and $\Delta T$.  Diamonds in 2d
become octahedra in the 3d lattice---an octahedron of radius $l_\ast$ being defined as the locus of points separated 
from
a given vertex by a distance in $(l_\ast-1,l_\ast]$ according to the ``$\ast$-metric.'' Note these octahedra are squashed
in the time direction by the factor $\alpha$.  The value of $\alpha$ should be chosen according to the frequency of
measurement errors relative to real errors.  The smaller the rate of measurement errors, the less probable it is to
generate age differences, so $\alpha$ should be higher.

At each step in a given round of recovery, scaled octahedra of fixed size
are extended around the birth sites of particles currently available.  
Once a given particle's octahedron encounters another particle's birth
site, the particles are paired and marked as ``unavailable.'' In 2d, a
recovery chain would be applied between every particle pair.  Now that is
inadvisable due to the presence of ghosts.  Having paired two particles
$r$ and $s$, we should determine whether it is more probable that either
(i) they are associated with two independent error chains whose other
endpoints may have been obscured by ghosts, or (ii) $r$ and $s$ are in
fact endpoints of the same chain. These probabilities are determined by
the number of defects necessary to account for $r$ and $s$ under the
assumption (i) or (ii), so we should find a 3d analog of the 2d result
that at least $l^\beta$ errors are necessary to generate a chain of length
$l$.  As we shall see, the obvious generalization is approximately
correct:  $l_\ast^\beta$ effective defects, accounting through $\alpha$
for the different occurrence probabilities of real errors and ghosts, are
necessary to generate a chain of length $l_\ast$ in the $\ast$-metric.  
In addition we will find, as would be expected, that at least $\sim
T^\beta$ effective defects are necessary to maintain a chain in existence
for $T$ rounds of recovery.  These two results allow us to compare the two
probabilities associated with cases (i) and (ii) above. This is done by
comparing the number of effective defects required for each case, which
are $T_r^\beta + T_s^\beta$ and $l_\ast^\beta$ respectively.  Here $T_r$
and $T_s$ are the ages of the two particles $r$ and $s$, and $l_\ast$ is
their $\ast$-metric separation.  Once the 3d algorithm is done pairing
particles, we apply a recovery chain between any given pair $r$ and $s$
whenever
\begin{equation}\label{eq:6.1}
l_\ast^\beta < T_r^\beta + T_s^\beta,
\end{equation}
which imposes a variable pairing-length cut-off on the algorithm.

Ambiguities in this 3d algorithm can arise in the process of identifying a
current particle with a particular birth 
site.
If errors occur on edges touching the original birth site, the particle's vertical world line may continue on a vertex
displaced from the original.  Also, ghosts may eclipse a particle in a given set of rounds, leaving holes in its world
line.  These difficulties may be overcome on a round-to-round basis by simply requiring that the age ascribed to a vertex 
be
conserved if its particle has been left over from previous rounds, i.e., has not yet been paired.  If a left-over
particle suddenly disappears, we probe with expanding diamonds around the eclipsed particle until we find an uneclipsed
particle who could inherit the lost age.  If the probe radius becomes large enough that the likelihood of eclipse due to
ghost overtakes the likelihood of eclipse due to real errors, we conserve age by manually adjusting our syndrome record 
as
if we had detected a particle at the vertex in question.  (We continue to alter the syndrome by hand, if need be, until 
it
becomes more likely that the hypothetical eclipsed particle is actually just a string of ghosts.)

Having now specified an algorithm in 3d, we must redo our failure rate analysis taking into account the time dimension 
and
the leaving over of chains from one round to the next.  Our method is basically the same as before, but the chain 
counting
function $h_2(n)$ must be generalized to $h_3(n,\bar n)$, where $n$ is still the number of real errors and $\bar n$ the
number of ghost errors involved in the chain.  The failure probability bound now becomes
\begin{equation}\label{eq:6.2}
F \le k^3\sum_{n,\bar n} h_3(n,\bar n)p^{n}q^{\bar n}
\end{equation}
where the sum is taken over all pairs $(n,\bar n)$ capable of generating
an NC loop.  Note that $h_3(n,\bar n)$ counts
chains involving errors which may have originated in previous rounds but have lasted through the present.  We again 
obtain
a recursion relation, now for $h_3(n,\bar n)$ in terms of $h_3(m \le n,\bar m \le \bar n)$, by considering all ways an
$[n,\bar n]$-chain could be broken into an $[m,\bar m]$-chain $S$ and an $[n-m,\bar n - \bar m]$-chain $S'$.  Recall that 
in 1d the 
coefficient $(m_<^{1/\beta}+1)$ in the recursion relation
(\ref{eq:2}) counted all the ways to choose the recovery chain connecting $S$ and $S'$.  In 2d this
coefficient became the area of a diamond of radius $m_<^{1/\beta}$.  And now in 3d it becomes the volume (in vertices) 
of
a $\ast$-metric octahedron with radius $l_\ast(m,\bar m)_<$, defined as the lesser of the two maximal $\ast$-metric 
lengths 
$l_\ast(m,\bar m)$ and 
$l_\ast(n-m,\bar
n - \bar m)$. The new recursion relation is
\begin{equation}\label{eq:7}
h_3(n,\bar n) \le \sum_{m=0}^{n}\sum_{\bar m=0}^{\bar n} V(l_\ast(m,\bar m)_<) h_3(m,\bar m) h_3(n-m,\bar 
n-\bar m).
\end{equation}

Again we want to bound the sum by finding the maximum $h \cdot h$ term, now varying both $m$ and $\bar m$. By 
generalizing 
the 1d/2d relation $l(n)=n^{1/\beta}$ we will later see that chains can 
grow
longest when ghosts are uniformly intermixed with real errors.  It turns out they work best by cooperating, as opposed 
to,
say, having all the real errors combine on one side of the chain and all the ghosts on the other.  So a maximal chain, 
hence the maximum $h\cdot h$ term, must have uniform composition, $m/\bar m = n/\bar n$.  Thus we can perform a 
calculation 
similar to that which gave (\ref{lss}), but with $l(m) \rightarrow l_\ast(m,(\bar n/n)m)$. In fact we need just observe, 
as 
will be shown later, that $l_\ast(m,(\bar n/n)m)$ grows faster with $m$ than does $l(m)$. Now $l_{SS'} \rightarrow 
l_{\ast 
SS'}$ is determined by the first equality in (\ref{lss}), which still
holds in 3d, so that if $l_{SS'}(m)$ was 
increasing 
on $1\le m \le n/2$ in 1d/2d, so must be $l_{\ast SS'}$ in 3d. Thus the maximum is located at $m = n/2$, hence $\bar m 
= 
\bar n/2$, and (\ref{eq:7}) becomes
\begin{equation}\label{eq:8}
h_3(n,\bar n) \le \Sigma_3(n,\bar n)h_3(n/2,\bar n/2)^2,
\end{equation}
where
\[
\Sigma_3(n,\bar n) \equiv \sum_{m=0}^{n} \sum_{\bar m=0}^{\bar n} V(l_\ast(m,\bar m)_<)
\]
and the octahedral volume is given by
\[
V(l_\ast) = \sum_{\Delta T=-[l_\ast/\alpha]}^{[l_\ast/\alpha]} 2\left[l_\ast-\alpha |\Delta 
T|\right] \left[l_\ast-\alpha |\Delta T| + 1\right]+1,
\]
with $[\cdots]$ denoting the greatest integer function.  The chain counting function, hence the critical probabilities we
shall soon bound, depend crucially on the function $l_\ast(n,\bar n)$ which bounds the $\ast$-metric length of a chain
containing $n$ real errors and $\bar n$ ghosts.  As there is no compact expression in general, we need to investigate
particular values of $n$ and $\bar n$.

The basic constraint on the length of an error chain is that none of its recovery chain components can be longer
($\ast$-metric) than either of the sub-chains which it joins.  Consider the case $\bar n=0$.  Even without any ghosts, 
the
chain has extra freedom in the 3d lattice.  Two purely spatial sub-chains may be joined by a purely spatial recovery 
chain
(not exceeding either of their lengths), or they may trade space for time so that one chain occurs in a recovery round
before the other.  But no extra $\ast$-metric length can be gained by trading space for time, because for any 
``time-like''
recovery chain there is always a ``space-like'' recovery chain of equal or greater length.  This implies $l_\ast(n,0) =
n^{1/\beta}$ just as in 2d.  Now consider adding one ghost to a pre-existing error chain.  If, for instance, a ghost at
$s$ is connected to one of the chain's endpoints, $s$ will show up as a new-born particle in the \emph{next} round (which
would not be the case had the ghost been a real particle, hence the endpoint of another chain).  Thus, the ghost 
generates
an age difference between the endpoints of that chain.  Depending on the value of $\alpha$, the algorithm might permit a
newborn ghost to join onto an older chain, causing a greater age difference.  To simplify analysis let us fix $\alpha$ so
that no newborn (i.e., age 1) particle may be joined with a particle of age greater than 2.  Consider two 
particles
of ages 1 and 3, separated by one edge.  The condition that they cannot be joined by the algorithm is given by the 
pairing
length cut-off (\ref{eq:6.1}) as $(1+2\alpha)^\beta > 3^\beta + 1^\beta = 3$, comfortably satisfied by choosing
$\alpha=2.4$.  From this it follows by checking cases that the most an add-on ghost can extend a chain is to add a 
spatial
separation of 2 edges and a final age difference of 2 rounds.  If $n \gg \bar n$, the maximal chain has a ``unit cell''
comprising $n/\bar n$ real errors arranged in a Cantor chain with one ghost added on to the end.  Unit cells are strung
together as singe error units in the Cantor pattern, giving
\[
l_\ast(n \gg \bar n) = n^{1/\beta} + (2\alpha + 2)\bar n^{1/\beta}
\]
which retains the basic $1/\beta$ scaling exponent.  Considering $n$ and $\bar n$ powers of two, one can check that this
relation holds for $n \ge 4 \bar n$.  If $n$ and $\bar n$ in this range are not powers of two, the above may be taken as 
a
bound.  Also, one may experiment with smaller unit cells to obtain
\[
l_\ast(n=2 \bar n) = (3+2\alpha)\bar n^{1/\beta} \;\;\;\;\;\; \mbox{and} \;\;\;\;\;\; l_\ast(n=\bar n) = 
(2+\alpha) \bar 
n^{1/\beta}.
\]
We may interpolate between the above three results by an appropriate step-function to achieve a bound on $l_\ast(n,\bar 
n)$
for all $n \ge \bar n$.  For $n=0$, one can investigate candidate maximal chains with a definite number of ghosts per 
unit
cell.  Each unit cell is gotten by saturating the cut-off (\ref{eq:6.1}).  It turns out the true maximal chain has six 
per
unit cell and scales according to
\[
l_\ast(0,\bar n) = (7+8\alpha)(\bar n/6)^{1/\beta}.
\]
For $n \ll \bar n$, the maximal chain unit cell has $\bar n/n$ ghosts and one real error.  It may be divided into many
sub-units, each comprising six ghosts, except for one odd sub-unit which also has the one real error.  Actually, we can
make the unit cell a bit longer by giving one of the six ghosts in the odd sub-unit to a different sub-unit.  This
construction gives
\[
l_\ast(n \ll \bar n) = (6+2\alpha)n^{1/\beta} + (7+8\alpha)(\bar n/6)^{1/\beta},
\]
which holds as a bound for $n < \bar n/6$.  The $\ast$-metric lengths of chains for $\bar n/6 \le n < \bar n$ may be
obtained by inspetion when $\bar n/n$ is integral.  We will not overtax the reader with all these formulas.  Again,
$l_\ast$ is bounded by interpolating to a step-function for intermediate cases.  Altogether we have a means of bounding
$l_\ast(n,\bar n)$ for any values of its arguments.  Note that these expressions for $l_\ast(n,\bar n)$ have been 
obtained
by mixing the $n$ real errors and $\bar n$ ghosts uniformly, hence the ``unit cells.'' That uniform mixture maximizes
$\ast$-metric length can be checked by comparison to the lengths, computed
with the above formulas, of segregated chains
comprising, e.g., one piece with only real errors and another with only ghosts.

Now that we have a handle on all the quantities involved in our recursion relation (\ref{eq:8}), let us use it to bound
$h_3(n,\bar n)$.  First consider the case $\hat n \equiv n/\bar n \ge 1$.  Recursion brings us down from $h_3(n,\bar n)$ 
to
the factor $h_3(n/\bar n,1)$, assuming both $n$ and $\bar n$ are powers of two.  And we have
\begin{equation}\label{red}
h_3(\hat n,1) \le 2 \, \Sigma_3(\hat n,0) \, h_3(\hat n/2,0) \, h_3(\hat n/2,1),
\end{equation}
expressing a division into two sub-chains of equal defect number, except that one has a ghost and the other does not.  
The
factor of 2 arises because the one ghost may be put in either of the two sub-chains.  We may recursively substitute for
$h_3(\hat n/2,1)$ in this relation to obtain an expression involving no $h$ factors other than $h_3(2^N,1)$ and those of 
the form $h_3(m,0)$. (The integer $N$ may be chosen freely.) The $h_3(m,0)$ terms may all be reduced to $h_3(2^N,0)$ 
using 
the original relation (\ref{eq:8}) with $\bar n=0$.
When account is taken of all the $\Sigma_3$ factors produced by these recursions, one finds the quantity $q^{\bar n} 
h_3(n
\ge 2^N \bar n)$ is bounded from above by
\begin{equation}\label{eq:9}
\left\{\left[\frac{\hat n q}{2^N} \frac{h_3(2^N,1)}{h_3(2^N,0)}
\prod_{L=N+1}^{\infty} \Sigma_3(2^L \hat n,2^L)^{2^{-L}} \right]^{\frac{1}{\hat n}}
h_3(2^N,0)^{2^{-N}} \prod_{L=N+1}^{\log_2 \hat n} \Sigma_3(2^L,0)^{2^{-L}} \right\}^n
\end{equation}
(For $\hat n \le 2^N$ the second product over $L$ should be set to unity.) Note that the bracketed expression
$\{\cdots\}$ has the form $g(\hat n,q)$, depending on $n$ and $\bar n$ only through $\hat n \equiv n/\bar n$.  Strictly, 
we
have obtained (\ref{eq:9}) only for $\hat n$ a power of two.  Calculating intermediate cases, we would expect to find a 
correction to (\ref{eq:9}) resembling the correction (\ref{l(n)}) to our 1d/2d scaling law $l(n) = n^{1/\beta}$.  We 
can now express the $n \ge \bar n$ part of the sum in (\ref{eq:6.2}) as
\[
\sum_{n \ge \bar n}(q^{\bar n}h_3(n,\bar n))p^n = \sum_{\hat n \ge 1}\sum_n (g(\hat n,q)\,p)^n.
\]
where the first sum is only over pairs $(n,\bar n)$ capable of generating
NC loops, which we have converted to a sum 
over
$\hat n,n$.  For a given $\hat n$ the sum over $n$ begins at a definite value $n=(k/2\gamma(\hat n))^\beta$, which is 
determined by
one of the $l_\ast(n \ge \bar n)$ formulas as the minimum $n$ such that $n$ real errors together with $\bar n = n/\hat n$
ghosts can generate a chain of spatial length $k/2$.  In particular, $\gamma(\hat n)$ is maximum at $\hat n = 1$ where 
the
$l_\ast(n=\bar n)$ formula gives $\gamma = 2 + \alpha$.  Considering the sum over $n$ as already performed, terms in the 
remaining sum 
over $\hat n$ depend on $p$,
$q$, and $\hat n$ alone.  Fixing $p$ and $q$, the maximum term will occur at some $\hat n = \hat n(q,p)$, which then
determines an asymptotic ($k\rightarrow \infty$) critical probability $p_c(q,p) = 1/g(\hat n(q,p),q)$ and scaling 
exponent 
through 
$\gamma(\hat n(q,p))$.

The case $n \le \bar n$ follows in the same way, and a bound is obtained
for the quantity $p^{n} h_3(n \le \bar n)$ which is exactly the expression
(\ref{eq:9}) with $q \leftrightarrow p$ so that the two arguments are
interchanged in all the $h$ and $\Sigma$ functions, $n \leftrightarrow
\bar n$, and $\hat n \rightarrow 1/\hat n$.  We may denote the resulting
expression inside $\{\cdots\}$ by $\bar g(\hat n,p)$, which gives rise to
a measurement error critical probability $q_c(q,p)=1/\bar g(\hat
n(q,p),p)$ and scaling exponent $(k/2\bar \gamma)^\beta$.  Thus the
failure rate bound in 3d is
\begin{equation}\label{eq:10}
F \;\; \le \;\; k^3\left[ \frac{p}{p_c(q,p)} \right]^{\left(\frac{k}{2\gamma}\right)^\beta} + 
k^3\left[ \frac{q}{q_c(q,p)} \right]^{\left(\frac{k}{2\bar \gamma}\right)^\beta}
\end{equation}
where $\gamma,\bar \gamma \le 2 + \alpha$ both depend on $(q,p)$.  The
accuracy threshold is no longer a single point $p_c$ as in the case of no
measurement errors, but is now a definite curve in the $qp$-plane---the
boundary of the region lying underneath the two curves given implicitly by
$p=p_c(q,p)$ and $q=q_c(q,p)$.  This is a sort of phase boundary between
the well-ordered, sub-threshold state where long error chains are
exponentially improbable and the disordered state where long chains occur
frequently and the encoded information is quickly corrupted.

To obtain this threshold curve one could calculate $p_c(q,p)$ and
$q_c(q,p)$ directly, or choose a far less intensive method which is to
calculate $g(\hat n \ge 1,q=1)$ and $\bar g(\hat n < 1,p=1)$ for a number
of values of $\hat n$ and use the fact that $g(\hat n,q) = g(\hat
n,1)q^{1/\hat n}$ and $\bar g(\hat n,p) = \bar g(\hat n,1)p^{\hat n}$ to
obtain curves in the $p$-$q$ plane corresponding to thresholds for
individual contributions to $F$ from chains with fixed error-ghost
composition $\hat n$.  The region underlying all these curves is exactly
the sub-threshold region.  We have numerically calculated $g(\hat
n=2^M,1)$ and $\bar g(\hat n=2^{-M},1)$ for $M = 0,1,\ldots,16,\infty$. In
these calculations we set the recursion limit $N$ in (\ref{eq:9}) by $N =
\min\{M,4\}$. This means we need to calculate bounds on $h_3(2^L,m)$ and
$h_3(m,2^L)$ for $L=0,1,2$; $m=0,1$. We reduce $h_3(4,1)$ by one
application of (\ref{red}), and $h_3(1,4)$ by one application of the
analogous relation with $q \leftrightarrow p$. Bounds on the remaining
$h_3$'s are again obtained by inspection of diagrams. For example, the
diagramatic break-down of $h_3(4,0)$ is almost identical to that of
$h_2(4)$ in 2d which we have already calculated. The only additions are
diagrams involving errors distributed over multiple rounds of recovery. In
particular, by reference to (\ref{lstar}) and (\ref{eq:6.1}) one finds
\[
h_3(4,0) = h_2(4) +
(\leftrightarrow\,\mbox{---}\,\leftrightarrow)(\leftrightarrow\,\leftrightarrow)
= 4997 + (4\cdot3\cdot3)\FR{1}{4}(4\cdot3) = 5105
\]
where the chain within the first parentheses occurs one round before the chain within the second.

\begin{figure} 
\centering
\scalebox{.5}{\includegraphics{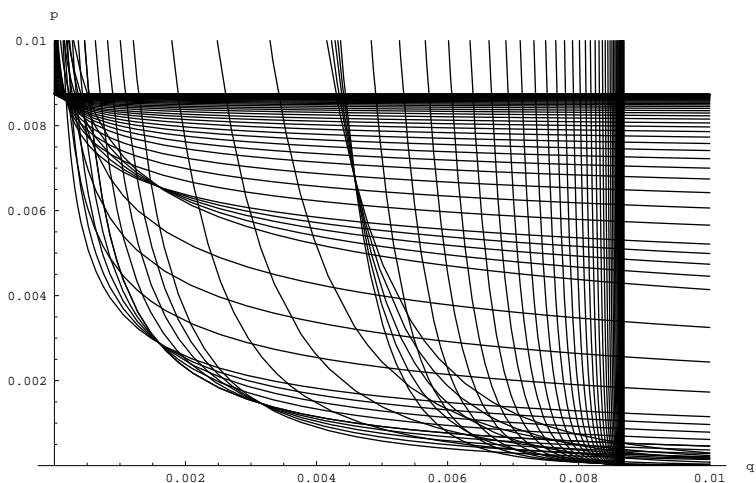}}
\caption{\small 
Accuracy threshold for recovery with faulty syndrome measurements
represented as a phase boundary in the $qp$-plane. Each curve correesponds
to error chains of a different composition $\hat n$, and the region
underneath them all is that in which recovery is stable.
}
\label{qp-plane}
\end{figure}

The values obtained for $\log(g)$ and $\log(\bar g)$ are roughly linear in
$1/\hat n$ and $\hat n$ respectively.  The points $\hat n = 2^{\pm M}$
chosen for these calculations possess high symmetry in the same way the
points $n = 2^M$ posses high symmetry in regard to the 1d/2d scaling
function $l(n)$.  Expecting for $g$ and $\bar g$ a similar kind of
peak-structure around points of high symmetry as was observed in
(\ref{l(n)}) for $l(n)$, we use linear interpolation for reasonable bounds
on points intermediate between $\hat n = 2^{\pm M}$ for
$M=0,1,\ldots,16,\infty$.  The resulting sub-threshold region is a
foot-like area with its heel at the origin of the $qp$-plane (see Fig.\
\ref{qp-plane}).  The phase boundary has three main parts, coming
respectively from the contributions to $F$ corresponding to $\hat n =
2,1,\mbox{ and 1/2}$, hence to chains of twice as many real errors as
ghosts, of equally as many, and of half as many.  The ankle is cut-off
around $p = 1/114.5$ by threshold curves for higher $\hat n$, hence more
real errors, and the toes are cut-off around $q = 1/115.3$ by curves for
lower $\hat n$, hence more ghosts.  These two values, then, represent the
limiting accuracy thresholds achieved by the 3d ``expanding octahedra''
algorithm.

Now these exponents $(k/2\gamma)^\beta$ and $(k/2\bar \gamma)^\beta$ are
over-estimates, just as in 2d, due to the conservative assumption that
any chain with a sufficient number of defects to form an NC loop will in
fact do so.  Recall in 2d consideration of the saturation point $n_0$ for
geodesic chains suggested a conservative estimate of $F$ with exponent
$(k/2)^\beta$ replaced by $k^\beta$ if $p \ge 0.0420\,p_c$.  The same
arguments apply in 3d, except geodesics may now move in the time
direction as well.  For a long geodesic chain, the 2+1 components of its
displacement will each average to the same $\ast$-metric length.  An NC
loop can be generated when this length is $k/2$, so the total
$\ast$-metric length of the chain is $3k/2$.  Consider a definite radial
line in the $(q,p)$ plane, on which $F$ is dominated by one particular
$\hat n$ threshold.  The saturation point analysis here is essentially the
same as in 2d, and one finds the 3d exponents are improved to
$(3k/2\gamma)^\beta$ and $(3k/2\bar \gamma)^\beta$ if $p/p_c \ge
2e^{-2/\beta} \approx 0.0840$.

\begin{figure} 
\centering
\scalebox{.5}{\includegraphics{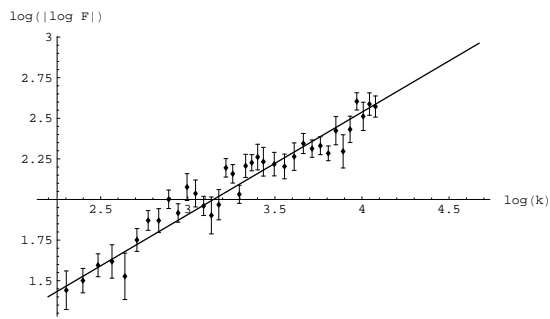}}
\caption{\small 
Recovery failure rates $F$ as a function of lattice size $k$ from Monte
Carlo simulations including measurement errors (``ghosts'') alongside the
theoretical prediction (line).
}
\label{ghosts}
\end{figure}

Numerical simulations with measurement errors are performed, leaving over
chains from one round to the next.  Recovery failure is assumed to occur
if either an NC loop occurs or an error chain's endpoints achieve a
spatial separation of $k$ or more (which would shortly lead to an NC
loop).  After a failure, the lattice is reset to an error-free state, and
recovery resumes.  The failure rate is calculated as the the number of
failures divided by the total number of rounds.  To speed up simulations
the expanding octahedral radii $l_\ast$ are incremented in steps larger
than one depending on the ages of current particles.  The increment is
chosen so that only five steps are necessary per round independent of $k$,
which was not observed to significantly affect performance.  In these
simulations we set $q = p/2$, with $k$ ranging from 10 to 60 for each
$(q,p)$.  The critical behavior is understood from our $qp$-plane
(Fig.\ \ref{qp-plane}) by walking out from the origin along the line $p =
2q$. This line happens to cross the phase boundary right near the edge of
the segment dominated by the $\hat n = 1$ threshold curve, corresponding
to $\gamma = 2 + \alpha$. Since this is the maximum possible $\gamma$, $F$
will indeed be dominated by the $\hat n = 1$ threshold in this edge
region. Thus our theoretical prediction here, with saturation improved
exponent, is
\[
F \sim \left[q\, p\, h_3(1,1) \prod_{L=1}^{\infty} \Sigma_3(2^L,2^L)^{2^{-L}}\right]^{(3k/2\gamma)^\beta}
\approx \left(\frac{p}{p_c}\right)^{1.0143\, k^\beta}
\]
neglecting a polynomial prefactor. Here $p_c$, specific to the case $p =
2q$, is at least $1/329.8$. This result for the exponent
$(3k/2\gamma)^\beta$ is shown as a line alongside the simulation data in
Fig.\ \ref{ghosts}.  The agreement here is not as good as in the absence
of measurement errors, but that is to be expected given the added
complications of the 3d algorithm.  These would presumably tend to
enhance the finite $k$ deviations from our asymptotic predictions.  
Still, our conservative estimate is good, with only a couple of the data
points completely below the line (indicating failure rates above the
estimate). 

Bravyi and Kitaev \cite{9} and, independently, Freedman and Meyer
\cite{FM} have exhibited lattice codes like TOR$(k)$ but using
simply connected lattices
with boundary in the plane.  This presents a major advantage for any
ultimate experimental implementation. A recovery algorithm appropriate to
these codes is just that given above, but modified to account for the
possibility of error chain endpoints being hidden on the boundary. Since
the boundary is asymptotically irrelevant to scaling properties of lattice
codes, the above results would seem to carry through.

\section{Lattice Codes on High Genus Surfaces}

In Kitaev's topological framework for quantum codes, the basic determiner
of a code's fidelity $1-\epsilon$ in maintaining encoded information is
the length $L$ of (i.e., the number of edges contained in) the shortest
possible
non-contractible loop on the lattice. In particular, we have seen
that the code's failure probability scales as
\begin{equation} \label{failprob}
\epsilon \sim (p/p_c)^{KL^\beta}
\end{equation}
where $p$ is an error probability for physical qubits, and $p_c$, $\beta$,
and $K$ are parameters depending on the particular error correction 
algorithm used.

$2N$ qubits may be encoded in $N$ separate lattices, each with fidelity
given above. What we will show here is that, if instead of $N$ separate
lattices we combine them into one large lattice on a high genus surface
constructed by a certain method, the information rate and/or fidelity may
be improved as the number of encoded qubits increases.

As a motivating example, consider joining two $L^\prime \times L^\prime$
toric lattices by removing a $L^\prime/2 \times L^\prime/2$ square from
each and sewing together the perimeters of the resulting square holes; it
is straightforward to define a code on this new lattice preserving the
total number, four, of encoded qubits. Taking $L^{\prime 2} \approx
4L^2/3$, the number of physical qubits is nearly the same as for two
separate $L \times L$ lattices, but the minimum length of non-contractible
loops is now $\approx 2L/\sqrt{3}$, improving the code's fidelity.

\begin{figure}[!hp]
\centering
\scalebox{.7}{\includegraphics{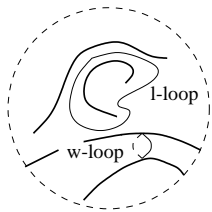}}
\caption{A simple w-loop and simple l-loop are shown on two different
handles.}
\label{loops}
\end{figure}

This suggests, given $N$ separate $L \times L$ toric lattices, we combine
the $2 L^2 N$ physical qubits into one large high-genus surface on which
each torus becomes one handle.  We can increase the code's fidelity by the
following construction.  Cut each handle through its width, along a
``w-loop'' (see Fig.\ \ref{loops}), giving two loose ends per handle. Then
randomly re-pair the set of $2N$ loose ends and rejoin each pair.

Before cutting and rejoining w-loops, the lengths of length-wise
``l-loops'' had been as small as $L$; now the shortest simple l-loops for
a typical handle will have length $\sim LN$. If each handle is made to
encode a qubit with $Z$ operator given by an l-loop, it appears the
chances of a $Z$ error to these encoded qubits has been markedly
diminished. One might like to say this error probability is now
exponentially small in $(LN)^\beta$. But there are more complicated l-type
loops with lengths much smaller than $LN$; each such loop involves many
handles. The minimal l-loops of this kind determine the encoded $Z$ error
probability of a lattice code based on this large surface. Let us first
determine the lengths of minimal l-loops, then symmetrize the construction
of our surface to handle encoded $X$ errors as well.

\begin{figure}[!hp]
\centering
\scalebox{.7}{\includegraphics{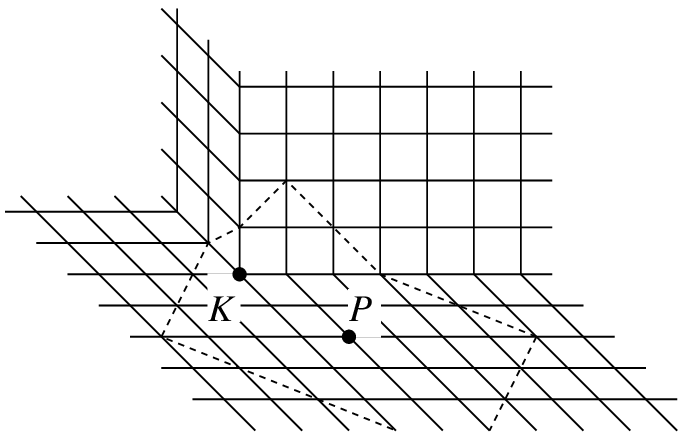}}
\caption{The kink $K$ appears at the base of a handle. A diamond
(dashed line) centered at $P$ and containing $K$ is shown.}
\label{kink}
\end{figure}

To characterize the minimal l-loops we will calculate some geometrical
properties of our high-genus surface, on which it will be convenient to
recall the square lattice of qubits. If each handle connects to the
surface through a square patch, the lattice will be locally identical to a
simple square lattice except at the corners of a square patch. Each corner
vertex has valence (number of edges containing it) equal to five not four,
giving five quadrants of 2d space (see Fig.\ \ref{kink}). The corner
constitutes a kink of negative curvature on the otherwise flat lattice.

Around some vertex $P$ draw a small diamond. As its radius $r$
increases, the diamond will encounter handles over which it must climb,
extending out to new places on the surface. From an intrinsic
perspective, the diamond merely sees an occasional kink, from which
emanates an extra quadrant of space. Having passed over a kink and
encroached into its extra quadrant, the diamond's perimeter will become
larger than it would be apart from the kink. It is not hard to see the
perimeter will contain an additional $r-r_\mathrm{k}$ vertices, where
$r_\mathrm{k}$ is the distance from $P$ to the kink.

As the diamond expands, it encounters more kinks and its perimeter grows
even faster. Moreover, all kinks contribute independently to the
perimeter. Assuming a constant density, $8/L^2$ per vertex, of kinks on
the lattice, the perimeter $c(r)$ approximately satisfies the recursion
relation
\begin{equation} \label{rec}
c(r) = 4r + \frac{8}{L^2} \sum_{r_\mathrm{k}=0}^r
c(r_\mathrm{k}) \cdot (r-r_\mathrm{k}), 
\end{equation}
obtained by adding independent contributions from all kinks within the
diamond; $c(r)=4r$ would be the result in flat space. The above relation
can be cast as a second order finite difference equation, with initial
conditions, whose solution is approximately
\begin{equation} \label{c(r)}
c(r) = L \sqrt{2} \sinh \left( \sqrt{8} \, r/L \right).
\end{equation}

To obtain the minimal l-loop length for a given handle $H$, consider the
set of open paths of length $r$ and starting at some vertex $P$ around the
base of $H$. As $r$ increases, the diamond forming the outer boundary of
this set will be pushed across various handles to random new places on the
surface, gradually filling it up. Once a path encounters the other end of
$H$, it can be closed across $H$ to form an l-loop. The chances there will
be such a path become significant only when the area of our $r$-diamond
approaches a significant fraction of the total surface area, $L^2 N$.
Obtaining the diamond's area by summing (\ref{c(r)}), this condition is
found to be $r \sim L \log N/\sqrt{8}$, which thus gives the minimal
l-loop length for almost all handles on the surface. As for the other
handles, one can either attempt to re-pair them or simply discard them
(close and eliminate them as handles).

The probability of encoded $Z$ errors associated with l-loops is thus
exponentially small in $(L \log N)^\beta$; however no improvement has been
achieved for $X$ error correction. To symmetrize the above construction,
we simply add an additional cut-and-pair step. Previously we had cut along
a w-loop on each handle and then re-paired all the cuts. Now we cut along
an l-loop, which may be as short as $\sim L \log N$, and randomly re-pair
as before. There is no longer a simple w-loop that can be drawn encircling
a given handle. A w-loop must proceed across many of these cuts before it
can close on itself non-contractibly. Topologically, this new step is
identical to the previous one but with the surface turned inside-out. The
effect of these new cuts and joins is just a doubling of the kink density
to $16/L^2$ in (\ref{rec}), giving the minimal loop length---now for
l-loops and w-loops---as $\sim L \log N / 4$.

This result can be applied either as an improvement to the
$L,N$-dependence of the lattice code's fidelity at fixed information rate,
an improvement to the information rate at fixed fidelity, or as a
simultaneous improvement to both. But there is another parameter to be
considered: the accuracy threshold $p_c$. Indeed, this convoluted surface
topology suggests a greater variety of possible catastrophic error
processes (long error chains) which would tend to worsen the threshold.

Nevertheless, as $L$ is increased at fixed $N$ the error processes
affected by the convoluted topology will only be those involving longer
and longer, hence less and less probable, error chains. As $L$ gets large
the threshold will then tend back to its original value. For the
quasi-local error correction algorithm presented above with $\beta =
\log_3 2$, the effect of convoluted topology is to multiply our bound on
the threshold $p_c$ by
\[ 
\sim \prod_{k=1}^{\log_3(L \log N)} 
\left[ \frac{2 (3^k)^2}{a(3^k)} \right]^{(1/3^k)^\beta}
\approx 8 e^{-12(\log N)^{1-\beta}/L^\beta}
\]
where $a(r)$ is the circular area obtained by summing (\ref{c(r)}). This
means that $N$ should not be increased faster than $\log N \sim
L^{\beta/(1-\beta)}$ or else the code's accuracy threshold may be greatly
diminished. Thus the fidelity $1-\epsilon$ from (\ref{failprob}) will
scale as
\[
-\log\epsilon \sim (L \log N)^\beta \sim L^{\beta/(1-\beta)} \, .
\]
Were we to naively put $\beta = 1$ above, we would find that there is no
restriction on how $N$ scales with $L$---unbounded gains could be had by
increasing $N$ at fixed $L$ without serious damage to the accuracy
threshold. In fact a different, global recovery algorithm does have $\beta
= 1$, and explicit consideration of its performance under convoluted
topology corroborates this result. Bounds on the threshold are obtained
here by counting certain classes of paths on the lattice \cite{4}. For
instance, the number of length $r$ paths with given starting point on a
flat square lattice is $4^r$. On our curved lattice, some vertices (the
kinks) have valence 5, so the number of walks is bounded by $v^r$ with
$4<v<5$. The effect of convoluted topology on a threshold bound based on
this counting would be to multiply it by the near-unity factor $4/v$.

We have thus shown how to achieve gains in the fidelity and/or efficiency
of storing quantum information by encoding many qubits in one block of a
topological quantum code. The code involves a lattice of qubits on a 2d
surface of highly convoluted topology. As more encoded qubits are added,
keeping fixed the number of physical qubits per encoded qubit,
asymptotically significant gains are obtained in the code's fidelity. This
is an economy of scale in the error correction hardware independent of any
software gains achieved by compressing redundancy within the encoded
information itself, as in Shannon's coding theorems and their quantum
equivalents \cite{5}, which rely on the encoded qubits' occupation of
``typical'' subspaces in the many-qubit Hilbert space.

One beneficial feature of the original topological codes is that error
correction operations are local on the lattice; however, this is also a
limitation. The convoluted topology of the above construction, which
effectively destroys the codes' locality, is a way of overcoming this
limitation (and sacrificing the associated benefit).

\chapter{Unconcatenated Quantum Computing}
\label{chap-toffoli}

\section{Encoded Computation: The Toffoli Gate}

So far we have mainly addressed error correcting codes in terms of storing
quantum information. To realize an actual quantum \emph{computer}, it is
necessary to show how quantum gates may be performed on encoded qubits.
While specific methods have been invented for a large class of
concatenated quantum codes, the lattice codes we have discussed were
designed specifically to obviate something inherent in the concatenation
process: the generic reliance on highly non-local gates. We must therefore
attempt to devise methods for performing encoded gates that do not rely on
the recursive structure of concatenated codes.

Most known quantum error correcting codes (concatenated and lattice codes
alike) can be defined by a set of operators, the stabilizer, each of which
fixes every codeword. For a number of stabilizer codes capable of simple
operations, like a bit flip $X_a$ or phase flip $Z_a$ on logical qubits
$a,b,\ldots$, it is also known how to perform any ``normalizer''
operation---i.e. one that can be composed from the C-NOT, which we will
now denote by $\dot X_{ab}$, the $\pi/2$ phase shift, and the Hadamard
rotation $R_a$.  Normalizer operations alone, however, are insufficient
for universal quantum computation; a quantum computer with only normalizer
operations can be simulated in polynomial time by a classical machine
\cite{6}.  A genuine quantum computer is realized either by the addition
of a non-trivial one or two-qubit gate, like a single qubit rotation by an
irrational multiple of $\pi$, or of a three-qubit gate like the Toffoli
(controlled-controlled-NOT), which flips the third qubit whenever the
first two are each $|1\rangle$. Because the necessary one or two-qubit
gates require rotations (e.g., $e^{\mri\theta X}$) involving SU(2) angles
that must be precisely tuned, they cannot be implemented fault-tolerantly.
Small errors in these angles are inevitable and will gradually accrue
until the computation veers completely off course. The Toffoli, on the
other hand, like the C-NOT, does not involve any such rotations and
therefore is suitable for fault-tolerant implementation.

Peter Shor has given a procedure \cite{1} for performing a Toffoli given
the ancilla state \begin{equation}\label{psi} |\psi_3\rangle \equiv
|000\rangle + |001\rangle + |010\rangle + |100\rangle, \end{equation}
which means essentially that possessing the tripartite entanglement of
this state is equally as powerful as being able to perform a Toffoli gate.
Let us review Shor's procedure (actually, a similar procedure that is
equivalent to Shor's).

As motivation, first consider the following construction, which uses one
ancilla bit $c$. Letting $c$ start as $|0\rangle$, suppose we could
perform a majority vote on $A\,B\,c$ so that, for example,
$|010\rangle \rightarrow |000\rangle$ and $|110\rangle \rightarrow
|111\rangle$. Equivalently, one might majority vote, but only carry out
the effect on $c$, leaving $A$ and $B$ unchanged, so $|010\rangle
\rightarrow |010\rangle$ and $|110\rangle \rightarrow |111\rangle$. Now
just C-NOT $c$ into $C$ and disentangle $c$ from $A\,B\,C$. The result is
exactly a Toffoli on $A\,B\,C$.

To majority vote on $A\,B\,c$, one measures $Z_A Z_B$ and $Z_B Z_c$.  If
both measurement results are $+1$, $A\,B\,c$ are already unanimous.  
Otherwise, the measurement results will reveal which bit is the
odd-one-out.  Unfortunately, these measurements have also revealed
information about the initial state $A\,B$, in general collapsing it,
inconsistent with the desired Toffoli gate, a linear operation. The
solution is to perform a majority vote not directly on $A\,B\,c$ but on
three ancilla qubits, which are first entangled with $A\,B$.  Here is
where $|\psi_3\rangle$ enters.

Given some arbitrary state of $A\,B\,C$, prepare $a\,b\,c$ in
$|\psi_3\rangle$ and perform the following operations: (I) C-NOT $A$ into
$a$ and $B$ into $b$, and (II) majority vote on $a\,b\,c$ (by measuring
$Z_a Z_b$ and $Z_b Z_c$ and flipping the odd-bit-out if necessary).  
Suppose, for example, the measurement results from (II) are $Z_a Z_b = -1$
and $Z_b Z_c = +1$.  All but 8 terms will be collapsed away of the total
$2^3 \times 4 = 32$ terms in the initial 6-qubit state. These 8 terms, as
they undergo (I) and (II), are (suppressing bra-ket notation):
\[
\begin{array}{rcccr}
                   &  \mbox{I}   &                     &       \mbox{II}                  \\ 
\smallskip
00C_0 100 & \rightarrow & 00C_0 100  & \rightarrow & 00C_0 000 \\
\smallskip
01C_1 001 & \rightarrow & 01C_1 011  & \rightarrow & 01C_1 111 \\
\smallskip
10C_2 000 & \rightarrow & 10C_2 100  & \rightarrow & 10C_2 000 \\
11C_3 010 & \rightarrow & 11C_3 100  & \rightarrow & 11C_3 000
\end{array}
\]
where $C_i = 0,1$. Note that all of the 8 possible bit values for $A\,B\,C$ are equally represented, so that 
all information in the initial superposition of $A\,B\,C$ is preserved (albeit decoherently).  Now C-NOT $c$ 
into $C$.  From the above table, this will flip $C$ iff $A\,B$ are $01$---not iff $A\,B$ are $11$, as 
desired for the Toffoli.  Applying $\dot X_{BC}$ then gives the desired result.  Finally, $A\,B\,C$ must be 
disentangled from the ancillas $a\,b\,c$ to restore the coherence of the original state.  This is 
accomplished by applying $\dot X_{ab}$ and $\dot X_{ac}$ and then measuring $X_a$.  If the result is $+1$, 
$A\,B\,C$ are disentangled.  If $-1$, a phase error on the $A\,B=01$ term has been introduced; it may be 
corrected by applying $X_A \dot Z_{AB} X_A$, where $\dot Z_{AB} \equiv R_{B} \dot X_{AB} R_{B}$ is the 
controlled-phase (C-PHASE) gate.

Had the measurement results for $Z_a Z_b$ and $Z_b Z_c$ been other than $-1$ and $+1$ respectively, as in 
the above example, it is straightforward to determine what gates must be applied in place of $\dot X_{BC}$ 
and $X_A \dot Z_{AB} X_A$.

The Toffoli now just requires preparation of the three-qubit state $|\psi_3\rangle$. First observe that if 
one can prepare
\[
|\psi_2\rangle \equiv |00\rangle + |01\rangle + |10\rangle,
\]
$|\psi_3\rangle$ may be obtained by preparing four qubits $a\,b\,c\,d$ in the state 
$|\psi_2\rangle|\psi_2\rangle$, measuring $Z_b Z_c$, and performing a few simple normalizer operations.  In 
particular, the measurement result $-1$ gives the state
\[
|0010\rangle + |0100\rangle + |0101\rangle + |1010\rangle,
\]
which can be turned into $|\psi_3\rangle|1\rangle$ by applying the C-NOTs:  $\dot X_{ac}$, $\dot X_{db}$, 
$\dot X_{ad}$, $\dot X_{bd}$, and $\dot X_{cd}$ in that order.

\section{Preparing the Two-Qubit Ancilla}

Let us define $\rho(\alpha_1,\alpha_2,\alpha_3)$ as the (unnormalized) mixed state
\[
\left[
\begin{array}{cc}
1 &  \alpha_1  \\
\alpha_2  &  \alpha_3 \\
\end{array}
\right]
\]
in the basis $\{|\psi_2\rangle,|11\rangle\}$, where $|\alpha_3| < 1$. It turns out, in the continuum of 
states $\rho(\alpha_i)$, there is nothing special about $|\psi_2\rangle$, obtained as $\alpha_i \rightarrow 
0$.  Being able to prepare any one state $\rho(\alpha_i)$ with $|\alpha_3| < 1$ is sufficient to prepare 
$|\psi_2\rangle$, hence to prepare $|\psi_3\rangle$ and construct a Toffoli gate.

\begin{figure}[!hp]
\centering
\scalebox{.7}{\includegraphics{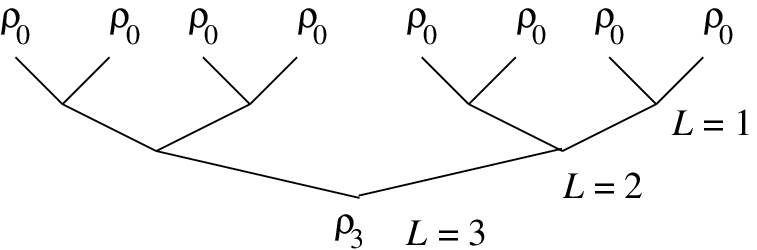}}
\caption{Combining $\rho_0$ states to prepare $\rho_N$ (above, $N=3$).}
\label{tree}
\end{figure}

The state $|\psi_2\rangle$ is prepared by combining two copies of $\rho(\alpha_i)$ through measurement to 
obtain a new mixed state which is closer to $|\psi_2\rangle$ than before, and combining two of these to get 
one still closer, etc., progressively \emph{purifying} $|\psi_2\rangle$
from the initial states. To 
start, prepare qubits $a\,b\,c\,d$ in the state $\rho_0 \otimes \rho_0$, where $\rho_0 = \rho(\alpha_i)$, 
and measure $Z_a Z_c$ and $Z_b Z_d$.  Suppose the results are $+1$ and $+1$.  Now perform $\dot X_{ac}$ and 
$\dot X_{bd}$ to disentangle $c\,d$.  For pure states, this whole process would give 
$|\psi_2\rangle|\psi_2\rangle \rightarrow |\psi_2\rangle|00\rangle$ and $|11\rangle |11\rangle \rightarrow 
|11\rangle|00\rangle$, while either of the initial states $|\psi_2\rangle |11\rangle$ or 
$|11\rangle|\psi_2\rangle$ are inconsistent with the assumed measurement results.  In terms of mixed states, 
this means $\rho_0 \otimes \rho_0 \rightarrow \rho_1 \otimes |00\rangle\langle 00|$ where $\rho_1$ is
\[
\left[
\begin{array}{cc}
1 &  \alpha_1^2  \\
\alpha_2^2  &  \alpha_3^2 \\
\end{array}
\right]
\]
which is exactly $\rho(\alpha_i^2)$. Prepare another $\rho_1$ from two new $\rho_0$ states, and combine the 
two $\rho_1$ states by again measuring $Z_a Z_c$ and $Z_b Z_d$.  Supposing the results are again $+1$ and 
$+1$, $c\,d$ are disentangled, leaving $a\,b$ in the state $\rho_2 = \rho(\alpha_i^4)$. Continuing this 
process through $N$ levels gives $\rho_N = \rho(\alpha_i^{2^N})$.  The whole procedure may be pictured as a 
tree of $\rho_L$ states, joining in pairs from level $L=0$ to $L=N$ (see Fig.\ \ref{tree}).  The 
recursiveness is reminiscent of concatenated codes, but here the complexity appears in the auxiliary 
purification process, not in the code itself.

The fidelity in preparing $|\psi_2\rangle$ is  
\begin{equation} \label{epsilon}
1-\epsilon \equiv
\frac{\mbox{tr}(\rho_N |\psi_2\rangle \langle \psi_2|)}{\mbox{tr}(\rho_N) \langle\psi_2|\psi_2\rangle} =
\frac{3}{3 + \alpha_3^{2^N}}
\end{equation}
very close to 1 if $|\alpha_3| < 1$.  The number of (logical) qubits used to achieve this fidelity is $\sim 
2^N$, which by (\ref{epsilon}) is $\sim \log \epsilon/\log |\alpha_3|$. This is the same kind of polylog 
scaling desired from the code itself (referring to the scaling of block size with desired failure rate 
$\epsilon$). Finding the number of operations on encoded qubits necessary to prepare $|\psi_2\rangle$ is not 
as easy, since the assumption that all $Z_a Z_b$, $Z_c Z_d$ measurement outcomes are $+1$,$+1$ requires 
repetition of the procedure a number of times before one expects such to occur.

To prepare a single $\rho_L$ state prepare two $\rho_{L-1}$ states and then combine them by measurements. If 
the measurement results are not $+1$,$+1$, just discard these states and keep trying. (This is not an 
optimal procedure, but it will suffice.)  Therefore, if the chances of any one attempt succeeding are 
$P(L)$, the expected number of logical operations $G(L)$ necessary to prepare $\rho_L$ is $\sim 2G(L-
1)/P(L)$. This assumes high confidence in the one pair of measurement results $+1$,$+1$, which should be the 
case since $a\,b\,c\,d$ are \emph{logical} qubits. But even if there is a significant probability 
$\epsilon_\mathrm{m} \gg \epsilon$ for any one measurement result to be in
error, the purification procedure 
can be made robust. Once a $+1$,$+1$ result is obtained, just repeat the measurements a number of times and 
accept the state only if, say, a majority of the results are $+1$,$+1$.  To get $1-\epsilon$ confidence in 
the measurement outcome, one must repeat $\sim \log \epsilon / \log \epsilon_\mathrm{m}$ times.  This 
implies 
\begin{equation} \label{GL}
G(L) \approx \frac{2}{P(L)} \, G(L-1) + \frac{\log \epsilon}{\log \epsilon_\mathrm{m}} \;\;.
\end{equation}

It is not hard to see that $P(L)$ must increase with $L$, since this recursion relation implies that either 
$|\psi_2\rangle$ will quickly begin to dominate successive $\rho_L$ states, in which case $P(L) \rightarrow 
1/3$, or $|11\rangle$ will dominate and $P(L) \rightarrow 1$.  Both of these values are larger than $P(1)$, 
which can be calculated as a function of $|\alpha_3|<1$ but is always bounded from below by 1/4. Iterating 
(\ref{GL}) with this bound gives 
\begin{equation} \label{GN}
G(N) \sim 8^N \frac{\log \epsilon}{\log \epsilon_\mathrm{m}} \sim
\frac{(\log \epsilon)^4}{(\log |\alpha_3|)^3 \log \epsilon_\mathrm{m}} \;.
\end{equation}
Note that $G(N)$ is the total number of logical operations, but these can be done in parallel so that the 
actual purification time is $\sim N \log\epsilon/\log \epsilon_\mathrm{m}
\sim 
\log(|\log\epsilon|)\log\epsilon /\log \epsilon_\mathrm{m}$.  The point is that even with the demand of a 
definite sequence of measurement results, time requirements still scale polylogarithmically with $\epsilon$.  
The crucial fact leading to this scaling is that the probability for getting the measurement results 
$+1$,$+1$ in combining two $\rho_L$ states is finite as $L \rightarrow \infty$.  Thus one can prepare 
$|\psi_2\rangle$, $|\psi_3\rangle$, and execute a Toffoli gate if one can prepare one of the mixed states 
$\rho(\alpha_i)$ with $|\alpha_3| < 1$.

There are multiple ways of obtaining a state $\rho(|\alpha_3|<1)$ for
codes which are not too large. In fact, Shor's own procedure for preparing
$|\psi_3\rangle$ can do so. An alternative method will be presented here,
applicable to codes possessing a non-trivial normalizer operation (here
C-NOT) that is \emph{transversal}, so the encoded operation factors into a
number of independent operations on physical qubits. The method works by
performing a very noisy measurement of the C-NOT operator.

\section{Noisy Measurement of C-NOT}

Were it possible to measure C-NOT with high fidelity, one could easily prepare $|\psi_2\rangle = 
\rho(\alpha_i=0)$. It turns out imperfect measurement of C-NOT is still capable of yielding 
$\rho(|\alpha_3|<1)$.

\begin{figure}[!hp]
\centering
\scalebox{.85}{\includegraphics{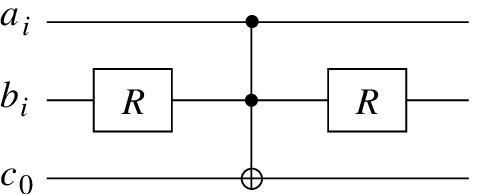}}
\caption{The operation $U$ on physical qubits $a_i\,b_i\,c_0$.}
\label{Ufig}
\end{figure}

For reference, the eigenstates of the C-NOT operator $\dot X_{ab}$ are
$|00\rangle$, $|01\rangle$, and $|10\rangle + |11\rangle$ with eigenvalue
$+1$, and $|10\rangle - |11\rangle$ with eigenvalue $-1$.  Let us first
describe a fault-\emph{in}tolerant measurement procedure, that is, one
which permits a single error to spread rampantly throughout a block.  
Prepare one physical ancilla bit $c_0$ as $|0\rangle$ and apply a certain
three-bit gate $U_{a_ib_ic_0}$ bitwise over physical bits $a_i$ and $b_i$
in the blocks encoding $a$ and $b$ (but always using the bit $c_0$). $U$
is shown in Fig.\ \ref{Ufig}.  The first Hadamard rotation causes the
Toffoli to flip $c_0$ just if $a_i b_i$ start in the $-1$ eigenstate
$|10\rangle - |11\rangle$ of $\dot X_{a_i b_i}$, and the second Hadamard
undoes the effect on $b_i$.

Apply $U$ bitwise over $a\,b$ and measure $Z_{c_0}$.  The result $Z_{c_0}=\pm 1$ is equivalent to the result 
that $\dot X_{ab} = \prod_{i}\dot X_{a_ib_i} = \pm 1$, so one has effectively measured $\dot X_{ab}$. To 
understand this process in detail, expand the initial state of $a\,b$ in eigenstates of the operators $\dot 
X_{a_ib_i}$ for $i=1,\ldots,n$:
\[
\sum_{\mathbf{x}} C_{\mathbf{x}} |\mathbf{x}\rangle = 
\left( \sum_{w(\mathbf{x})=0} + \sum_{w(\mathbf{x})=1} \right) C_{\mathbf{x}} |\mathbf{x}\rangle
\]
where $|\mathbf{x}\rangle = |x_1\cdots x_n\rangle$ and each $|x_i\rangle$ is one of the four eigenstates 
($x_i = 1,2,3,4$) of $\dot X_{a_i b_i}$.  The right hand sum is rearranged to segregate strings of even and 
odd weight. The weight function $w(\mathbf{x})$ equals the number (mod 2) of ``4''s occurring in the string 
$\mathbf{x}$, $x_i = 4$ corresponding to the $-1$ eigenstate $|10\rangle - |11\rangle$ of $\dot X_{a_i 
b_i}$. Using the transversality of C-NOT and the definition of $w(\mathbf{x})$, one finds $\dot X_{ab} 
|\mathbf{x}\rangle = (-1)^{w(\mathbf{x})} |\mathbf{x}\rangle$.
Thus the sum over strings with $w(\mathbf{x})=0$ is the projection onto the $+1$  eigenspace of $\dot 
X_{ab}$, and the sum with $w(\mathbf{x})=1$ is the projection onto the $-1$ eigenspace. It follows that the 
action of $U=\prod_i U_{a_ib_ic_0}$ is
\[
U |\mathbf{x}\rangle_{ab} |0\rangle_{c_0} = 
|\mathbf{x}\rangle_{ab} |w(\mathbf{x})\rangle_{c_0},
\]
which means measuring $Z_{c_0}$ is equivalent to measuring $\dot X_{ab}$.

This method of measurement is highly sensitive to errors; just one physical bit error can change 
$w(\mathbf{x})$ for an entire string of bits, making the measurement result erroneous.  As the block size 
$n$ gets large, the chances of an  even number of such errors occurring becomes nearly equal to the chances 
of an odd number occurring. Thus the  measurement result tells very little about whether a $+1$ eigenstate 
or a $-1$ eigenstate of $\dot X_{ab}$ has been obtained. This little bit of information, however, turns out 
to be important for preparing $|\psi_2\rangle$. 

As mentioned the above procedure is fault-intolerant, since one physical bit phase error may infect $c_0$ 
and thus spread rampantly throughout the block.  It can be made fault-tolerant by using an ancilla $c$, 
which is not just one bit, but a superposition of $n$ physical bits over all even weight strings (``weight'' 
is now in the sense of counting ``1''s). Such a superposition is prepared as
\[
|\mathrm{even}\rangle_c = \left( \prod_i R_{c_i} \right)(|0\cdots0\rangle_c + |1\cdots1\rangle_c).
\]
The gate $U_{a_ib_ic_i}$ will be applied bitwise across $a\,b\,c$ so that a single error in one block can at 
most spread to one bit in each of the other two blocks.  Acting bitwise on $|x_1\rangle|\;\rangle_{c_1}$ 
through $|x_n\rangle|\;\rangle_{c_n}$, $U$ will flip a number of bits in the initial $c$ state equal (mod 2) 
to exactly $w(\mathbf{x})$.  Thus 
\begin{equation} \label{Ugate}
U |\mathbf{x}\rangle_{ab} |\mathrm{even}\rangle_c =
\left\{
\begin{array}{ll}
|\mathbf{x}\rangle_{ab} |\mathrm{even}\rangle_c & w(\mathbf{x})=0 \\
|\mathbf{x}\rangle_{ab} |\mathrm{odd}\rangle_c & w(\mathbf{x})=1
\end{array}
\right.
\end{equation}
Measuring $Z_{c_i}$ bitwise over $c$ with the result $\prod_{i}Z_{c_i} = \pm 1$ is now equivalent to 
measuring $\dot X_{ab}$ with the result $\dot X_{ab} = \pm 1$.  Note that a single phase error in the 
$n$-bit cat state, or equivalently a bit flip in the sum over even weight strings, will change this sum into 
one over odd weight strings, again altering the measurement result while still projecting the state onto one 
of the eigenspaces of $\dot X_{ab}$. So the measurement procedure is now fault-tolerant, but the measurement 
\emph{result} is still highly sensitive to single bit errors, giving little information about which 
eigenspace the state $|\;\rangle_{ab}$ collapses into.

One can also perform a noisy measurement of the C-PHASE operator $\dot Z_{ab}$.  The action of $\dot Z_{ab}$ 
is just to apply a minus sign if $a\,b$ are in $|11\rangle$, which is unitarily equivalent to $\dot X_{ab}$ 
through the basis change $R_b$.  To measure $\dot Z_{ab}$ first apply $R_b$, then measure $\dot X_{ab}$ by 
the above method, and reapply $R_b$.  These procedures may be adapted, by changing the bitwise operation 
$U$, to noisy measurement of such operators as $\dot X_{ab} \dot X_{cd}$, $\dot Z_{ab} \dot Z_{cd}$, and 
$\dot Z_{ab} \dot  Z_{bc}$.

\section{Preparing the Two-Qubit Mixed State}

First prepare two logical qubits $a\,b$ as $(|0\rangle + |1\rangle)^2$ and measure $\dot Z_{a b}$ by the 
method given above, making use of an ancilla block $c$. If the measurement result were +1 and all qubits 
were error-free, one would have prepared exactly $|\psi_2\rangle$. But this will be changed by errors 
(i.e. decoherence, gate errors, or measurement errors) occurring either to the bits encoding $a$ and 
$b$ or to those of the cat-like ancilla $c$ used in the noisy measurement procedure. In fact $c$ is 
especially vulnerable because it is not protected by any code at all---a single bit error anywhere in $c$ 
can reverse the observed measurement result for $\dot Z_{a b}$.
 
Depending on whether errors are unitary or decoherent, this yields a cohererent or incoherent superposition 
of $|\psi_2\rangle$ and $|11\rangle$, which will be shown to be of the form $\rho(|\alpha_3|<1)$ in either 
the unitary or decoherent case, hence a candidate for purification.

Phase errors to the bits of $a\,b$ cannot be transmitted to $c$ by the above procedure, so are irrelevant. 
Bit flip errors to $a\,b$ can be transmitted but are equivalent to bit flip errors occurring to the bits of 
$c$ so all errors can be effectively regarded as occurring to $c$ alone. Let us first consider the case of 
(uncorrelated) errors purely decoherent in the Pauli basis $\sigma^m$, so that each qubit $c_i$ suffers no 
error, a phase error, a bit error, or both errors---each with some fixed classical probability.

Phase errors in $c$ can affect only the relative sign of terms in
$|\mathrm{even}\rangle_c$ and $|\mathrm{odd}\rangle_c$ of (\ref{Ugate}),
hence are extinguished once the $Z_{c_i}$ measurements are made.  
Depending on whether $c$ is attacked by an even or odd number of bit
errors, the measured eigenvalue of $\dot Z_{ab}$ will be inferred either
rightly or wrongly from the outcome of the $Z_{c_i}$ measurements. So
given the result $\prod_{i}Z_{c_i} = +1$, an even number of bits errors
will yield $|\psi_2\rangle$ as desired; however, an odd number will yield
$|11\rangle$ unbeknownst to us.

If each $c_i$ suffers a bit error with probability $p_i$, the difference between the chances of an even 
number of bit errors and of an odd number is
\begin{equation} \label{prodp}
\left(\prod_{i=1}^n \sum_{x_i = 0,1}\right) (-1)^{x_i} p_i^{x_i} (1-p_i)^{1-x_i}  =  \prod_i (1-2p_i).    
\end{equation}
Given that these two probabilities sum to 1, this implies the preparation procedure will yield not exactly 
$|\psi_2\rangle$, but the state $\rho(0,0,\alpha_3)$ with
\[
\alpha_3= \frac{1-\prod_i (1-2p_i)}{1+\prod_i (1-2p_i)} \approx 1 - 2\prod_i (1-2p_i)
\]
where the last expression holds for large $n$. It thus appears that one cannot tell whether or not 
$|\alpha_3|<1$ for a given ancilla block $c$ if even a few of its bits might have $p_i > 1 /2$. This is true 
even though current codes themselves are completely robust to these ``defective'' bits so long as their 
distribution is suitably uncorrelated and infrequent at the level of the code's threshold error rate. Still 
what has to be considered for the purification process is not a single
ancilla block $c$, giving rise to one 
$|\psi_2\rangle$-like state, but a sequence of such blocks, each with its own set of defective bits and 
consequent value of $\alpha_3 = \alpha_3^{(m)}$, where $m$ runs from $1$ to $2^N$, the number of 
$|\psi_2\rangle$-like states input to the purification process.

The fidelity in purifying $|\psi_2\rangle$ is that given by
(\ref{epsilon}) with $\alpha_3^{2^N}$ replaced 
by
\[
\prod_m \alpha_3^{(m)} \approx
e^{ -2^{N+1} \left\langle \prod_i (1-2p_i) \right\rangle },
\]
where $\langle\cdots\rangle$ is an average over the ensemble of $c$ blocks. Assuming errors are uncorrelated 
between different $c$ blocks, the average factorizes and the purification 
fidelity is
\[
1-{\textstyle\frac{1}{3}} \, e^{ -2^{N+1} \prod_i (1-2\langle p_i\rangle) }.
\]
Here only the ensemble averaged bit flip error rates appear. Assuming the locations of defective qubits are 
uncorrelated between different $c$ blocks, their $p_i > 1/ 2$ contributions are simply weighted out in the 
average. This means infrequent defective bits no longer pose a problem, owing to the distributed nature of 
the purification process. Defining an average error rate $p = (1/n)\sum_i
\langle p_i\rangle$, the above 
product is approximately $e^{-2pn}$. In order that the resulting fidelity be comparable to that of the code 
itself, which is $\sim 1 - \exp(-K n^\beta)$ for some power $\beta$ and constant $K$, the number of physical 
qubits used in purification is roughly
\[
n 2^N \sim n^{1+\beta} e^{2pn},
\]
which puts a limit on the block size $n$ of the code being used, since the number of qubits used in 
purification should not grow exponentially with block size. Thus $n$
cannot be larger than 
\begin{equation} \label{np}
n \sim \frac{1}{p}\log\frac{1}{p}.
\end{equation}

The opposite case of purely unitary errors is the same in its result. Here, errors comprise a set of unitary 
operators which act on the $c_j$ respectively. Each such operator can be written in the Pauli basis and put 
in the form 
\[
E_j = (A_j\mathbf{1} + \mri B_j\sigma^z)+ i \sigma^x (C_j\mathbf{1} +
\mri D_j\sigma^z), 
\]
where $A_j,B_j,C_j,D_j$ are real. Assuming low error rates, so $\prod_i |A_i| \gg \prod_i |B_i|$ and 
likewise with $C_i$ and $D_i$ in place of $B_i$, one can show that these errors take $c$ from its prepared 
state $|\mathrm{even}\rangle$ to 
\[
\prod_i E_i|\mathrm{even}\rangle = |\mathrm{even}\rangle +
\mri\tan(\Sigma_C)|\mathrm{odd}\rangle,
\]
where $\Sigma_C \approx \sum_i \tan^{-1}(C_i/A_i)$. This leads to preparation of the state $|\psi_2\rangle + 
\mri\tan(\Sigma_C)|11\rangle$ in place of $|\psi_2\rangle$. But this state
is precisely $\rho(\alpha_i)$ with 
$\alpha_{1,2} = \mri\tan(\Sigma_C)$ and $\alpha_3 = -\tan^2(\Sigma_C)$,
which can be purified to 
$|\psi_2\rangle$ if $\tan^2(\Sigma_C)<1$. For large $n$, $\Sigma_C$ will be very sensitive to the error 
amplitudes $C_i$, and in practice one would have no way of knowing whether $\tan(\Sigma_C)<1$ or not. This 
is the same problem noted above in the case of pure decoherence, and it too disappears when one realizes 
that the purification input is not a single state $|\psi_2\rangle +
\mri\tan(\Sigma_C)|11\rangle$ but an 
ensemble of such states each generated by a different set of blocks
$a\,b\,c$. The purification fidelity is 
now given by (\ref{epsilon}) with $\alpha_3^{2^N}$ replaced by 
\[
\prod_{m=1}^{2^N} \tan^2(\Sigma_C) \approx e^{2^{N+1} \langle \log(\tan\Sigma_C) \rangle}.
\]
where $\langle\cdots\rangle$ again averages over the ensemble of $c$ blocks. Using this, the definition of 
$\Sigma_C$, and expanding the logarithm gives $\langle \log(\tan\Sigma_C) \rangle$ as
\begin{equation} \label{ksum}
-\sum_{k=0}^\infty \frac{2}{2k+1} \prod_i \langle
e^{-\mri 2(2k+1)\tan^{-1}(C_i/A_i)} \rangle.
\end{equation}
The factorization follows assuming independence of errors between different qubits in each $c$ block. Now 
expand the exponential in a power series. If the error distributions are all such that the two bit flip 
amplitudes $C_i$ and $-C_i$ are equally likely to occur, the expectation values of the odd terms in the 
power series vanish. (If the distributions are otherwise, one might use the computational basis 
$\{|0\rangle,-|1\rangle\}$ instead of $\{|0\rangle,|1\rangle\}$ for the qubits in half the $c$ blocks, so 
that the same physical error would correspond to the bit flip amplitude $-C_i$ as often as it would to 
$C_i$.) The even terms may then be resummed and the expectation value in (\ref{ksum}) becomes
\[
\left\langle \cos\left( 2(2k+1)\tan^{-1}(C_i/A_i) \right) \right\rangle \equiv 
\cos\left( 2(2k+1)\sqrt{\langle p_i \rangle} \right),
\]
where $\langle p_i \rangle = \langle C_i^2/A_i^2 \rangle$ to lowest order in $C_i/A_i$, hence $\langle p_i 
\rangle$ can be taken roughly as a bit flip error rate. For small $k$ the cosine functions will be close to 
1, and the product over $i$ will be greatest. As $k$ gets large, the cosines will sample their full range 
and the product will be highly suppressed. Therefore the cosine above can be replaced by $\exp(-2(2k+1)^2 
\langle p_i \rangle)$, as if $k$ were always small, giving the main contribution to the sum:
\[
\langle\log(\tan\Sigma_C)\rangle \; \sim \;
-\sum_{k=0}^\infty \frac{2}{2k+1} e^{ -2(2k+1)^2 \sum_i \langle p_i \rangle } 
\; < \; {-2}e^{-2pn},
\]
where $p$ is again the average of $\langle p_i \rangle$ over $i=1,\ldots,n$; this is basically the same 
result as obtained for purely decoherent errors. Thus again (\ref{np}) gives the largest allowed block size 
in the regime where the resources needed for purification scale
polynomially with block size.

\section{Progressive Concatenation}

In case higher fidelity is desired of the code than (\ref{np}) allows, the above methods by themselves are 
insufficient and one must resort to concatenation. However, in conjunction with these methods, an 
unconventional, exponentially weaker form of concatenation can be used. A usual concatenated code is self-
similar, the same abstract code (e.g. the 7-qubit code) being used at each level in its recursion. 
Here one is free to increase the block size at each level, as long as (\ref{np}) is satisfied level-by-
level. In these ``progressive'' concatenated codes, many fewer levels are necessary given a desired fidelity 
$1-\epsilon$.

In particular, for one error correction algorithm \cite{11} in the context of lattice codes, some reasonable 
parameters are $p = p_c/10 \lesssim 10^{-3}$, so that $n = 1000$ is acceptable by (\ref{np}). Here $\epsilon 
\sim (p/p_c)^{n^\beta}$ where $\beta = \log_9 2 \approx .315$, which gives $\epsilon \sim 10^{-9}$. So, if 
the desired fideltiy is below $1-10^{-9}$, \emph{no concatenation is necessary}. Otherwise, one can begin 
concatenating. 

Consider a single concatenation of a chosen code. Physical qubits with error rate $\epsilon_0=p$ are 
arranged in code blocks of size $n_1$, and these blocks are themselves arranged in blocks of size $n_2$. The 
effective error rate at this higher level is just the failure rate of blocks at the lower level: 
\begin{equation} \label{eps1}
\epsilon_1 \sim (\epsilon_0/p_c)^{Kn_1^\beta},
\end{equation}
where $p_c$, $K$, and $\beta$ come from details of the code being concatenated. The code as a whole has 
failure rate
\begin{equation} \label{eps2}
\epsilon_2 \sim (\epsilon_1^\ast/p_c)^{Kn_2^\beta}.
\end{equation}
where $\epsilon_1^\ast$ includes the effect of storage errors described by $\epsilon_1$ and also gate errors 
associated with the operations necessary to perform error correction. Thus $\epsilon_1^\ast$ will have the 
same form as $\epsilon_1$ in (\ref{eps1}) but with $\epsilon_0=p$ replaced by a physical qubit error rate 
$\epsilon_0^\ast$ including the effect of these additional errors. In other words one must deal not only 
with a storage error threshold but also with a gate error threshold associated with the computations 
necessary for error correction. Of course, if one intends to use the logical qubits stored by the code for 
actual computations, this would be necessary anyway. 

Assuming the effective error rate $\epsilon_0^\ast$ is still below threshold, a single concatenation of the 
code in the above example gives
\[
\epsilon_2 = \left( \frac{(\epsilon_0^\ast/p_c)^{K n_1^\beta} }{ p_c } \right)^{K n_2^\beta} \sim 10^{-830}
\]
where the values $K=1$, $\beta = \log_9 2$, and $\epsilon_0^\ast = p_c/5$ have been used. The block sizes 
$n_1 = 1000$ and $n_2 = 2\cdot10^7$ were chosen to be consistent with $n_L \sim (1/\epsilon_{L-1}^\ast) 
\log(1/\epsilon_{L-1}^\ast)$, i.e. the condition (\ref{np}) applied at each level. Thus, as long as 
one does not require a fidelity better than $1-10^{-830}$, a single concatenation is sufficient given the 
above parameters.

Because the block sizes $n_L$ may increase so rapidly, the number $N$ of
levels necessary for a desired fidelity $1-\epsilon$ scales differently
than in usual concatenation, in which $N \sim \log(\log(1/\epsilon))
\equiv \log^{(2)}(1/\epsilon)$. For these progressive concatenated codes,
$N$ is determined self-consistently by $N \sim \log^{(N)}(1/\epsilon)$.
One might wonder about the asymptotic behavior of thresholds and
fidelities as $N \rightarrow \infty$, taking into account the reciprocal
effects between progressive concatenation and purification; however, this
seems irrelevant given the smallness of $\epsilon$ already at $N=2$.

\newcommand{\mbm}[1]{\mbox{\boldmath $ #1 $}}
\newcommand{\mre}[0]{e}
\newcommand{\tsum}[1]{{\textstyle \sum #1}}
\newcommand{\bin}[2]{
	\left(
		\begin{array}{@{}c@{}}
			#1 \\ #2
		\end{array}
	\right)
}

\chapter{Path Integrals and Beable Trajectories}
\label{chap-path}

The strategy employed in the last chapter to achieve a desired quantum
state (in our case a certain two or three-qubit entangled state) by an
iterative process of purification appears somewhat narrow in its
application in the context of certain quantum error correcting codes.  
However it turns out this underlying idea can be translated into classical
terms and proves valuable in defining a purely classical algorithm for
the simulation of quantum systems---meaning, that the goal of the
algorithm will be to simulate quantum systems but its implementation will
be entirely concerned with classical computations on a normal computer.

As we have discussed, numerical simulation of quantum systems by classical
algorithms is limited by the fact that the dimension of a system's Hilbert
space grows exponentially with the number of physical degrees of freedom.
One general approach to this problem is to obtain a correspondence between
the desired quantum system and a statistical ensemble of classical systems
that are easier to simulate. Green function Monte Carlo \cite{GFMC} and
coherent state representations in quantum optics \cite{creps} both derive
Fokker-Planck equations, which can be realized through stochastic
trajectories in an associated classical configuration space. Another
category of methods well known in condensed matter and particle theory is
that of path integrals with pseudo-dynamical importance sampling
\cite{Negele}. It turns out these configuration space methods are closely
related to well known beable\footnote{John Bell used the term ``beables''
rather than the misnomer ``hidden variables'' to distinguish them from
observables in quantum theory.} models of quantum mechanics.

Understanding these relationships offers another perspective on existing
computational methods and also points the way to new such methods. In
particular Monte Carlo, path integral, and coherent state representation
approaches have mainly been applied by exploiting their connection to
diffusion processes, with the result that calculations have been limited
to ground state or thermal properties, or else very restricted classes of 
Hamiltonians. In the case of coherent state methods, like the $P$ 
or positive $P$ representation, Hamiltonians with higher than quadratic 
interaction terms give rise to master equations that cannot be cast as 
stochastic differential equations over an associated phase space. In the 
case of Monte Carlo and path integral methods, a connection to diffusion 
processes can only be made by going to imaginary time.

However, the beable methods to be presented here are naturally formulated
in real time and are not restricted to narrow classes of Hamiltonians.  
Whatever their computational efficacy, they are thus at least
well-defined, general methods for studying dynamics.

\section{Langevin Method and Nelson's Mechanics}

The thermal expectation value of an operator $\mathcal{O}$ is often given
in the form of an imaginary time path integral:
\begin{equation}\label{O}
\langle \mathcal{O} \rangle = 
	\int \mathcal{D}x \mathcal{O}(x)e^{-S_\mathrm{E}(x)}
\end{equation}
where $S_\mathrm{E}(x)$ is some Euclidean action over the degrees of
freedom $x=(x^1,\ldots,x^d)$. To compute this path integral we need a
method to select paths over which the action may be sampled according to
the weight $e^{-S_\mathrm{E}(x)}$. The methods to be considered
here generate paths as the solutions of initial value problems in the
classical configuration space $\{x\}$ of the system.

In the Langevin or ``stochastic quantization'' approach
\cite{Negele}\cite{PW}, $\langle \mathcal{O} \rangle$ is computed as a
long-time average over a path $x(\tau)$ generated by the stochastic
differential equation
\begin{equation}\label{langevin}
dx^i = -\partial_i S_\mathrm{E} d\tau + \sqrt{2}dW^i
\end{equation}
where $\hbar=1$, $\partial_i \equiv \partial/\partial x^i$, and $dW^i$ is
a Wiener process, i.e.\ an independent Gaussian random variable at each
$\tau$ with mean $\langle dW^i \rangle = 0$ and variance $\langle (dW^i)^2
\rangle = d\tau$. Here, $\tau$ is not a time (real or imaginary) but an
auxiliary variable introduced to parameterize paths. And one can show that
an ensemble of trajectories evolving by (\ref{langevin}) samples
$e^{-S_\mathrm{E}(x)}$ in the limit $\tau \rightarrow \infty$

Nevertheless, observe the similarity between the Langevin equation
(\ref{langevin}) and the stochastic process involved in the generalized
Nelson hidden variable theory \cite{Davidson} for particles of mass $m=1$
under a potential $V(x)$:
\begin{equation}\label{nelson}
dx^i = 
	\left(\partial_i S + \alpha\frac{\partial_i R}{R}\right)d\tau +
	\sqrt{\alpha} dW^i
\end{equation}
where $dW^i$ is also a Wiener process, and $\alpha>0$ is a free parameter.
Note that Nelson's theory \cite{Nelson} results if $\alpha=1$ and Bohm's
deterministic theory \cite{Bohm} if $\alpha=0$. While (\ref{nelson}) is
normally defined in real time, we need the imaginary time analog for
comparison with the path integral (\ref{O}). This is easily obtained by
taking $\psi(x,\tau) = Re^{S}$ as a solution of the imaginary
time ($\tau = -it$) Schrodinger equation
\begin{equation}\label{itschrod}
\frac{\partial\psi}{\partial \tau} = 
	-\frac{1}{2}\sum_i \partial_i^2\psi + V\psi
\end{equation} 
which will hold if we evolve $R$ and $S$ by
\begin{equation}\label{R}
\frac{\partial R^2}{\partial \tau} +
	\sum_i \partial_i\left(R^2\partial_i S\right) = 0 
\end{equation}
\begin{equation}\label{S}
\frac{\partial S}{\partial \tau} + \frac{1}{2}\sum_i(\partial_i S)^2 -
	V(x) - V_\mathrm{q}(x) = 0
\end{equation}
where
\begin{equation}\label{qpotential}
V_\mathrm{q} = -\frac{1}{2}\sum_i\frac{\partial_i^2 R}{R}
\end{equation}
is Bohm's quantum potential. 

The crux of this kind of hidden variable theory is that an ensemble of
particles evolving by (\ref{nelson}) will be distributed as $R(x,\tau)^2$
at time $\tau$ if they are initially distributed as $R(x,0)^2$ at time 0.
This follows from the fact that the Fokker-Planck equation 
\begin{equation}\label{fp}
\frac{\partial P}{\partial \tau} + \sum_i \partial_i\left[
\left(\partial_i S + \alpha\frac{\partial_i R}{R}\right)P - 
	\frac{\alpha}{2}\partial_i P\right] = 0
\end{equation}
for the distribution $P(x,\tau)$ generated by (\ref{nelson}) becomes
identical to the conservation equation (\ref{R}) if we put $P=R^2$.
Note that (\ref{R}) and (\ref{fp}) are unchanged from the case of real
time.

The effect of going to imaginary time is just to invert $V(x)$ and
$V_\mathrm{q}(x)$ in (\ref{S}), which resembles the classical
Hamilton-Jacobi equation for the Euclidean action $S_\mathrm{E} = \int H
d\tau$, where $H$ is the classical Hamiltonian. In fact, (\ref{S}) implies
that $S = \int (H+V_\mathrm{q})d\tau$, where the integral is taken along a
trajectory defined by the ``current velocity'' $v^i = \partial S/ \partial
x^i$, which means a Bohm trajectory.

Therefore, choosing $\alpha=2$ in (\ref{nelson}) gives the time-reverse of
the Langevin equation (\ref{langevin}) except for extra $\psi$-dependent
terms in the drift. The result is that, while the Langevin trajectories
sample $e^{-S_\mathrm{E}(x)}$ in the limit $\tau \rightarrow
\infty$, the Nelson trajectories generate a particle distribution that
samples $R^2$ itself for all $\tau > 0$. This suggests that we
look for a Langevin-like method of importance sampling in which the drift
is somehow guided by information we might have about $\psi(x,\tau)$.

\section{Guided Random Walks}

Indeed such a method already exists in the literature. It relies on a
trial wavefunction $\psi$ to generate trajectories useful for the
evaluation of observables, and these trajectories become exactly Nelson
trajectories as $\psi$ becomes more accurate.

The ground state expectation value of some observable $\mathcal{O}$ is
given by
\begin{equation}\label{Olim}
\langle \mathcal{O} \rangle =
\lim_{\tau \rightarrow \infty} 
\frac{ \langle \psi_0| \mre^{-\tau H} 
		\mathcal{O} \, \mre^{-\tau H} |\psi_0 \rangle }
	{ \langle \psi_0| \mre^{-2\tau H} |\psi_0 \rangle }
\end{equation}
where $|\psi_0\rangle$ is our trial evaluated at $\tau=0$. The exponential
factors in the numerator serve to dampen out any excited state components
in $\psi_0$ as $\tau \rightarrow \infty$. Identifying
\[
\langle \psi_0|\mre^{-\tau H}|x\rangle = \bar\psi(x,\tau)
\]
as a corresponding trial wavefunction for the imaginary time-reverse
Schrodinger equation, (\ref{itschrod}) with $\tau \rightarrow -\tau$, and
breaking the other $\mre^{-\tau H}$ factor into $N$ pieces
$\mre^{-\epsilon H} \cdots \mre^{-\epsilon H}$, the numerator in
(\ref{Olim}) can be written as a path integral:
\begin{equation}\label{pathint}
\int dx_0 \cdots dx_N \,
\bar \psi(x_0,0) \psi(x_0,0) \mathcal{O}(x_N)
\prod_{p=0}^{N-1} U_\epsilon(x_{p+1},x_p)
\end{equation}
\begin{equation}\label{U}
U_\epsilon(x_{p+1},x_p)
\equiv 
\langle x_{p+1}| \mre^{-\epsilon H} |x_p\rangle
\frac
	{\bar\psi(x_{p+1},\tau_{p+1})}
	{\bar\psi(x_p,\tau_p)}
\end{equation}
where $x_p=(x_p^0,\ldots,x_p^d)$ denotes a single classical configuration
of the system at time $\tau_p = p\epsilon$, and we have assumed that
$\mathcal{O}$ is a local observable with matrix elements $\langle
x| \mathcal{O} |x^\prime\rangle = \delta(x-x^\prime) \mathcal{O}(x)$.
The ratio of $\bar\psi$ terms in (\ref{U}) produces a sequence of
cancellations that allows us to have $\bar\psi$ evaluated at $\tau=0$ in 
(\ref{pathint}). To lowest order in $\epsilon$, it can be shown
\cite{Moskowitz} that
\begin{equation}\label{matrixel}
U_\epsilon(x_{p+1},x_p)
=
(2\pi\epsilon)^{-\frac{d}{2}} \; 
	\mre^{ -\sum_i\frac{1}{2\epsilon}\left(x_{p+1}^i-x_p^i -
		D_p^i\right)^2 }
	\mre^{-\Delta S_p}
\end{equation}
where the drift and residual action are given by
\begin{equation}\label{drift}
D_p^i = \epsilon 
\left[ 
	\frac{\partial}{\partial x^i} \log \bar\psi
\right]_{x_p}
\end{equation}
\[
\Delta S_p = \epsilon \left[H\bar\psi - \frac{\partial}{\partial\tau}
		\bar\psi \right]_{x_p} \, .
\]

In the method of guided random walks \cite{Negele}, the path integral
(\ref{pathint}) is sampled by stochastic trajectories over configuration
space $\{x\}$. In particular, the Gaussian factor in (\ref{matrixel}) is
interpreted as the probability to go from $x_p$ to $x_{p+1}$ over the
$p$-th time step, and $\mre^{-\sum_p\Delta S_p}$ gives the ``score'' of
the path $(x_0,\ldots,x_N)$ in the stochastic average.

Now, if $\bar\psi(x,\tau)$ were chosen to exactly satisfy the time-reverse
Schrodinger equation, we see that $\Delta S_p$ would vanish and the
trajectories would optimally sample (\ref{pathint}). But then, as
$\epsilon \rightarrow 0$, the jump probability in (\ref{matrixel}) and the
drift (\ref{drift}) exactly match those of Nelson's theory, (\ref{nelson})
with $\alpha=1$. In other words, the guided random walk method of
importance sampling can be seen as an attempt to generate Nelson
trajectories (for the time-reverse problem) with a trial wavefunction
defining the drift.

The fact that the generalized Nelson theory (\ref{nelson}) matches the
quantum evolution for any $\alpha>0$ suggests a new one-parameter family
of guided random walk algorithms for path integral evaluation, with
$\alpha$ controlling the strengths of both the diffusion and the drift.
This family is similar to the family of hybrid Langevin molecular dynamics
algorithms introduced by Kogut and Duane \cite{kogut}, with the pure
molecular dynamics algorithm in the Kogut family corresponding to Bohm
trajectories ($\alpha=0$) in our family. While both of these last
algorithms are deterministic, there does not appear to be any quantitative
relationship between them.

\section{Iterated Bohm Trajectories}

The effectiveness of importance sampling in a guided random walk algorithm
depends on the accuracy of the trial wavefunction. It would therefore be
desirable if one could use information gained from the trajectories
themselves to improve the trial and generate new trajectories. Iterating
the process may then produce a sequence $(\psi^{(0)},\psi^{(1)},\ldots)$
of successively better trial wavefunctions.

The following method in the case of pure Bohm trajectories is due to
Goldstein \cite{Goldstein}. Since Bohm's theory is equally applicable to
real time $t$, let us return to that case. Bohm's equations of motion,
obtained by using (\ref{nelson}) with $\alpha=0$ and differentiating the
real time version of (\ref{S}) are
\begin{equation}\label{bohm}
\frac{d}{dt} v^i = -\frac{\partial}{\partial x^i}
\left[V(x)+V_\mathrm{q}(x,t)\right]
\end{equation}
with the minus sign replaced by a plus sign in the imaginary time case of
(\ref{itschrod}). Here the Bohm particle velocity $v^i$ is identified with
$\partial S/\partial x^i$ as usual in Hamilton-Jacobi theory, except that
this $S$, from (\ref{S}), includes the effect of $V_\mathrm{q}$ in
addition to the classical potential $V$. Also, here, 
\[
\frac{d}{dt} = \frac{\partial}{\partial t} + 
	\sum_i v^i\frac{\partial}{\partial x^i}
\]
defines the the convective or ``along-the-trajectory'' derivative.

Writing $\psi^{(0)} = R^{(0)}e^{\mathrm{i}S^{(0)}}$, we evolve
trajectories according to (\ref{bohm}) with $R^{(0)}$ used to the
calculate $V_\mathrm{q}$ in (\ref{qpotential}). Were $\psi^{(0)}$ an exact
solution, (\ref{R}) and (\ref{S}) would ensure that these trajectories
generate a particle distribution equal to $(R^{(0)})^2$. However, with an
inexact trial wavefunction, the actual distribution will differ from
$(R^{(0)})^2$. We can thus use this actual distribution to define the next
iterate $R^{(1)}$, which can then be used to obtain a new $V_\mathrm{q}$,
new trajectories from (\ref{bohm}), and a new phase $S^{(1)}$ if desired.
Iterating this procedure yields a sequence $(\psi^{(0)}, \psi^{(1)},
\ldots)$ that has the exact solution $\psi$ as a fixed point; however,
convergence is not guaranteed for any fixed number of trajectories being
propagated.

There is also another technical issue. The equations of motion
(\ref{bohm}) will generate unique non-crossing trajectories when
$V_\mathrm{q}$ is obtained exactly from $\psi(x,t)$, but no such guarantee
exists for an inexact iterate $\psi^{(i)}$. Viewing (\ref{bohm}), with
$\psi^{(i)}$ used to calculate $V_\mathrm{q}$, as defining a map from
configuration space at time $0$ to itself at time $t$, one thus finds that
the map will generically be many-to-one. In effect, for $t$ large enough,
$\psi^{(i)}(x,t)$ would be a multivalued function of $x$. Either dealing
with this multivalued-ness or eliminating it by fiat poses a significant
problem in defining the algorithm.

Goldstein's iterative algorithm can also be compared to moving grid
methods for quantum dynamics that rely on Bohm trajectories to define the
grid \cite{Wyatt} \cite{Hersch2}. These methods propagate a swarm of Bohm
trajectories similar to the way a fluid dynamics algorithm propagates a
large number of fluid elements, and therefore they do not require any
trial wavefunction to get started. They are PDE propagator algorithms that
are truly local in time and (configuration) space.

\section{Discrete Beables}

Let us now consider methods posed explicitly for a finite dimensional
Hilbert space spanned by the basis states $|n\rangle$ with $n=1,\ldots,N$
and with Schrodinger equation
\begin{equation}\label{schro}
\frac{d}{dt}|\psi(t)\rangle = -\mathrm{i}H|\psi(t)\rangle \, .
\end{equation}

Originally for the purpose of attacking conceptual problems in quantum
field theory, John Bell found an analog of Bohm's model in which
trajectories are generated by ``beables'' (i.e. classical-like particles
in state space) stochastically jumping between states connected by
non-zero Hamiltonian matrix elements \cite{Bell}. In place of
(\ref{nelson}), Bell takes the probability for a trajectory to jump from
state $m$ to a distinct state $n$, sometime in the interval
$(t,t+\epsilon)$, as
\begin{equation}\label{T}
T_{nm}(t)\,\epsilon = \left\{
	\begin{array}{ll}
		2\,\mathrm{Re}\{z_{nm}(t)\}\,\epsilon & 
			\mbox{if Re$\{z_{nm}(t)\} > 0$} \\
		0 & \mbox{if Re$\{z_{nm}(t)\} \le 0$}
	\end{array}
\right.
\end{equation}
where we define
\begin{equation}\label{z}
z_{nm}(t) = -\mri H_{nm} \frac{\psi_n(t)^\ast}{\psi_m(t)^\ast} \, .
\end{equation}
and $\psi_n = \langle n|\psi\rangle$, etc.  To ensure normalization, the
probability for a trajectory to stay at $m$ is thus given by
$1-\sum_n^\prime T_{nm}(t)\epsilon$, where the primed sum excludes the
diagonal term $n=m$. Note the similarity between (\ref{z}) and (\ref{U}).
But here jumping from $m$ to $n$ is allowed only if there is a non-zero
matrix element $H_{nm}$ for the transition.

From (\ref{z}), we find
\begin{equation}\label{Rez}
\mathrm{Re}\{z_{nm}\} = -\mathrm{Re}\{z_{mn}\}
	\frac{|\psi_n|^2}{|\psi_m|^2} \, ,
\end{equation}
which implies that either $T_{nm} = 0$ or $T_{mn} = 0$. Together with
(\ref{schro}) this gives
\begin{equation}\label{modpsi}
\frac{d}{dt} |\psi_n|^2 = \sum_m 2\,\mbox{Re}\{z_{nm}|\psi_m|^2\}
= \sum_m (T_{nm}|\psi_m|^2 - T_{mn}|\psi_n|^2)
\end{equation}
in analogy with (\ref{R}). Note that the $T_{nm}$ term contributes when
Re$\{z_{nm}\}>0$, and the $T_{mn}$ term contributes when Re$\{z_{nm}\}<0$.

Now consider the distribution $P_n(t)$ of beables in state space, jumping
in accordance with (\ref{T}). The analog of the Fokker-Planck equation
(\ref{fp}) is
\begin{equation}\label{dfp}
\frac{d}{dt}P_n = \sum_m (T_{nm}P_m - T_{mn}P_n) \, .
\end{equation}
And this becomes identical to (\ref{modpsi}) if we put $P_n =
|\psi_n|^2$. Thus, provided $P_n(0) = |\psi_n(0)|^2$, we are guaranteed
$P_n(t) = |\psi_n(t)|^2$ for all $t>0$, just as in the continuous case.

\section{A New Guided Random Walk}\label{newwalk}

Can we find a stochastic method for path integral evaluation resembling
Bell trajectories? Consider the real-time analog of (\ref{pathint}) for a
finite dimensional Hilbert space, so that the continuous variables $x_p$
are replaced by discrete ones $n_p$. We can make a connection to Bell's
theory by expanding $\mre^{-\mri\epsilon H}$ to first order in $\epsilon$
and evaluating matrix elements. Assuming a diagonal observable $\langle n|
\mathcal{O} |m\rangle = \delta_{nm} \mathcal{O}_n$, we get
\begin{equation}\label{bellpath}
\langle \mathcal{O} \rangle_t = 
\sum_{\mathcal{P}(N)} 
	\psi_{n_N}(t)^\ast \psi_{n_0}(0) \, \mathcal{O}_{n_N}
	\prod_{p \in J} (-\mri\epsilon H_{n_{p+1}n_p})
\end{equation}
where $\mathcal{P}(N) = (n_0,\ldots,n_N)$ specifies a path in which
$H_{n_{p+1}n_p} \ne 0$ for each $n_{p+1} \ne n_p$. The jump set is defined
as $J[\mathcal{P}] = \{p\; |\; n_{p+1} \ne n_p\}$. Now we can use the same
trick of introducing $\psi$ ratios to write
\begin{equation}\label{zprod}
\psi_{n_N}(t)^\ast \prod_{p \in J} (-\mri H_{n_{p+1}n_p}) =
\psi_{n_0}(0)^\ast \prod_{p \in J}
	z_{n_{p+1}n_p}(t_p)
	\prod_{p \notin J} 
	\frac
		{\psi_{n_p}(t_{p+1})^\ast}			
		{\psi_{n_p}(t_p)^\ast}			
\end{equation}
with $z_{nm}$ as in (\ref{z}).

To realize (\ref{bellpath}) as a weighted average over stochastic
processes, we need to choose the jump rates $T_{nm}$ so that the
probability
\begin{equation}\label{probP}
\mathrm{Prob}[\mathcal{P}(N)] = \prod_{p \in J} \epsilon T_{n_{p+1}n_p}
	\prod_{p \notin J} (1-\epsilon\tsum{_n^\prime} T_{n n_p})
\end{equation}
of realizing a given path $\mathcal{P}(N)$ resembles the double product in
(\ref{zprod}) as closely as possible. $\sum_n^\prime$ is a sum
excluding
the diagonal term ($n = n_p$ here).

Consider the choice would be ($n_{p+1} \ne n_p$):
\begin{equation}\label{Tpath}
T_{n_{p+1}n_p} = |z_{n_{p+1}n_p}| a_p
\end{equation}
where the $z_{nm}$ are given by (\ref{z}) with our trial wavefunction
$\psi$, and the $a_p$ are to be determined. The remaining phase factors in
(\ref{zprod}) not assimilated into (\ref{probP}) will then constitute the
complex ``score'' of the path $\mathcal{P}(N)$ in the stochastic average.
That is, the stochastic average that computes (\ref{bellpath}) is obtained
by propagating trajectories according to (\ref{Tpath}), calculating the
score of each trajectory, and adding up all the scores.

One can show that apart from endpoint contributions around $p=0,N$, the
double products are matched in absolute value with error O$(N\epsilon^2)$
if the $a_p$ satisfy

\begin{equation}\label{ap}
\sum_p \left(
	\frac{\log a_{j(p)}}{N_p\epsilon} - 
	a_p \tsum{_n^\prime} |z_{n n_p}| - 
	\frac{\partial}{\partial t} \log |\psi_{n_p}(t_p)|
\right) = 0
\end{equation}
where $j(p)$ is the largest member of $J$ less than $p$, and $N_p$ is the
number of time steps between the $j(p)$-th step and the next jump.
Unfortunately we cannot make the sum vanish term by term. When we need to
choose $a_p$ at the $p$-th time step, we do not yet know the value of
$N_p$ for the trajectory, because we do not yet know when the next jump
will occur. It is therefore necessary to choose $a_p$ dynamically. At each
time step $p$ we can evaluate the sum (\ref{ap}) up to $p$ and choose
$a_p$ to stabilize it back toward 0. 

In particular, if we change $a_p$ only just after a jump, so that $a_p =
a_{j(p)+1}$ for all $p$, we can explicitly calculate the expectation value
of the sum (\ref{ap}) with respect to variations in $N_p$ produced by the
the jump probabilities (\ref{Tpath}). Setting this expectation value to
zero, term by term, and solving for $a_p$ yields a critical value such
that choosing $a_p$ above (below) the critical value will tend over time
to increase (decrease) the sum. This enables us to stabilize the sum and
achieve a correspondence between, on the one hand, a stochastic average
over trajectories defined by (\ref{Tpath}) and, on the other hand, the
path sum (\ref{bellpath}).

We have thus obtained a stochastic algorithm for evaluation of the
discrete path integral that resembles Bell's discrete hidden variable
theory. As opposed to the standard random walk algorithms, this one
involves trajectories that jump from one site to another only if there is
a non-zero matrix element for the transition. In particular, with a
spatial basis $|n\rangle$ and a local Hamiltonian, the path space is
reduced considerably. The price is that the trajectory scores become
complex in the case of imaginary time as well as in real time.

If the matrix elements $H_{nm}$ are bounded and independent of $t$, a
generic estimate may be obtained for the importance of different classes
of paths. Going back to (\ref{bellpath}) and performing the sum over paths
comprising the same sequence of jumps, differing only by when each jump
occurs, we have
\begin{equation}\label{bellpath2}
\langle \mathcal{O} \rangle_t = 
\sum_{M=0}^N 
	\mathcal{O}_{n_M} \bin{N}{M} \zeta_M[\psi]
\end{equation}
\begin{equation}\label{ZM}
\zeta_M[\psi] \equiv \sum_{\bar\mathcal{P}(M)} 
	\psi_{n_M}^\ast \psi_{n_0}
	\prod_{p=0}^{M-1}(-\mri \epsilon H_{n_{p+1}n_p})
\, .
\end{equation}
$\zeta_M$ is a kind of partition function summing over all paths
$\bar\mathcal{P}(M) = (n_0,\ldots,n_M)$ without stops: $n_{p+1} \ne n_p$
for all $p=0,\ldots,M-1$. The binomial coefficient counts the number of
paths $\mathcal{P}(N)$ from (\ref{bellpath}) that correspond to one path
$\bar\mathcal{P}(M)$ in (\ref{bellpath2}). 

Using Stirling's formula, (\ref{bellpath2}) may be approximated as
\begin{equation}\label{Osum}
\langle \mathcal{O} \rangle_t = \sum_M \mathcal{O}_{n_M} \mre^{S(M)}
\end{equation}
\begin{equation}\label{SM}
S(M) = M\log\left(\frac{\mre N}{M}\right)
	- \frac{M^2}{N} + \log \zeta_M 
\, .
\end{equation}
Now, $\log |\zeta_M| \sim M\log\epsilon$ can be interpolated into a smooth
function of $M$ for any finite $\epsilon$. However, from (\ref{ZM}) we
can write
\begin{equation}\label{ZM1}
\zeta_{M+1} = 
\sum_{\bar\mathcal{P}(M)} 
\psi_{n_M}^\ast \phi_{n_0}
\prod_{p=0}^{M-1}(-\mri \epsilon H_{n_{p+1}n_p})
\left[
	\tsum{_n^\prime} (-\mri\epsilon H_{n n_M})
	\frac{\psi_n^\ast}{\psi_{n_M}^\ast}
\right] \, ,
\end{equation}
which means that $\zeta_{M+1}$ is obtained from $\zeta_M$ by multiplying
each
term
in the path sum by a different factor with magnitude O$(\epsilon)$ and
unconstrained phase. This suggests that $\arg(\zeta_{M+1})$ will differ
randomly from $\arg(\zeta_{M})$, so that $\mathrm{Im}(\log \zeta_M) =
\arg(\zeta_M)$
can not be interpolated into a smooth function of $M$. Thus we expect
$\mathrm{Re}\{S\}$, not $\mathrm{Im}\{S\}$, to control which terms
are dominant in (\ref{Osum}).

To find these terms we need to evaluate
\[
\frac{d}{dM}\mathrm{Re}\{S\} = \log\left(\frac{N}{M}\right) - \frac{2M}{N}
	+ \frac{d}{dM} \log |\zeta_M| 
\, .
\]
Care must be taken in defining the derivative on the right, since $\log
|\zeta_M|$ diverges as $\epsilon \rightarrow 0$. We use the finite
difference
\[
\frac{d}{dM} \log |\zeta_M| \; \approx \; \log |\zeta_{M+1}| - \log
|\zeta_M| 
\]
together with (\ref{ZM1}) and the definition (\ref{z}) to get
\[
\frac{d}{dM} \log |\zeta_M| \; \approx \;
\log\left(
	\epsilon 
		\left|
		\left\langle \tsum{_n^\prime} z_{n n_M} \right\rangle_M
		\right|
\right)
\]
where $\langle \cdots \rangle_M$ averages over all paths
$\bar\mathcal{P}(M)$ with complex weight as in (\ref{ZM}).

Solving the saddle point equation $dS/dM = 0$ will determine which
path lengths $M$ are most important for computing (\ref{bellpath2}).
This equation has the form
\[
\log x + 2x + A = 0
\]
where $A \rightarrow \infty$ as $\epsilon \rightarrow 0$, and one can
verify that
\[
x = \mre^{-A - 2\mre^{-A - 2\mre^{-A-\cdots}}}
\]
is an explicit solution. When $A$ is large, $x = \mre^{-A}$ is a very good
approximation, which gives the saddle point equation
\begin{equation}\label{saddle}
M = t 
\left| 
	\left\langle \tsum{_n^\prime} z_{n n_M} \right\rangle_M
\right|
\, ,
\end{equation}
whose solution we denote $M=M_\ast$. Note that the right side of
(\ref{saddle}) itself depends on $M$. However, assuming $H_{nm}$ are
bounded, $\tsum{_n^\prime} \langle z_{n n_M} \rangle_M$ will be bounded,
and the solution $M_\ast$ will not be very sensitive to its
$M$-dependence.

To determine the spread of dominant terms around $M=M_\ast$, we evaluate
\begin{equation}\label{ddS}
\frac{d^2}{dM^2}\mathrm{Re}\{S\} = 
	-\frac{1}{M} - \frac{2}{N} + \frac{d^2}{dM^2} \log |\zeta_M|
\end{equation}
by putting
\begin{eqnarray*}
\frac{d^2}{dM^2} \log |\zeta_M| 
	& \approx & 
	\log |\zeta_{M+2}| - 2\log |\zeta_{M+1}| + \log |\zeta_M| \\
	& = & 
	\log\left( \frac{|\zeta_{M+2}/\zeta_M|}{|\zeta_{M+1}/\zeta_M|^2}
\right)	
\, .
\end{eqnarray*}
Separating out two extra matrix elements of $H$ in $\zeta_{M+2}$ as we had
separated out one matrix element in (\ref{ZM1}), we obtain
\begin{equation}\label{ddZ}
\frac{d^2}{dM^2} \log |\zeta_M| \; \approx \; 
\log
\left(
\frac
{\left|
	\left\langle \tsum{_{mn}^\prime} z_{mn} z_{n n_M} \right\rangle_M
\right|}
{\left|
	\left\langle \tsum{_n^\prime} z_{n n_M} \right\rangle_M
\right|^2}
\right)
\end{equation}
which is independent of $\epsilon$. If this term dominates in (\ref{ddS}),
it will control the spread of terms around $M=M_\ast$ that contribute to
(\ref{Osum}). Otherwise, $1/M_\ast$ will set the scale of
(\ref{ddS}), and the spread will be $\Delta M \sim \sqrt{M_\ast}$.

Consider a simple limiting case where the diagonal matrix elements $H_{nn}
= E_0$ are all equal, and we choose $|\psi\rangle$ in (\ref{bellpath2})
such that $H|\psi\rangle = E|\psi\rangle$. Now it is easy to verify that
$\tsum{_n^\prime} z_{n m} = E - E_0$ for any $m$, so that (\ref{saddle})
gives $M_\ast = |E-E_0|t$. Also, (\ref{ddZ}) vanishes, leaving the spread
as $\Delta M \sim (|E-E_0|t)^{1/2}$.

\section{Iterated Bell Trajectories}

While we have succeeded in re-analyzing the path sum in a way motivated by
Bell's jump rule (\ref{T}), the new rule (\ref{Tpath}) is somewhat
cumbersome to implement and gives rise to complex path scores. A more
natural approach is found by going the other way: re-analyzing Bell's jump
rule in a way motivated by path integral methods.

Consider the observable $\mathcal{O}=|n\rangle \langle n|$, so that
$\langle \mathcal{O} \rangle_t = |\psi_n(t)|^2$. (\ref{bellpath}) now
gives the path integral representation of a kind of Green function for
$|\psi_n(t)|^2$ rather than $\psi_n(t)$. The classical analog of this in
probability theory is just the equation
\begin{equation}\label{pathsm} P_n(t) =
\sum_{\mathcal{P}(N-1)} P_{n_0}(0) \, \mathrm{Prob}[\mathcal{P}(N)]
\end{equation} 
where the sum is taken over all paths $\mathcal{P}(N)$ with $n_N=n$ fixed.
This gives the expectation value of a classical state function 
$\mathcal{O}$ as
\begin{equation}\label{statefun}
\langle \mathcal{O} \rangle_t = \sum_{\mathcal{P}(N)}
\mathcal{O}_{n_N} P_{n_0}(0) \, \mathrm{Prob}[\mathcal{P}(N)]
\, .
\end{equation}
What makes this merely an analogy to the quantum case is that in the
latter the amplitudes involved in the path sum are generally complex,
seemingly a necessity to achieve the affects of quantum interference. But
Bell has shown us that we need not be content with just an analogy. The
integral formulation of Bell's model based on (\ref{probP}) re-expresses
the quantum path sum with a real, strictly positive path amplitude,
exactly as in (\ref{pathsm}) and (\ref{statefun}). This is accomplished
simply by choosing the $T_{nm}$ according to (\ref{T}), with the result
that $P_n(t) = |\psi_n(t)|^2$ for all $t>0$ if we set $P_n(0) =
|\psi_n(0)|^2$. Quantum interference effects, here, are not manifested by
the contributions from certain paths cancelling those from other paths,
but rather by the propensity of beable trajectories to follow or not to
follow certain paths in the first place. The sum over histories is
manifestly undemocratic here.

To implement this as a stochastic algorithm we need to use a trial
wavefunction for $\psi$ in (\ref{z}). The beable distribution $P_n(t)$ of
the resulting trajectories then serves to compute $|\psi_n(t)|^2$ and
$\langle \mathcal{O} \rangle_t$ as stochastic averages for (\ref{pathsm})
and (\ref{statefun}) respectively.

In analogy with Goldstein's algorithm, we might then attempt to use these
trajectories to generate a new trial $\psi^{(1)}$, which could be used to
generate new trajectories, etc. The problem is that, while in the
continuous case knowledge of $P^{(i)}=|\psi^{(i)}|^2$ is enough to compute
new trajectories via (\ref{bohm}), here we need $\mathrm{Re}\{z^{(i)}\}$
to compute new trajectories via (\ref{T}). But $\mathrm{Re}\{z^{(i)}\}$
depends on the phase as well as the amplitude of $\psi^{(i)}$.

What must be done is to derive an evolution equation for
$\mathrm{Re}\{z_{nm}\}$ from (\ref{schro}) in the discrete case, just as
Bohm obtained (\ref{bohm}) from the continuous Schrodinger equation. Using
(\ref{schro}) to differentiate the definition (\ref{z}) gives
\begin{equation}\label{dzdt}
\frac{d}{dt}z_{nm} = z_{nm} \sum_k (z_{km} - z_{kn})
\end{equation}
which contains the same information as (\ref{schro}) itself. Notice that
the (time-independent) Hamiltonian does not enter as a parameter in this
equation, but only in the initial conditions when the $z_{nm}(0)$ are
given in terms of the $\psi_n(0)$. An additional $dH/dt$ term would appear
in (\ref{dzdt}) for the time-dependent case.

Now consider propagating (\ref{dzdt}) along a trajectory $\mathcal{P}$ at
the $i$-th stage of iteration. One scheme would be to evolve
$z_{n_{p+1}n_p}^{(i)}$ along $\mathcal{P}$ by evaluating the right hand
side of (\ref{dzdt}) with $z^{(i-1)}$. But we might as well take the
$z_{nm}$ factor on the right hand side as the current iterate $z^{(i)}$,
since we can do so with the information obtained just from propagating
$\mathcal{P}$. Thus we propagate
\begin{equation}\label{dzdti}
\frac{d}{dt}z_{nm}^{(i)} = z_{nm}^{(i)} 
	\sum_k (z_{km}^{(i-1)} - z_{kn}^{(i-1)})
\end{equation}
along beable trajectories generated by the jump rule $T_{nm}^{(i)} =
2\mathrm{Re}\{z_{nm}^{(i-1)}\}$ ($\mathrm{Re}\{z\}>0$). An alternative
scheme can be obtained by defining
\begin{equation}\label{defZ}
Z_m \equiv \sum_n z_{nm} = 
\sum_n -\mri H_{nm} \frac{\psi_n(t)^\ast}{\psi_m(t)^\ast} =
-\frac{d}{dt}\log\psi_m^\ast(t)
\end{equation}
and summing (\ref{dzdt}) over $n$, which gives
\begin{equation}\label{derZ}
\frac{d}{dt}Z_m = Z_m^2 - \sum_{n} Z_n\,z_{nm} \, .
\end{equation}
Using (\ref{defZ}) in (\ref{dzdt}), we obtain the time-dependence $z_{nm} 
\propto \exp[\int(Z_m-Z_n)dt]$ and can thus eliminate the $z_{nm}$ from
(\ref{derZ}). After extracting the $n=m$ term from the sum, we have
\begin{equation}\label{dZdt}
\frac{d}{dt}Z_m = Z_m^2 + \mri H_{mm} Z_m - 
\sum_{n(\ne m)} 
Z_n\,z_{nm}(0)\exp\left(\int_0^t \left[Z_m(s) - Z_n(s)\right] ds \right)
\end{equation}
as a replacement for (\ref{dzdt}). This can be cast for an iteration
scheme by taking the $Z$'s in the sum as $(i-1)$-th iterates, hence known
functions in the above equation, and the other $Z$'s as $i$-th iterates.
It would also be possible to keep $Z_m$ in the sum as an $i$-th iterate,
but this would result in a complicated integro-differential equation.

Having propagated the $i$-th ensemble of trajectories and calculated
$Z^{(i)}$---$z^{(i)}$ in the former scheme---along each of them, we still
have to determine $Z^{(i)}$ for the sites in state space that neither
contain or neighbor any members of the beable ensemble at each time step.
The natural choice in the present context of a discrete state space is
simply to retain the $Z^{(i-1)}$ values at these sites. Since trajectories
are attracted to sites with large $|\psi^{(i)}|^2$, this will cause the
algorithm to concentrate on these sites and not on the low probability
ones.

In essence, this scheme is constructed so that the iterate $Z^{(i-1)}$
guides the $i$-th set of trajectories, which then serve as an importance
sampling prescription to calculate the next iterate $Z^{(i)}$. In this way
an initial trial $Z^{(0)}$, or equivalently $\psi^{(0)}$, may be
iteratively improved as the algorithm (hopefully) converges in computing
the dynamical expectation value $\langle \mathcal{O} \rangle_t$ for a
given observable $\mathcal{O}$.

The effectiveness of this method of importance sampling, here in the
context of sampling entire dynamical evolutions as opposed to simply
ground state distributions, is a delicate matter above and beyond
questions of the stability and accuracy of the iteration scheme itself.

In implementing this algorithm, one must be cautious of potential
divergences of the $Z_m$, since this will generally occur if
$\psi_m=0$. Fortunately, we will not have to control these divergences in
a very sensitive manner when solving the dynamical equations (\ref{dZdt})
themselves because we expect successive iterates to iron out any initial
imperfections. Our goal is then merely to prevent catastrophic events that
might corrupt computations occuring on neighboring sites. We can
accomplish this much with a simple implicit method (supressing the site
index):
\[
Z(t+\Delta t) = Z(t) + Z^\prime(t)\Delta t + \mathrm{O}(\Delta t^2)
	      = \frac{Z(t)}{1 - \frac{Z^\prime(t)}{Z(t)}\Delta t}
			+ \mathrm{O}(\Delta t^2) \, .
\]
In fact, for better accuracy, we adopt a second order implicit method
based on the relation:
\[
Z(t+\Delta t) = \frac{Z(t)}{1 - \frac{Z^\prime(t)}{Z(t)}\Delta t
	+ \left( 
		\frac{Z^\prime(t)^2}{Z(t)^2} - 
		\frac{1}{2}\frac{Z^{\prime\prime}(t)}{Z(t)}
	\right)\Delta t^2
	}
	+ \mathrm{O}(\Delta t^3) \, .
\]
The form of the denominator has been chosen to reproduce the appropriate
Taylor expansion up to second order and requires a computation of
$Z^{\prime\prime}(t)$, which can be accomplished by differentiating
(\ref{dZdt}) once to get
\[
\frac{d^2}{dt^2}Z_m = (2Z_m + \mri H_{mm})\frac{dZ_m}{dt} - 
\sum_{n(\ne m)} Z_n\,z_{nm} \left[\frac{dZ_n}{dt} + Z_n(Z_m-Z_n)\right] \, .
\]
To implement this in our iteration scheme, we must remember that terms in
the sum are to be taken as $(i-1)$-th iterates and terms outside the sum
as $i$-th iterates. 

In order to propagate (\ref{dZdt}) along trajectories we need to compute
both $Z$ and $\int Z dt$ at each step. The latter must also be done to
second order by expanding the integrand in $\int_t^{t+\Delta t} Z(s) ds$
around $s=t$:
\[
\int_t^{t+\Delta t}\left[Z(t) + Z^\prime(t)(s-t) + \cdots \right] ds = 
	Z(t)\Delta t + {\textstyle\frac{1}{2}} Z^\prime(t)\Delta t^2
	+ \mathrm{O}(\Delta t^3) \, .
\]
There is nothing in principle to prohibit an implicit third or higher
order method for propagating $Z$ and $\int Z dt$ along these lines; each
higher order would simply require another differentiation of (\ref{dZdt}).

The distinguishing characteristic of an iterative algorithm such as the
one presented here is that while the solution $Z$ at some time $t$ will be
directly fixed in any one iteration by $Z$ at times just prior to $t$,
successive iterates will bring in information from much further back
before $t$. The global nature of this process is paid for by the
computational cost of propagating a whole new time-development for each
iterate.

However, it is possible to taylor this trade-off by subdividing the total
simulation interval $(0,T)$ into windows $(t_{w-1},t_w)$ each
of some fixed duration $\Delta t_W$ so that iteration will proceed in each
window until convergence is obtained before moving on to the next window.
Thus, in the first window, the initial conditions $Z(t=0)$, together with
some initial guess for $Z(0<t<t_1)$, will be used to begin iteration, and
after convergence is reached the solution endpoint $Z(t_1)$ will provide
the initial condition for the next sequence of iterations over
$(t_1,t_2)$, etc.. We will determine that convergence has been reached
by the $i$-th iterate in a given window $(t_{w-1},t_w)$, if the
correlation coefficient of the two data sets $\{Z_m^{(i-1)}(t_w)\}$ and
$\{Z_m^{(i)}(t_w)\}$ exceeds some threshold value very close to 1.

As with the guided random walk methds, we also need some kind of trial
wavefunction---here, as the initial iterate $\psi^{(0)}(t)$ or
$Z^{(0)}(t)$ for each window $(t_{w-1},t_w)$.  One may obtain this
$\psi^{(0)}$ from a completely separate method, which would likely rely on
considerations more specific to the Hamiltonian in question. However, in
order to concentrate on the iterative algorithm alone, we will use a
rather simple trial: 
\begin{equation}\label{Z0}
Z^{(0)}(t) = Z^{(f)}(t_{w-1}) \;\;\;t \in (t_{w-1},t_w)
\end{equation}
where $f$ indicates the final iteration of the previous window. The
benefit of this is to make the algorithm totally self-contained. The costs
are an increase in the number of iterations necessary for convergence and,
most probably, decreases in the stability of the algorithm as
$T$ grows.

All that remains is to specify a physical system and initial condition,
and then we can investigate the performance of this iterative trajectory
algorithm as a function of its various paramters. We will consider a 1d
ferromagentic lattice of spins $\mbm{\sigma}_n$ with Heisenberg
Hamiltonian
\begin{equation}\label{heis} 
H = -\frac{1}{2}\sum_{n=1}^N \mbm{\sigma}_n \cdot\mbm{\sigma}_{n+1} 
\end{equation}
where we have taken the prefactor $1/2$, together with $\hbar=1$, as
defining our unit of time. Periodic boundary conditions are imposed so
that $\mbm{\sigma}_{N+1} = \mbm{\sigma}_1$. Since Bell's theory operates
in a preferred basis that defines the (discrete) classical configuration
space over which trajectories are propagated, let us select the tensor
product basis of $\sigma^z$ eigenstates for each spin. In particular, we
will confine ourselves to the (dynamically invariant) subspace of a single
spin excitation, i.e. the eigenspace of $\sum_n \sigma_n^z$ with second
lowest eigenvalue. This implies an $N$-element set of classical
configurations indexed by $n$, corresponding to the possible locations of
a single spin excitation on the 1d circular lattice.

With the above definitions, we see that $\langle n|H|m\rangle$ vanishes,
hence so does $z_{nm}$, unless $|n-m|\le 1(\mbox{mod}N)$. The jumping
rates determined by (\ref{T}) then imply that particles can jump only
between neighboring sites on the lattice (in a single time-step $\Delta
t$). 

Exact energies and eigenstates for (\ref{heis}) may be obtained in the
single spin excitation subspace \cite{feynspin}; the eigenstates are found
to be \emph{spin waves} on the lattice:
\[
|\phi_a\rangle = \sum_{n=1}^N e^{2\pi\mri a/N}|n\rangle
\]
where $a$ is an integer, with $-N/2 \le a \le N/2$, characterizing the
energy $E_a = 4\sin^2(\pi a/N)$ and direction of the spin wave. One can
form quasi-coherent states of left/right-moving spin waves, which we will
approximate as 
\[
|\alpha\rangle_{L/R} =
	\sum_a \exp\left[\frac{(a-|\alpha|^2)^2}{2|\alpha|^2} \pm 
	\mri\, a\arg\alpha \right] |\phi_a\rangle
\]
where left-movers ($L$) take the $+$ sign and a sum over $0\le a\le N/2$,
and right-movers ($R$) take the $-$ sign and a sum over $-N/2 \le a \le
0$.

We can thus define our initial state at $t=0$ as a superposition
\[
|\psi(0)\rangle = |\alpha\rangle_L +
	{\textstyle\frac{1}{2}}|\!-\!\alpha\rangle_R \, .
\]
We will take $\alpha=\sqrt{6}$, which corresponds to coherent state
distributions peaked around six spin wave quanta. The left-moving state is
centered around the position $n=0$ at $t=0$, while the extra $\pi$ phase
of the right-mover $|\!-\!\alpha\rangle_R$ shifts its center to $n=N/2$.
The two first make contact around $t=7$ in our units, at which point they
begin to interfere with each other.

\begin{figure} 
\centering
\scalebox{.75}{\includegraphics{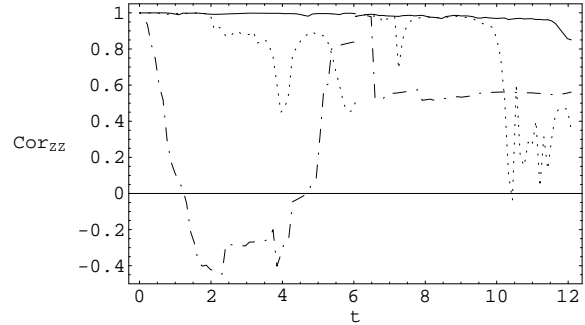}} 
\caption{\small
The correlation coefficient $\mathrm{Cor}_{ZZ}$ between
calculated and exact values of Im$\{Z^{(i)}_m(t)\}$ at iteration $i=1$
(dahsed), $i=120$ (dotted), and $i=1360$ (solid) for the first window $t
\in (0,6.05)$---and at $i=1$ (dashed), $i=120$ (dotted), and $i=1230$ for
the second window $t \in (6.05,12.1)$.}
\label{cor}
\end{figure}

We illustrate the action of our algorithm, with two large windows of size
$\Delta t_W = 6.05$, $N_\mathrm{traj}=50$ trajectories, and a time step
$\Delta t = 0.018$.  Plotting the time dependence of the correlation
coefficient between calculated and exact values of Im$\{Z_m(t)\}$ in Fig.\
\ref{cor} at various stages in the iteration, we see how our inaccurate,
static initial iterate is gradually ironed-out over the course of many
iterations. (We have used Im$\{Z\}$ and not Re$\{Z\}$ here because the
Im$\{Z^{(0)}\}$ values are more stark in their deviation from the exact
values.) The poor quality of the static initial iterate $Z^{(0)}$ together
with the large window size used in this illustration necessitate many
iterations ($\sim 1300$ per window) to achieve the level of accuracy shown
in the figure. Shorter windows, relative to the natural system time-scale,
will likely be preferable in most calculations.

\begin{figure} 
\centering
\scalebox{.75}{\includegraphics{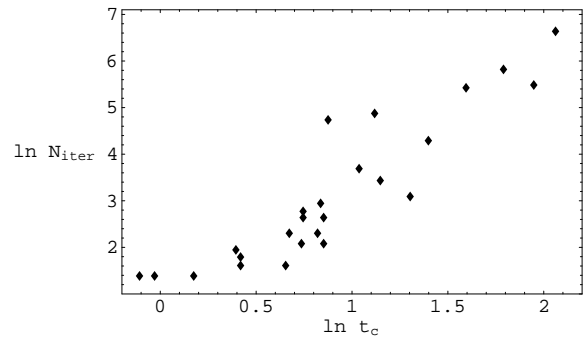}} 
\caption{\small
The number $N_\mathrm{iter}$ of iterations (convergence required at the
$10^{-8}$ level) necessary to achieve a given stability time $t_c$, as we
vary the window size $\Delta t_W$. We have used $N_\mathrm{traj}=10$
trajectories, and a time step $\Delta t = 0.018$ here.
}
\label{Dt}
\end{figure}

In general, the larger we make $\Delta t_W$, the more (temporally) global
the algorithm can be, and the more accurate the algorithm is in
calculating $Z$ at some fixed time $t$. In the absence of an accurate,
long-time zeroth iterate $Z^{(0)}$, we can expect that lack of exact
convergence over successive windows (and over longer times within each
window) will cause an accumulation of errors that ultimately destabilize
the algorithm by some time $t_c$. We can somewhat arbitrarily peg this
stability time as that at which the correlation coefficient between
calculated and exact values of Im$\{Z_m(t)\}$ first dips below some
threshold, say 0.75. Increasing $\Delta t_W$ will tend to increase $t_c$,
but it will also require more iterations within each window to achieve a
desired level of convergence, say to one part in $10^8$, as measured by
the correlation between Re$\{Z^{(i-1)}_m(t_w)\}$ and Re$\{Z^{(i)}_m
(t_w)\}$. We thus face a trade-off between the required number of
iterations $N_\mathrm{iter}$ for some fixed $\Delta t_W$ and the stability
time $t_c$ (see Fig.\ \ref{Dt}).

\begin{figure} 
\centering
\scalebox{.75}{\includegraphics[bb = 20 20 200 200]{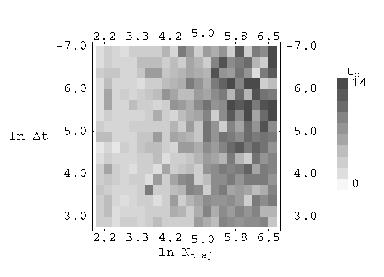}}
\caption{\small
The stability time $t_c$ as a function of the time step $\Delta t$ and
number $N_\mathrm{traj}$ of trajectories. We have used a window size
$\Delta t_W=1.0$ here.
}
\label{dtrep}
\end{figure}

Similar trade-offs exist between $t_c$ on the one hand and on the other:
(i) the number $N_\mathrm{traj}$ of trajectories used in our ensemble, and
(ii) the size of the time step $\Delta t$. Fig.\ \ref{dtrep} gives an idea
of the interplay between these two. The general lack of smoothness in
these plots is the result of the algorithm's sensitivity to small changes
in certain of its parameters. For instance, small changes in $\Delta t_W$
may shift the window edges into or out of resonances in the dynamics;
difficult moments in the simulation may respond differently depending on
where they occur relative to these window edges.

The results presented here are limited by the choice of a static initial
iterate (\ref{Z0}). Ultimately this purification scheme will depend on the
quality of that with which we begin the purification. What we have then is
at least one stage in a migration of configuration space methods to
the problem of simulating dynamics.

\newcommand{\mrf}[0]{\mathrm{f}} 
\newcommand{\mcM}[0]{\mathcal{M}} 

\chapter{Beables for Quantum Control}
\label{chap-control}

Beable methods offer two potential benefits in simulating quantum
systems. We have already discussed how beables can help to define
iterative numerical algorithms for simulating dynamics over an associated
classical state space. But the beale framework also offers the possibility
of a more concrete physical interpretation of simulation results
themselves. Moreover, when we are interested not only in properties of
some final state but also questions concering exactly \emph{how} that
final state was arrived at, beables can be of use. One current research
area in particular is prone to such questions: quantum control of
molecular systems via pulse-shaped lasers.

Advances in amplitude and phase modulation for ultrafast lasers, fast
detection techniques, and their integration via closed-loop algorithms
have made it possible to control the dynamics of a variety of quantum
systems in the laboratory. Excitation may be either in the strong or weak
field regime, with the goal of obtaining some desired final state. Success
in achieving that goal is gauged by a detected signal (e.g., the mass
spectrum in the case of selective molecular fragmentation), and this
information is fed back into a learning algorithm \cite{control}, which
alters the laser pulse shape for the next round of experiments. High duty
cycles of $\sim 0.1$ seconds or less per control experiment make it
possible to iterate this process many times and perform efficient
experimental searches over a control parameter space defining the laser
pulse shape.

As an example of this process, experiments have employed closed-loop
methods for selective fragmentation and ionization of organic
\cite{organic} and organometallic \cite{metal} \cite{Na2K} compounds, as
well as for enhancing optical response in solid-state and other chemical
systems \cite{AlGaAs} \cite{Xray} \cite{dye} \cite{CH3OH}. Yields of
targeted species are typically enhanced considerably over those obtained
by non-optimized methods. It is found that the optimal pulse shapes
achieving these enhancements can be quite complicated, and understanding
their physical significance has proven difficult. The same general
observations also apply to the many optimal control design simulations
carried out in recent years \cite{Dahleh} \cite{Kosloff} \cite{Zhu}
\cite{Rice}.

This chapter will address the identification of control mechanisms in
theoretical calculations as well as for direct application in the
laboratory by offering a definition of mechanism in terms of beable
trajectories.

\section{Beables and Quantum Theory}\label{beables}

Consider a control problem posed in terms of the quantum evolution
\begin{equation}\label{schrod} 
\mri \hbar \frac{d}{dt}|\psi(t)\rangle = (H_0-\mu E(t))|\psi(t)\rangle 
\end{equation} 
over a finite dimensional Hilbert space with basis $|n\rangle$ where
$n=0,1,2,\ldots$. Here $H(t)=H_0-\mu E(t)$ incorporates the effect of the
control field $E(t)$ via the dipole moment operator $\mu$, and we can
explicitly follow the evolution of $|\psi\rangle$ into a desired final
state $|\psi(t_\mathrm{f})\rangle$.

We are concerned with the question: what is the importance of a given
sequence $n_1 \rightarrow n_2 \rightarrow \cdots$ of actual
transitions---or, more specifically, of a given trajectory defined as a
function $n(t)$ of time---in achieving the desired final state
$|\psi(t_\mathrm{f}) \rangle$? In other words, it is clear that the system
\emph{is} being driven into a desired state, but can we find a physical
picture of \emph{how} this is being accomplished?

A conventional answer to the question raised above, essentially that given
by Bohr \cite{Bohr} on first seeing Feynman's path integral, is to reject
the question as ill-posed because quantum mechanics is said to forbid
consideration of precisely defined trajectories over the classical state
space $\{n\}$. Nevertheless, it is well established that there exist
dynamical models generating an ensemble of trajectories $n(t)$ whose
statistical properties exactly match those associated with
$|\psi(t)\rangle$ at each $t$. In the case of a continuous state space,
the first such model (as sketched in Chapter \ref{chap-path}) was that of
de Broglie \cite{Broglie}, later rediscovered and completed by Bohm
\cite{Bohm}. They reintroduce classical-like particle trajectories into
quantum theory by taking the probability current $\mathbf{J}[\psi]$ to
describe a statistical ensemble of real particles. So,
\begin{equation}\label{v}  
\mathbf{v} = \frac{\mathbf{J}}{|\psi|^2}
= \mathrm{Re} \left\{
	-\mri \frac{\hbar}{m} 
	\frac{\nabla\psi(\mathbf{x},t)}{\psi(\mathbf{x},t)}
\right\}
\end{equation}
gives the velocity of a particle with mass $m$ and position $\mathbf{x}$
at time $t$, in de Broglie-Bohm (dBB) theory. The physical particle is
taken to exist independently of, but also to have its motion determined
by, the wavefunction $\psi$. The time evolution of $\psi$ itself is just
given by the Schrodinger equation.

Bohm developed a full account of how ensembles of such classical-like
particles could reproduce the predictions of quantum mechanics. A basic
issue is to compare $\psi(\mathbf{x},t)$ with the statistical distribution
$P(\mathbf{x},t)$ describing an ensemble of particles evolving by
(\ref{v}). One can show that if the initial distribution of particles
satisfies $P(\mathbf{x},0) = |\psi(\mathbf{x},0)|^2$, then
$P(\mathbf{x},t) = |\psi(\mathbf{x},t)|^2$ will hold for all $t>0$. That
is, if the ensemble is initially in the ``quantum equilibrium''
distribution given by $|\psi(\mathbf{x},0)|^2$, the dynamics---(\ref{v})
for the particles, and the Schrodinger equation for $\psi$---will preserve
this equilibrium, consistent with the predictions of standard quantum
theory \cite{Bohm}\cite{Nino}. The result is easily generalized to
arbitrary interacting $N$-particle systems by taking $\mathbf{x}$ as a
point in the $3N$ dimensional configuration space.

In dBB theory the position representation has a special status. While one
may still regard $\psi$ as a basis-independent object, the particle
dynamics is given by (\ref{v}) specifically in terms of $\langle
\mathbf{x} | \psi \rangle$ rather than $\langle \mathbf{p} | \psi \rangle$
or some other representation. But, it is easy to formulate analogs of dBB
theory in different bases. For instance, one might choose the momentum
values $\mathbf{p}$ as the beables of the theory, and the dBB trajectories
$\mathbf{x}(t)$ would be replaced by momentum space trajectories
$\mathbf{p}(t)$.

In the context of a finite dimensional Hilbert space with basis
$|n\rangle$, the beables can be taken as the sites $n$ of the classical
state space $\{n\}$ analogous to $\{\mathbf{x}\}$ or $\{\mathbf{p}\}$.
Some law analogous to (\ref{v}) must be given to generate beable
trajectories $n(t)$ over the state space. Such trajectories would provide
a physical picture of the quantum transitions induced by a control field
$E(t)$. John Bell's definition (\ref{T}) of stochastic trajectories over
$\{n\}$ is one such law. Moreover, it can be shown \cite{Guido} to be
minimal in the sense that any alternative law will require higher jump
rates---in fact, these higher rates are such that the increased flux
associated with jumping from $n$ to $m$ is found to exactly counterbalance
that in the opposite direction.

The answer to the initial question regarding the importance of a given
trajectory in achieving the desired state $|\psi(t_\mathrm{f})\rangle$ is
quite simple in Bell's theory. The importance may be taken as just the
path probability (\ref{probP}).

The argument given in Chapter \S\ref{chap-path} for the equivalence of
Bell's theory and ordinary quantum mechanics ensures that the path
probabilities $\mathrm{Prob}(\mathcal{P})$ are consistent with the quantum
distribution $|\psi_n(t)|^2$ governing observables. But, it should be
noted that Bell's theory is not unique in this regard. The rule (\ref{T})
may be altered in non-trivial ways while preserving the master equation
(\ref{modpsi}) \cite{Guido}, although Bell's rule is minimal in the sense
mentioned above. The definition (\ref{T}) might even be changed in ways
that do not preserve (\ref{modpsi}), if one is willing to relinquish a
strict probability interpretation for the trajectories \cite{Mitra1}.

In general, there are many different ways to assign probabilities to
trajectories that all result in the same time-dependent occupation
probabilities $P_n(t)$. The predictions of quantum mechanics, therefore,
cannot select a single assignment. This non-uniqueness at the root of
quantum mechanism identification can be dealt with only by reference to
the simplicity and explanatory power of a given mechanism definition.
Below we adopt the definition (\ref{T}).

\section{Simulating Beables in Quantum Control}\label{simulating}

Our ultimate goal is to obtain dynamical mechanism information directly
from experimental data associated with the closed-loop control field
optimization, without pre-existing knowledge of the system Hamiltonian or
wavefunction. Methods employing Bell's theory for this purpose are
presented in \S\ref{mechanism}, but first we shall study control
mechanisms for a model system whose Hilbert space and quantum evolution
are given explicitly in numerical simulations.

Consider a quantum-optical system with level energies $\hbar \omega_n$ and
dipole moments $\mu_{nm}$.  Applying an external laser field $E(t)$, the
Hamiltonian in the interaction picture is
\begin{equation}\label{HI}
H_I = E(t) \sum_{n\,m} \mu_{nm} \mre^{\mri \omega_{nm} t}
					|n\rangle \langle m|
\end{equation}
where $\omega_{nm} \equiv \omega_n - \omega_m$. We will drop the subscript
$I$ from now on. The definition here of $|n\rangle$ as interaction picture
states has the affect of eliminating larger contributions to the jump
probabilities $T_{nm}$ from the $\hbar \omega$ terms, hence reducing the
overall frequency of jumps. $E(t)$ is assumed to be given by an
independent optimization algorithm designed to, for example, maximally
transfer population from $|n_\mri\rangle$ to $|n_\mathrm{f}\rangle$.

A simple second-order Schrodinger propagator was used to solve
(\ref{schrod}) in the interaction picture, relying on a factorization of
the evolution operator as
\begin{equation}\label{evol}
\mathcal{T}\left\{\mre^{-\frac{\mri}{\hbar}\int_0^t H(s)ds}\right\} = 
\prod_{p=0}^{N-1} \mathcal{T}\left\{
	\mre^{-\frac{\mri}{\hbar}\int_{t_p}^{t_{p+1}} H(s)ds}
\right\}
\end{equation}
where $t_p = p\epsilon \equiv pt_\mathrm{f}/N$ and $\mathcal{T}$ is the
time-ordering symbol. Choosing a time step $\epsilon \ll \hbar/\mu E$, we
can approximate (\ref{evol}) by dropping the $\mathcal{T}$ operations on
the right hand side and computing the integrals directly. In doing this an
error is accrued per time step given by the Baker-Hausdorf identity
$\mre^{A+B} = \mre^A\mre^B\mre^{-\frac{1}{2}[A,B]+\cdots}$ as
\begin{equation}\label{com}
\frac{1}{\hbar^2} \int_{t_p}^{t_{p+1}} \int_{t_p}^{t_{p+1}}
	 [H(r),H(s)]\,dr\,ds \; \sim \;
\left(\frac{\mu E}{\hbar}\right)^2 \epsilon^3 \omega
\, .
\end{equation}
The right hand estimate is obtained by expanding $H(r)$ to first order
about $r=s$ and noticing that the $E^\prime(s)$ term in $H^\prime(s)$
commutes with $H(s)$. The error (\ref{com}) would generally dominate third
order terms like $(\mu E \epsilon/\hbar)^3$.

If the control field is given as $E(t) = \mathrm{Re}\{\sum_i
\alpha_i E_i(t)\}$,
where
\[
E_i(t) = A(t) \mre^{\mri(\phi(t) + \omega_i^\mathrm{c}t)} 
\]
with $A(t)$ and $\phi(t)$ possibly adiabatic, we can evaluate $\int
H(s)ds$ by writing
\begin{equation}\label{intA}
\int_{t_p}^{t_{p+1}} \mu E_i(s) \mre^{\mri \omega s} ds
\; \approx \;
\frac{ \mu A(t_p) \mre^{\mri\phi(t_p)} }
	{ \mri ( \omega + \omega_i^\mathrm{c} ) }
	\left( 
	\mre^{ \mri (\omega + \omega_i^\mathrm{c}) t_{p+1} } 
	- \mre^{ \mri (\omega + \omega_i^\mathrm{c}) t_p } 
	\right)
\; .
\end{equation}
(Simply writing $\int H(s)ds \approx \epsilon H(t_p)$ is not appropriate
because we do not want to exclude weak field excitation, i.e.\ $\mu
E \ll \hbar \omega$, so that $\omega \epsilon \sim 1$ may hold.) Thus in
the adiabatic case $|\psi(t)\rangle$ can be propagated in steps determined
by $A(t)$ and $\phi(t)$ rather than the phase factors $\mre^{\mri\omega
t}$.

Consider the evolution of beable trajectories according to (\ref{T}),
which appears to require a time step small enough that each part of $H$,
including the $\mre^{\mri\omega t}$ terms, not vary much over the step.
Nevertheless, the total probability of jumping from $m$ to $n$ over
$(t_p,t_{p+1})$ is given by the integral $\int T_{nm}(s)ds$ over that
range with $\sim (\mu E\epsilon/\hbar)^2$ corrections. Thus we can take an
effective jump probability for the interval $(t_p,t_{p+1})$ as given by
(\ref{T}) with
\begin{equation}\label{zint}
z_{nm}(t_p) \; \approx \;
-\frac{\psi_n(t_p)^\ast}{\psi_m(t_p)^\ast}
	\frac{\mri}{\hbar \epsilon} \int_{t_p}^{t_{p+1}} H_{nm}(s) ds
\end{equation}
evaluated using (\ref{intA}). Note that we have included a factor of
$1/\hbar$
explicitly into the definition (\ref{z}) of $z_{nm}$. If $\omega\epsilon
\ll 1$ does not hold, care must be taken to extend the integration in
(\ref{zint}) only over $t \in (t_p,t_{p+1})$ for which
$\mathrm{Re}\{z_{nm}(t)\}>0$, leading to additional boundary terms in the
phase difference part of (\ref{intA}). Moving the $\psi^\ast$ ratio
outside the integral in (\ref{zint}) produces an error per time step of
order
\[
\frac{\epsilon^2 H}{\hbar} 
\frac{\partial \psi}{\partial t} 
\; \sim \;
\left(\frac{\mu E \epsilon}{\hbar}\right)^2
\]
which is again comparable to (\ref{com}). Therefore beable trajectories
may be propagated in steps determined by the possibly adiabatic amplitude
$A(t)$ and phase $\phi(t)$, i.e.\ synchronously with the Schrodinger
propagator.

\section{Mechanism Analysis for a Model 7-Level System}
\label{model}

\begin{figure}
\centering
\scalebox{.5}{\includegraphics{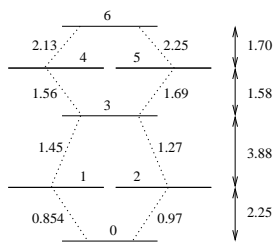}}
\caption{\small The model 7-level system $|n\rangle$ with $n
=0,1,\ldots,6$. The transition frequencies $\omega_{nm}$ in units of
fs$^{-1}$ are shown on the right, and non-zero dipole matrix elements
$\mu_{nm}$ in units of $10^{-30}$ C$\cdot$m are indicated by dotted lines.} 
\label{7levels}
\end{figure}

\begin{figure}
\centering
\scalebox{.75}{\includegraphics{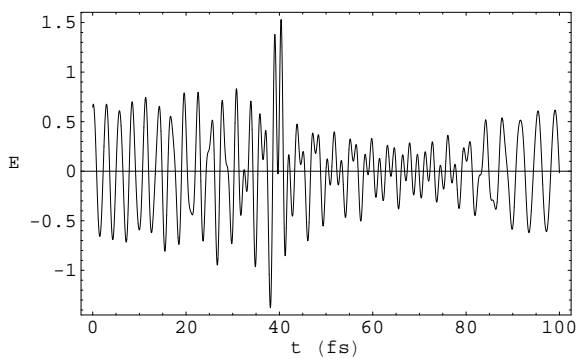}}
\caption{\small 
Electric field $E(t)$ in $V/\mathrm{\AA}$ obtained from an
optimization algorithm for population transfer from $|0\rangle$ to
$|6\rangle$ \cite{Mitra2}.
}
\label{Efield}
\end{figure}

\begin{figure}
\centering
\scalebox{.75}{\includegraphics{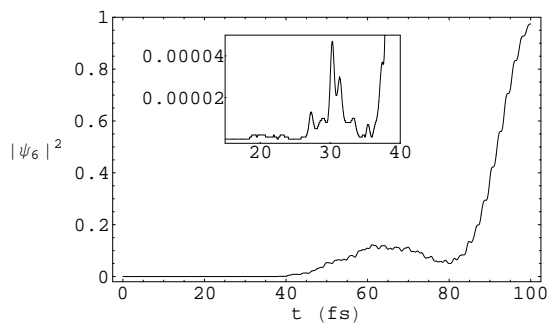}}
\caption{\small 
Population $|\psi_6(t)|^2$ as a function of time (fs). Detail for
small $t$ is shown in the inset (same units).
}
\label{psi6}
\end{figure}

The beable trajectory methodology for identification of control mechanisms
will be illustrated with a 7-level system where $\omega_n$ and $\mu_{nm}$
are given in Fig.\ \ref{7levels}. The (non-adiabatic) control field $E(t)$
shown in Fig.\ \ref{Efield} is obtained from a steepest descents algorithm
over the space of field histories \cite{Mitra2}. It is optimized to
transfer population from the ground state $|0\rangle$ to the highest
excited state $|6\rangle$. By $t = 100$ fs, the transfer is found to be
completed with approximately 97\% efficiency (see Fig.\ \ref{psi6}).

\begin{figure}
\centering
\scalebox{.75}{\includegraphics{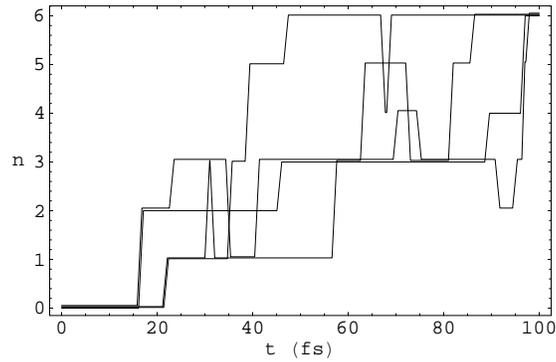}}
\caption{\small 
One each of the 4, 6, 8, and 10-jump trajectories generated by the jump
rule (\ref{T}) are shown with their sites $n$ plotted against time. For
viewing purposes, we have displaced them a small amount vertically from
each other and tilted the jump lines slightly away from the vertical.
} 
\label{4traj}
\end{figure}

Together with the second-order Schrodinger propagator, using time step
$\epsilon = .025$ fs, an ensemble of $N_{\mathrm{traj}} = 10^5$ beable
trajectories is evolved, all starting in the ground state $n=0$ at $t=0$.
At each time step, a given beable at site $m$ is randomly made either to
jump to a neighboring site $n \ne m$ according to the probabilities
$T_{nm}\epsilon$ given by (\ref{T}) with (\ref{zint}), or else stay at
$m$. Four sample trajectories are shown in Fig.\ \ref{4traj}. As a check,
one can count the number of beables residing on each site $n$ at time $t$
to estimate the occupation probabilities $P_n(t)$ and verify that they
match the quantum prescriptions $|\psi_n(t)|^2$. The finite-ensemble
deviations are observed to be consistent with a
$(N_{\mathrm{traj}})^{-1/2}$ convergence law.

About 60\% of the trajectories generated are found to involve four jumps,
and of these the trajectories passing through sites $n=2,5$ are noticeably
more probable than those passing through $n=1,4$. 6-jump trajectories
comprise about 30\% of the ensemble. And it becomes increasingly less
likely to find trajectories with more and more jumps. The largest number
of jumps observed in a single trajectory was 14. Three such trajectories
occurred out of the ensemble total $10^5$.

A natural expectation is that the optimal field $E(t)$ would concentrate
on the higher probability trajectories and not waste much effort on
guiding highly improbable trajectories, such as the 14-jumpers, to the
target state $n=6$, as the latter have essentially no impact on the
control objective (final population of the target state). Interestingly,
though, the vast majority of even the lowest probability trajectories are
still guided to $n=6$. Apparently, the optimal field is able to coordinate
its effect on low probability trajectories with that on other trajectories
at no real detriment to the latter. We shall come back to this point
later.

One way to conveniently categorize the large set of trajectories, each
expressible as a sequence of time-labeled jumps $(t_1,n_1) \rightarrow
(t_2,n_2) \rightarrow \cdots$, is to drop the time labels, leaving only
the ``pathway'' $n_1 \rightarrow n_2 \rightarrow \cdots$. The importance
of a given pathway is then computed as the frequency of trajectories
associated with that pathway. Table \ref{tab} lists some important and/or
interesting pathways and their probabilities.  

\begin{table}
\begin{center}
\begin{tabular}{|l|l|}
\hline
probability & pathway \\
\hline
0.19  & 0 2 3 5 6 \\
0.16  & 0 2 3 4 6 \\
0.14  & 0 1 3 5 6 \\
0.12  & 0 1 3 4 6 \\
0.018  & 0 2 3 5 6 5 6 \\
\hline
0.005  & 0 2 \\
\hline
0.0007  & 0 2 3 5 6 4 3 5 6 \\
\hline
\end{tabular}
\caption{\small The five most probable pathways, followed by the highest
probability pathway failing to reach $n=6$ at $t=100$ fs, and then the
highest probability pathway involving a topologically non-trivial cycle
in state space. The fractional error in the pathway probability $P$ is
given roughly by $(10^5 P)^{-1/2}$.}
\label{tab}
\end{center}
\end{table}

\begin{figure}
\centering
\begin{tabular}{cc}
\scalebox{.6}{\includegraphics{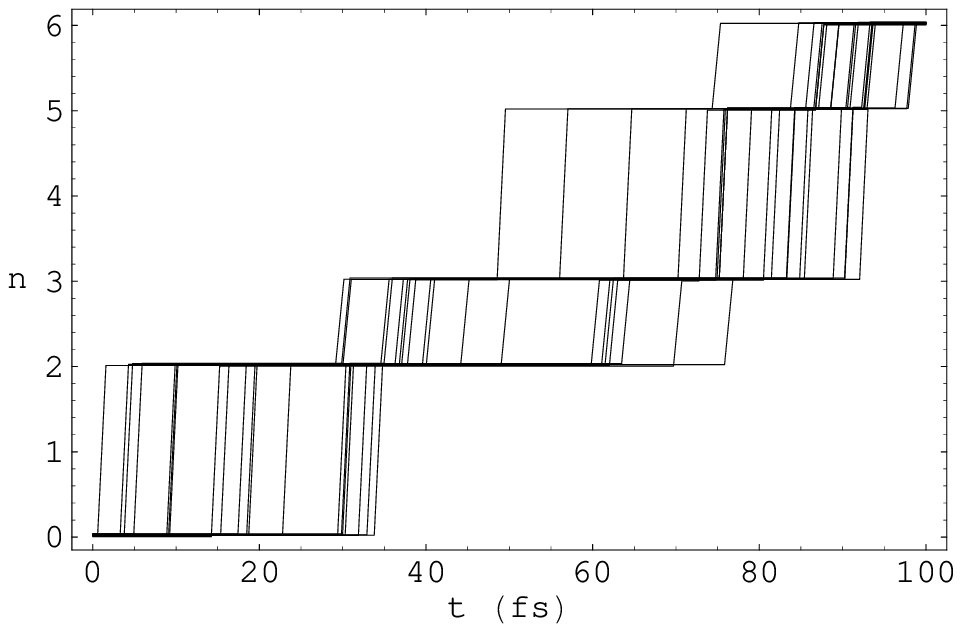}} &
\scalebox{.6}{\includegraphics{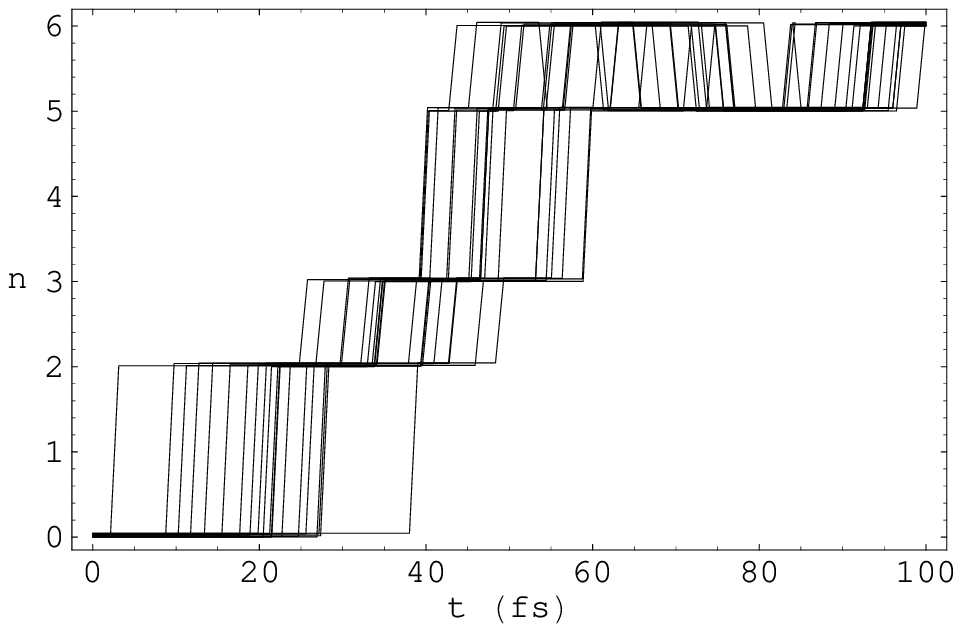}} \\
\end{tabular}
\caption{\small 
A sample of 20 trajectories each from the pathways 0~2~3~5~6 and
0~2~3~5~6~5~6.
} 
\label{46jumpers}
\end{figure}

Fig.\ \ref{46jumpers} shows some typical trajectories associated with the
first and fifth pathways listed in Table \ref{tab}---involving 4 and 6
jumps respectively. $E(t)$ guides the 4-jumpers upward in energy, and they
begin to arrive at $n=6$ around $t = 80$ fs, early enough that stragglers
can catch up but too late for the over-achievers of the group to head off
elsewhere. This corresponds to the onset of heavy growth for
$|\psi_6(t)|^2$ around $t = 80$ fs (see Fig.\ \ref{psi6}). The 6-jumpers
first reach $n=6$ around $t = 50$ fs, but almost all fall back to $n=5$ by
$t=80$ fs, reuniting with the 4-jumpers just as they begin to jump up to
$n=6$. These 6-jumpers, along with other high-order contributions, thus
explain the small surge in $|\psi_6(t)|^2$ between 50 and 80 fs. Another
much smaller surge around $t = 30$ fs and one still smaller around $t =
20$ fs (see inset of Fig.\ \ref{psi6}) are attributable to 8-th and higher
order trajectories ``ringing'' back and forth on $5 \leftrightarrow 6$.

For $t \in (70\mbox{ fs},80\mbox{ fs})$, many of the 6-jumpers are at
$n=6$ and need to be de-excited on the $6 \rightarrow 5$ transition before
they can jump back up to $n=6$. Simultaneously, many of the 4-jumpers are
at $n=5$ and should not be prematurely excited on $5 \rightarrow 6$, lest
they not remain at $n=6$ through $t=100$ fs. The optimal field thus faces
a conundrum: how to stimulate the $2 \leftrightarrow 6$ transition
preferentially for the 6-jumpers (in $n=6$) over the 4-jumpers (in $n=5$).
The means by which this feat is accomplished may be understood by
reference to the jump rule (\ref{T}). $E(t)$ induces jumps through the
explicit $H_{nm}(t)$ factor but also through the $\psi^\ast$ quotient,
which depends on $E(t)$ through $(\ref{schrod})$. In particular,
(\ref{Rez}) implies that at any one time $t$ jumps on this transition must
be either all upward or all downward. The active direction is switched
back and forth according to the sign of $\mathrm{Re}\{z_{65}(t)\}$.

\begin{figure}
\centering
\scalebox{.75}{\includegraphics{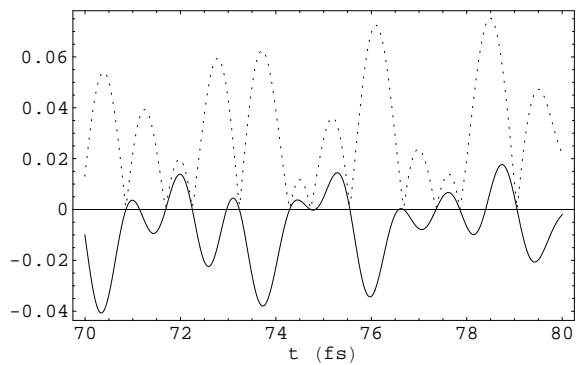}}
\caption{\small 
The optimal field modulus $|E(t)|$ (dotted line) and
$\mathrm{Re}\{z_{65}(t)\}$ (full line) in $\mbox{fs}^{-1}$ over the range
$(70\mbox{ fs}, 80\mbox{ fs})$. Their anticorrelation causes beables to be
preferentially selected for the downward transition $6 \rightarrow 5$ over
the upward transition $5 \rightarrow 6$.
} 
\label{RezEfield}
\end{figure}

Fig.\ \ref{RezEfield} plots $|E(t)|$ and $\mathrm{Re}\{z_{65}(t)\}$, which
controls the upward jump rate $T_{65}(t)$. For $t \in (70\mbox{
fs},80\mbox{ fs})$ one sees that when $|E(t)|$ is large, most often
$\mathrm{Re}\{z_{65}(t)\}$ dips below zero, disallowing any upward jumps.
The correlation coefficient between $|E(t)|$ and
$\mathrm{Re}\{z_{65}(t)\}$ in this range is $-0.4955$. On the other hand,
the correlation between $|E(t)|$ and $\mathrm{Re}\{z_{56}(t)\}$, which
controls downward jumping, is $+0.4475$ over the same range.

\begin{figure}
\centering
\scalebox{.75}{\includegraphics{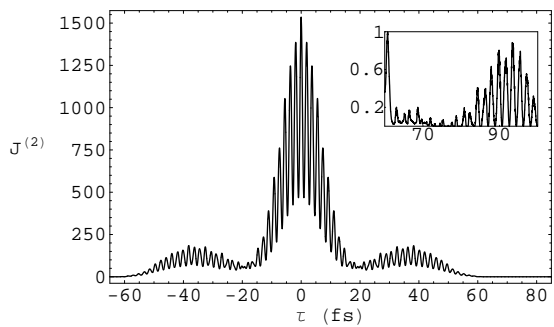}}
\caption{\small 
Jump correlation function $J_{\Omega}^{(2)}(\tau)$ associated with jumps
on $5 \rightarrow 6$, plotted against the delay time $\tau$ (fs) for the
ensemble of $10^5$ trajectories. Detail for large $\tau$ is
shown in the inset (same units).
} 
\label{J2}
\end{figure}

Looking at the trajectories in more detail, one notices a distinct
bunching of jumps. Beables tend to jump together in narrow time bands, or
else to abstain in unison from jumping. This behavior can be gauged
by calculating the two-time jump-jump correlation function:
\[
J_{\Omega}^{(2)}(\tau) \equiv 
\frac{1}{N}\sum_{p=0}^{N-1} J_{\Omega}(t_p)J_{\Omega}(t_p + \tau)
\]
where $J_{\Omega}(t)$ is the number of jumps of type $\Omega$ occurring in
$(t,t+\epsilon)$, and $\Omega$ is a subset of the entire ensemble of
trajectories. For instance, the two-time function with $\Omega$ taken as
the set of jumps on the $5 \rightarrow 6$ transition is plotted in Fig.\
\ref{J2}. The fs time-scale oscillations correspond to the level
splittings $\omega_{nm}$ and the dominant frequency components of $E(t)$.
Enhanced correlations around $\tau=0$ correspond to the jump bunching
noticeable in the trajectories. Two side-bands around $\tau = \pm 40$~fs
are associated with 6-jump and higher order trajectories that go up, down,
and up again on $5 \leftrightarrow 6$ over the approximate time window
$(50\mbox{ fs},90\mbox{ fs})$. This conclusion can be verified by
computing two-time functions with $\Omega$ specialized to particular
pathways. Other much smaller features for $|\tau| > 60$ fs (see inset of
Fig.\ \ref{J2}) are attributable to higher order trajectories ringing on
$5 \leftrightarrow 6$.

In general, the fs oscillations characteristic of these two-time functions
show that $E(t)$ works in an essentially discrete way, turning on the flow
of beables over a given transition and then turning it off with a duty
cycle of $\approx 2$~fs. The associated bandwidth of $\approx 0.5 \mbox{
fs}^{-1}$ is small enough to discriminate between all non-degenerate
$\omega_{nm}$ except between $\omega_{35} (=\omega_{34})$ and $\omega_{56}
(=\omega_{46})$, which differ by only $0.12\mbox{ fs}^{-1}$. This
circumstance leaves effectively three distinguishable transitions. With a
total time of 100 fs, the control field $E(t)$ can potentially enact
roughly $150$ separate flow operations. The fact that trajectories with
pathway probability $\ll 1$\% are still almost always guided successfully
to $n=6$ suggests that these $\sim 150$ operations are more than necessary
to obtain the 97\% success rate achieved by the optimal control algorithm
in this simulation. It appears that the algorithm actively sweeps these
aberrant trajectories back into the mainstream so as to maximize even
their minute contribution to the control objective.

\section{Control Mechanism Identification in The
Laboratory}\label{mechanism}

Using these beable trajectory methods to extract mechanism information
directly from closed-loop data is complicated by the fact that we cannot
assume knowledge of a time-dependent wavefunction, Hamiltonian, or
possibly even the energy level structure of the system. Frequently in the
laboratory, the only available information consists of final state
population measurements and knowledge of the control field $E(t)$.

The following analysis aims to show how a limited statistical
characterization of beable trajectories may be generated from laboratory
data associated with a given optimal control field. In particular, we will
show how to extract $j_\mathrm{min}$, the minimum number of jumps
necessary to reach the final state $n_\mathrm{f}$ from the initial state
$n_\mathrm{i}$; also $\langle j_\mathcal{P} \rangle$, the average number
of such jumps over an ensemble of beable trajectories; and possibly higher
moments $\langle (j_\mathcal{P})^k \rangle$ as well. After a general
formulation of this analysis is presented, it will be applied to simulated
experimental data in the case of the model 7-level system considered
above.

We propose to obtain mechanism information by examining the effect on the
final state population $|\psi_{n_\mrf}(t_\mrf)|^2$ of variations in the
control field \emph{away from} optimality. Detecting only this one
population, i.e.\ the control objective itself, limits how much
mechanism information we can gain. It may be possible to obtain a more
detailed understanding of a given mechanism by probing additional aspects
of the system at times other than just $t_\mrf$, and this may be done
either in our present framework or through extensive spectroscopic
methods. One expects there will always be a trade-off between the
complexity of such methods and the level of mechanistic detail they
reveal.

Consider a particularly simple scheme wherein the amplitude of the control
field is modulated by a constant $\mcM$ independent of time:
\[
E(t) \rightarrow \tilde E(t) = \mcM E(t)
\]
giving rise to a new time-dependent solution $|\tilde \psi(t)\rangle$---in
particular, a new final state population $|\tilde \psi_{n_\mrf}
(t_\mrf)|^2$ and new path probabilities P$\widetilde{\mathrm{rob}}
(\mathcal{P})$. These quantities are obtained by taking $T_{nm}
\rightarrow \tilde T_{nm}$ in (\ref{probP}), which is to say using
$\tilde E(t)$ and $\tilde \psi_n(t)$ in the jump rule (\ref{T}).

To express P$\widetilde{\mathrm{rob}}(\mathcal{P})$ in terms of 
Prob$(\mathcal{P})$, we can write
\begin{eqnarray}
\prod_{p \in J} \tilde T_{n_{p+1}n_p} &=& 
\nonumber
	\mcM^{j_\mathcal{P}} \prod_{p \in J} T_{n_{p+1}n_p}
	\prod_{p \in J} \frac{\cos \tilde \phi_p}{\cos \phi_p} \; \times\\
\label{prodTx}	&& \; \;
	\prod_{p \in J} 
		\frac{|\tilde \psi_{n_{p+1}}(t_p)|}{|\tilde
		\psi_{n_p}(t_p)|} 
	\left(\prod_{p \in J} 
		\frac{|\psi_{n_{p+1}}(t_p)|}{|\psi_{n_p}(t_p)|}
	\right)^{-1}
\end{eqnarray}
where $j_\mathcal{P}$ is the number of jumps in $\mathcal{P}$ and 
\[
\tilde \phi_p \; \equiv \; \arg\left(
	-\mri H_{n_{p+1}n_p} 
	\frac{\tilde \psi_{n_{p+1}}(t_p)}{\tilde \psi_{n_p}(t_p)} 
\right) \, .
\]
To simplify (\ref{prodTx}), note that if $j_\mathcal{P}$ were very large,
then successive terms in each of the last two products would tend to
cancel, leaving only endpoint contributions. Making the reasonable
approximation that they do completely cancel yields
\begin{equation}\label{prodT}
\prod_{p \in J} \tilde T_{n_{p+1}n_p} \; \approx \; 
	\mcM^{j_\mathcal{P}} \,
	\frac{|\tilde \psi_{n_\mrf}(t_\mrf)|}{|\psi_{n_\mrf}(t_\mrf)|} \,
	\prod_{p \in J} T_{n_{p+1}n_p}
	\prod_{p \in J} \frac{\cos \tilde \phi_p}{\cos \phi_p} \, .
\end{equation}
Further, we can make the expansion
\[
-\ln \prod_{p \in J} \frac{\cos \tilde \phi_p}{\cos \phi_p} 
\; = \;
a_\mathcal{P}^{(1)}(\mcM-1) + a_\mathcal{P}^{(2)}(\mcM-1)^2 + \cdots
\]
about $\mcM=1$, where the $a_\mathcal{P}^{(i)}$ depend on the path
$\mathcal{P}$ but not on $\mcM$. And similarly:
\begin{eqnarray*}
-\ln \prod_{p \notin J} \left(
	1-\epsilon\tsum{_n^\prime} \tilde T_{n n_p}
\right) 
&\approx&
\epsilon \tsum{_{p \notin J}} \tsum{_n^\prime} \tilde T_{n n_p}\\
&=&
\epsilon \tsum{_{p \notin J}} \tsum{_n^\prime} T_{n n_p} +
b_\mathcal{P}^{(1)}(\mcM-1) + \cdots
\end{eqnarray*}
Combining these expansions gives a relationship between the path
probabilities P$\widetilde{\mathrm{rob}}(\mathcal{P})$ in the modulated
case and those, Prob$(\mathcal{P})$, in the unmodulated case, which are
the ones containing mechanism information regarding the actual optimal
control field $E(t)$. We can thus write the final population as
\begin{eqnarray*}
|\tilde\psi_{n_\mrf}(t_\mrf)|^2 &=&
\sum_\mathcal{P} |\psi_{n_0}(0)|^2
\, \mbox{P$\widetilde{\mathrm{rob}}(\mathcal{P})$}\\
& \approx &
\frac{|\tilde \psi_{n_\mrf}(t_\mrf)|}{|\psi_{n_\mrf}(t_\mrf)|}
\sum_\mathcal{P} |\psi_{n_0}(0)|^2 \mcM^{j_\mathcal{P}}
\mre^{-a_\mathcal{P}(\mcM-1)} \, \mathrm{Prob}(\mathcal{P})
\end{eqnarray*}
where $a_\mathcal{P} \equiv a_\mathcal{P}^{(1)} + b_\mathcal{P}^{(1)}$,
and higher order terms in the expansion have been dropped. (This
approximation is not as crude as it might seem, since for small $\mcM$
away from 1, the behavior of $|\tilde\psi_{n_\mrf}(t_\mrf)|^2$ is
dominated by the $\mcM^{j_\mathcal{P}}$ factor.) Cancelling one power of
$|\tilde \psi_{n_\mrf}(t_\mrf)|$, and recalling that the sum is taken only
over paths ending on $n=n_\mrf$ so that $|\psi_{n_\mrf}(t_\mrf)|^2 =
\sum_\mathcal{P} \mathrm{Prob}(\mathcal{P})$, we have
\begin{equation}\label{M^j}
|\tilde\psi_{n_\mrf}(t_\mrf)|
\; \approx \;
|\psi_{n_\mrf}(t_\mrf)| \left\langle 
	\mcM^{j_\mathcal{P}} \mre^{-a_\mathcal{P}(\mcM-1)}
\right\rangle
\end{equation}
where $\langle \cdots \rangle$ denotes an average over the trajectory
ensemble generated by the (unmodulated) optimal field $E(t)$. Beables in
this ensemble are taken as initially distributed at $t=0$ according to
$|\psi_{n}(0)|^2$, and only trajectories that successfully reach
$n=n_\mrf$ at $t=t_\mrf$ are counted.

\begin{figure}
\centering
\scalebox{.75}{\includegraphics{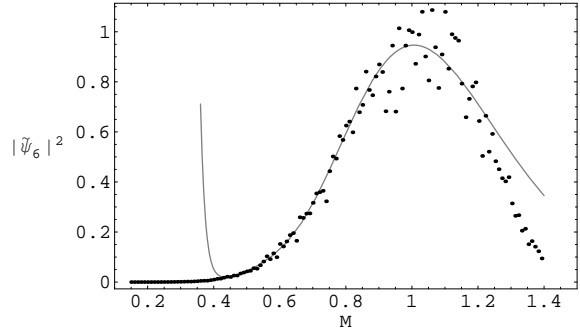}}
\caption{\small 
The best fit of (\ref{jlogM}) to the simulated $|\tilde
\psi_{n_\mrf}(t_\mrf)|^2$ data (10\% noise) as a function of $\mcM$; it
occurs over the fitting range $\mcM \in (.44,.92)$.
} 
\label{fit}
\end{figure}

Note that for $\mcM$ close enough to 0, the minimum value $j_\mathrm{min}$
taken on by $j_\mathcal{P}$ will dominate the expectation value in
(\ref{M^j}), and
\begin{equation}\label{jminlogM}
\ln |\tilde\psi_{n_\mrf}(t_\mrf)| = j_\mathrm{min} \ln\mcM +
\mathrm{O}(1)
\end{equation}
gives the dominant behavior independent of $a_\mathcal{P}$. If we suppose
that $a_\mathcal{P}$, where it is relevant, depends primarily on the
endpoints of $\mathcal{P}$, which are fixed, and only weakly on the rest
of the path, then $a_\mathcal{P}$ can be approximated by some
characteristic value $a$. Putting $\mcM^{j_\mathcal{P}} =
\mre^{j_\mathcal{P} \ln\mcM}$ in (\ref{M^j}) and expanding in powers of
$\ln\mcM$ now gives
\begin{equation}\label{jlogM}
|\tilde\psi_{n_\mrf}(t_\mrf)| 
\; \approx \;
|\psi_{n_\mrf}(t_\mrf)| \, \mre^{-a(\mcM-1)} 
\sum_{k=0}^\infty \frac{\langle (j_\mathcal{P})^k \rangle}{k!}
	(\ln\mcM)^k
\end{equation}
for the final state population under a modulated field, expressed
in terms of the desired statistical properties of the trajectory ensemble 
under the optimal field itself. Here, $a$ enters as an additional
parameter that must be extracted from the data. Equations (\ref{jminlogM})
and (\ref{jlogM}) form the working relations to extract mechanism
information from laboratory data.

\section{Simulated Experiments on a 7-Level System}\label{simulated}

In order to extract quantities like $\langle j_\mathcal{P} \rangle$ using
the results (\ref{jminlogM}) and (\ref{jlogM})  data must be generated for
the final state population $|\tilde \psi_{n_\mrf}(t_\mrf)|^2$ at many
values of the modulation factor $\mcM$ over some range
$(\mcM_\mathrm{min},\mcM_\mathrm{max}) \sim (0,1.5)$. The desired
quantities are obtained as parameters in fitting (\ref{jminlogM}) and
(\ref{jlogM})  to the data as a function of $\mcM$.

One set of simulated data for the above 7-level system is shown in Fig.\
\ref{fit}; the sampling increment is $\Delta M = .01$. Noise has been
introduced by multiplying the exact $|\tilde \psi_{n_\mrf}(t_\mrf)|^2$
values by an independent Gaussian-distributed random number for each value
of $\mcM$, where the distribution is chosen to have mean 1, and various
standard deviations $\sigma$ have been sampled.

\begin{figure}
\centering
\scalebox{.75}{\includegraphics{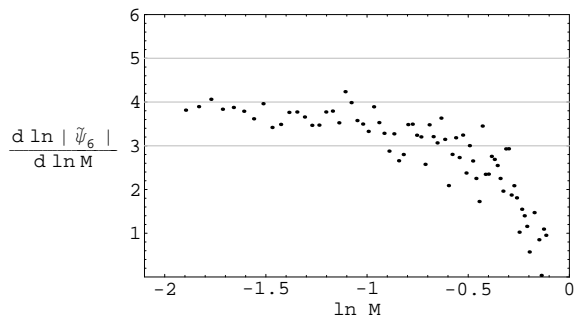}}
\caption{\small 
The derivative is calculated from simulated data with
noise level $\sigma=.1$; its limiting value as $\ln\mcM \rightarrow
-\infty$ gives $j_\mathrm{min}$. 
} 
\label{jmin}
\end{figure}

We can determine $j_\mathrm{min}$ from the data using (\ref{jminlogM}),
which implies 
\begin{equation}\label{dlogpsidlogM}
j_\mathrm{min} = \lim_{\mcM \rightarrow 0} 
	\frac{d \ln |\tilde \psi_{n_\mrf}(t_\mrf)|}{d\ln\mcM} \, .
\end{equation}
For instance, Fig.\ \ref{jmin} plots the derivative in
(\ref{dlogpsidlogM}), calculated with finite differences from the $|\tilde
\psi_{n_\mrf}(t_\mrf)|^2$ simulated data for $\sigma=.1$, which correctly
gives $j_\mathrm{min}=4$ as the limiting value. Determination of
$j_\mathrm{min}$ proved robust to multiplicative Gaussian noise up to the
40\% level ($\sigma=.4$).

The quantity $\langle j_\mathcal{P} \rangle$ is more difficult to extract,
because while the sum in (\ref{jlogM}) converges to 0 as $\mcM \rightarrow
0$, the terms of the sum individually diverge and must cancel in a
delicate manner. Therefore truncating the sum to an upper limit
$k_\mathrm{max}$ becomes a very bad approximation near $\mcM=0$. This
unstable behavior can be controlled by carefully setting the range
$(\mcM_\mathrm{min}, \mcM_\mathrm{max})$ of data to be fitted, given a
choice of $k_\mathrm{max}$.

It is also convenient to constrain the fit by the previous determination
of $j_\mathrm{min} = 4$. We have done this by noting that if $E(t)$ is
truly optimal, then $|\tilde \psi_{n_\mrf}(t_\mrf)|$ must have a maximum
at $\mcM=1$, which implies that $a = \langle j_\mathcal{P} \rangle$. This
can be used as a weaker constraint on the auxiliary parameter $a$ by just
requiring $a > j_\mathrm{min} = 4$ in the fit without necessarily
supposing that $E(t)$ is exactly optimal. We then check that $a \approx
\langle j_\mathcal{P} \rangle$ is satisfied in the fit. Fig.\ \ref{fit}
shows one such fit where the fitting range is $\mcM \in (.44,.92)$. One
can see that the fit closely tracks the data for $\mcM$ in this range but
quickly diverges from the data just below $\mcM = .44$ (and, less
severely, above $\mcM = .92$) due to the sum-truncation instability
mentioned previously.

In order to identify appropriate ranges in general, we have searched over
all combinations such that 
\begin{equation}\label{range}
\begin{array}{c}
	\begin{array}{lcccr}
	.2 &<& \mcM_\mathrm{min} &<& .8 \\ 
	.7 &<& \mcM_\mathrm{max} &<& 1.6 \\
	\end{array}\\
\mcM_\mathrm{max}-\mcM_\mathrm{min}>10
\end{array}
\end{equation}
Mathematica's implementation of the Levenberg-Marquardt non-linear fitting
algorithm was used on simulated data for each value of $\sigma$ between 0
and .5 with a .01 increment. The best fit at each $\sigma$ was used to
determine the value of $\langle j_\mathcal{P} \rangle$ most consistent
with the simulated data at the given noise level. 

For this analysis $k_\mathrm{max}=4$ was chosen somewhat arbitrarily
to balance computational cost and precision. In practice it is likely that
the moments $\langle (j_\mathcal{P})^k \rangle$ for lower $k$ values will
be most reliably extracted from the data, especially considering the
laboratory noise. In the simulations it was found that $\langle
j_\mathcal{P} \rangle$ could be reliably extracted, but higher moments
were unstable and unreliable. For example, $\langle (j_\mathcal{P})^2
\rangle$ was frequently found to lie slightly below the corresponding fit
values for $\langle j_\mathcal{P} \rangle^2$, which is inconsistent with
the interpretation of these values as statistical moments of an underlying
random variable $j_\mathcal{P}$. Further constraints could be introduced
to attempt to stabilize the extraction of higher moments, but care is
needed so as not to overfit the data.

\begin{figure}
\centering
\scalebox{.75}{\includegraphics{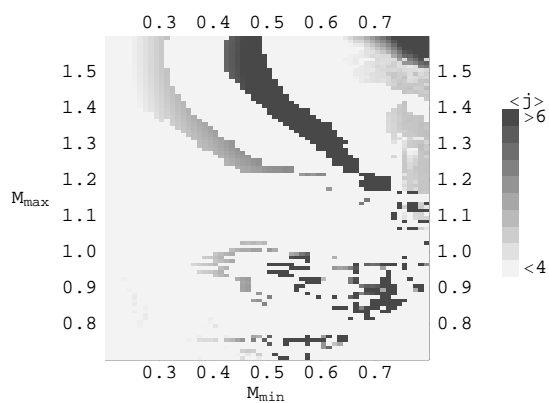}}
\caption{\small 
Fit values for $\langle j_\mathcal{P} \rangle$ over a set of different
fitting ranges $(\mcM_\mathrm{min},\mcM_\mathrm{min})$; $\sigma=.1$ here. 
}
\label{javg}
\end{figure}

\begin{figure}
\centering
\scalebox{.75}{\includegraphics{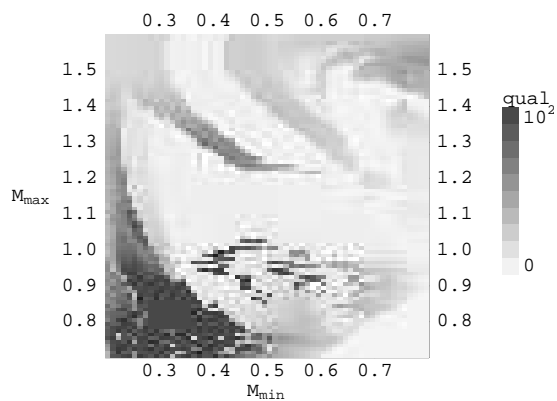}}
\caption{\small 
Fit qualities as measured by the inverse of the mean squared deviations
between the simulated data and the fit; $\sigma=.1$ here.
The highest fit quality appears at $(.44,.92)$. 
} 
\label{fq}
\end{figure}

Fig.\ \ref{javg} shows the $\langle j_\mathcal{P} \rangle$ values obtained
by fitting data with $\sigma=.1$ for each choice of $(\mcM_\mathrm{min},
\mcM_\mathrm{max})$ and Fig.\ \ref{fq} shows the corresponding quality of
each fit as measured by its mean squared deviations. In Fig.\ \ref{javg},
as well as in the corresponding plots for all other values of $\sigma$
studied, two diagonal strips emerge running above a set of smaller
islands. The surrounding white ``sea'' comprises fits that give $\langle
j_\mathcal{P} \rangle < 4$, which we know to be ruled out by the
determination of $j_\mathrm{min}$. 

A virtually identical pattern arises in the fit quality plots. The two
strips and underlying islands are seen to give much better fits than the
white sea. An additional connected region of good fits is found to extend
across the lower-left corner of Fig.\ \ref{fq}, nearly all of which are
ruled out by $j_\mathrm{min}=4$. This connected region is somewhat
pathological because much of it corresponds to fitting ranges that fail to
capture the important behavior of $|\tilde \psi_{n_\mrf}(t_\mrf)|^2$ near
$\mcM=1$, and therefore can be ignored. Then the best fits for all values
of $\sigma$ sampled are found to come from the cluster of islands at
$\mcM_\mathrm{max} \approx .95$. As $\sigma$ is increased from 0 to .5,
these islands flow from $\mcM_\mathrm{min} \approx .5$ to
$\mcM_\mathrm{min} \approx .3$, carrying with them the best fit site. Note
that the small triangular area in the lower right corner, most noticeable
in Fig.\ \ref{fq}, is a region excluded from consideration by the third
constraint in (\ref{range}).

\begin{figure}
\centering
\scalebox{.75}{\includegraphics{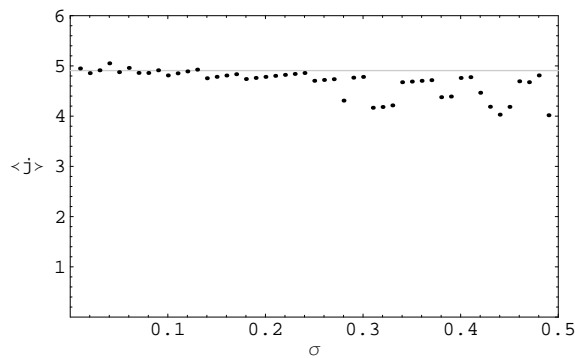}}
\caption{\small 
Best fit values for $\langle j_\mathcal{P} \rangle$ as a function of the
noise level $\sigma$, compared to the exact value $\langle j_\mathcal{P}
\rangle = 4.907$ (gray line).
} 
\label{javgsigma}
\end{figure}

The best fit values of $\langle j_\mathcal{P} \rangle$ are shown as a
function of $\sigma$ in Fig.\ \ref{javgsigma}. These values are to be
compared with the exact result $\langle j_\mathcal{P} \rangle = 4.907$
obtained from the trajectory ensemble calculations in \S\ref{model}, which
require explicit knowledge of the level structure and dipole moments
$\mu_{nm}$ of the system. The ramping behavior in Fig.\ \ref{javgsigma}
results from the sampling increment $\Delta\mcM$ of the simulated data.
Transitioning between one ramp and another corresponds to the shifting of
the best fit location by one or two units of $\Delta\mcM$.

These $\langle j_\mathcal{P} \rangle$ values are in good agreement (3\%
discrepancy) with the exact value for noise at the level of 0--25\%. It
should be noted that a qualitative change occurs in the case of no noise
($\sigma=0$), where the islands all disappear and the strips become
extended much further on the downward diagonal. Inspecting the fits
individually indicates that mean squared deviation does not give an
adequate measure of fit quality in this special case. This anomaly seems
due to the fact that, in the absence of Gaussian noise from experimental
statistics, systematic deviations from (\ref{jlogM}) associated with the
approximation (\ref{prodT}) become important.

Since Bell's model can be defined for any choice of basis $|n\rangle$,
there is a more general question of how the above kind of mechanism
analysis might vary with the choice of basis. Beyond that, Bell's jump
rule (\ref{T}) itself permits generalization \cite{Guido}, providing
additional freedom over which trajectory probability assignments may vary.
The import of this freedom for mechanism identification remains to be
determined.


\begin{thebibliography}{}

\bibitem{ether} R. Laughlin and D. Pines, PNAS \textbf{97} 28-31 (2000).

\bibitem{feynman} R. Feynman, ``Simulating physics with computers,'' Int.
J. Theor. Phys. {\bf 21} 467 (1982).

\bibitem{lloyd} S. Lloyd, ``Universal Quantum Simulators,'' Science
\textbf{273} 1073 (1996).

\bibitem{shor} P. Shor, ``Polynomial-Time Algorithms for Prime
Factorization and Discrete Logarithms on a Quantum Computer,'' SIAM J.
Sci. Statist. Comput. \textbf{26} 1484 (1997).

\bibitem{hersch} A. Peirce, M. Dahleh, and H. Rabitz, ``Optimal Control of
Quantum Mechanical Systems: Existence, Numerical Approximations, and
Applications,'' Phys. Rev. A \textbf{37} 4950 (1988).

\bibitem{preskill} see Chapter 6 of John Preskill's quantum computation
notes on the web at
http://www.theory.caltech.edu/people/preskill/ph229/index.html.

\bibitem{atac} ``Quantum computation with quantum dots and terahertz
cavity quantum electrodynamics,'' Phys. Rev. A \textbf{60} 3508 (1999),
quant-ph/9904096.

\bibitem{landauer} R. Landauer, ``Is quantum mechanics useful?'' Phil.
Tran. R. Soc. Lond. \textbf{353} 367 (1995).

\bibitem{9qubit} P. Shor, ``Scheme for reducing decoherence in quantum
computer memory,'' Phys. Rev. A \textbf{52} R2493 (1995).

\bibitem{FM} M. H. Freedman and D. A. Meyer, ``Projective plane and planar
quantum codes,'' quant-ph/9810055 (1998).

\bibitem{1} P. Shor, ``Fault-tolerant quantum computation,'' \emph{Proc.
37th Ann. Symp. on the Foundations of Computer Science}, (IEEE Computer
Society Press, Los Alamitos, CA, 1996), quant-ph/9605011.

\bibitem{reliable} J. Preskill, ``Reliable Quantum Computers,''
Proc. Roy. Soc. Lond. A \textbf{454} 385 (1998).

\bibitem{2} E. Knill and R. Laflamme, ``Concatenated quantum codes,''
quant-ph/9608012.

\bibitem{3} E. Knill, R. Laflamme, and W. Zurek, ``Threshold accuracy for
quantum computation,'' quant-ph/9610011; E. Knill, R. La amme, and W. H.
Zurek, ``Resilient quantum computation,'' Science \textbf{279} 342 (1998).

\bibitem{4} D. Aharonov and M. Ben-Or, ``Fault-tolerant quantum
computation with constant error,'' \emph{Proc. 29th Ann. ACM Symp. on
Theory of Computing}, (ACM, New York, 1998), quant-ph/9611025.

\bibitem{5} C. Zalka, ``Threshold estimate for fault-tolerant quantum
computing,'' quant-ph/9612028.

\bibitem{6} D. Gottesman, ``Stabilizer codes and quantum error
correction,'' quant-ph/9705052 (1997); ``Theory of fault-tolerant quantum
computation,'' Phys. Rev. A \textbf{57} 127 (1998), quant-ph/9702029.

\bibitem{7} J. Preskill, ``Fault-tolerant quantum computation,'' in
\emph{Introduction to Quantum Computation and Information}, Hoi-Kwong Lo,
Sandu Popescu, and Tim Spiller, (World Scientifc, New Jersey, 1998),
quant-ph/9712048.

\bibitem{8} A. Kitaev, ``Fault-tolerant quantum computation by anyons,''
quant-ph/9707021 (1997).

\bibitem{9} S. Bravyi and A. Kitaev, ``Quantum codes on a lattice with
boundary,''
	quant-ph/9811052 (1998).

\bibitem{10} M. Freedman and D. Meyer, ``Projective plane and planar
quantum codes,''
	quant-ph/9810055 (1998).

\bibitem{11} E. Dennis, A. Kitaev, A. Landahl, J. Preskill, ``Topological
quantum memory,'' J. Math. Phys. \textbf{43} 4452-4505 (2002),
quant-ph/0110143.

\bibitem{05} H. Barnum \emph{et al.}, Phys. Rev. A \textbf{57} 4153
(1998).

\bibitem{GFMC} D. Ceperley, M. Kalos, in \emph{Monte Carlo Methods in
Statistical Mechanics}, edited by K. Binder, Springer Verlag (1979).

\bibitem{creps} H. Carmichael, \emph{Statistical Methods in Quantum Optics
1}, Springer Verlag (1999).

\bibitem{Negele} J. Negele, H. Orland, \emph{Quantum Many-Particle
Systems}, Perseus (1998).

\bibitem{PW} G. Parisi, Y. Wu., Sci. Sin. \textbf{24} 483 (1981).

\bibitem{Davidson} M. Davidson, Lett. Math. Phys. \textbf{3} 271 (1979),
quant-ph/0112063.

\bibitem{Nelson} E. Nelson, Phys. Rev. \textbf{150} 1079 (1966);
\emph{Dynamical Theories of Brownian Motion}, Princeton University Press
(1967).

\bibitem{Bohm} D. Bohm, Phys. Rev. \textbf{85} 166 (1952); Phys. Rev.
\textbf{85} 180 (1952).

\bibitem{Moskowitz} J. Moskowitz \emph{et al.}, J. Chem. Phys. \textbf{77}
349 (1982).

\bibitem{kogut} S. Duane and J. Kogut, Phys. Rev. Lett. \textbf{55} 2774
(1985).

\bibitem{Goldstein} S. Goldstein, private communication.

\bibitem{Wyatt} C. Lopreore and R. Wyatt, Phys. Rev. Lett. \textbf{85} 895
(2000).

\bibitem{Hersch2} B. Dey, A. Askar, and H. Rabitz, J. Chem. Phys.
\textbf{109} 8770 (1998); X.-G. Hu, T.-S. Ho, H. Rabitz, Phys. Rev. E
\textbf{61} 5967 (2000).

\bibitem{Bell} J. Bell, ``Beables for quantum field theory,''
\emph{Speakable and unspeakable in quantum mechanics}, Cambridge
University Press (1987).

\bibitem{feynspin} R. Feynman, \emph{Statistical Mechanics: A Set of
Lectures}, Addison-Wesley (1972), cf. 202.

\bibitem{control} R. Judson, H. Rabitz, Phys. Rev. Lett. \textbf{68} 1500
(1992).

\bibitem{organic} R. Levis, G. Menkir, H. Rabitz, Science \textbf{292}
709 (2001).

\bibitem{metal} A. Assion, T. Baumer, M. Bergt, T. Brixner, B. Kiefer, V.
Seyfried, M. Strehle, G. Gerber, Science, \textbf{282} 919 (1998).

\bibitem{Na2K} S. Vajda, A. Bartelt, E. Kaposta, T. Leisner, C. Lupulescu,
S. Minemoto, P. Francisco, L. Woste, Chem. Phys. \textbf{267} 231 (2001).

\bibitem{AlGaAs} J. Kunde, B. Baumann, S. Arlt, F. Morier-Genoud, U.
Siegner, U. Keller, Appl. Phys. Lett. \textbf{77}  924 (2000).

\bibitem{Xray} R. Bartels et al., Nature \textbf{406} 164 (2000).

\bibitem{dye} C. Bardeen et al., Chem. Phys. Lett. \textbf{280} 151
(1997).

\bibitem{CH3OH} T. Weinacht, J. White, P. Bucksbaum, J. Phys. Chem. A
\textbf{103} 10166 (1999).

\bibitem{Dahleh} M. Dahleh, A. Peirce, H. Rabitz, Phys. Rev. A
\textbf{37} 4950 (1988).

\bibitem{Kosloff} R. Kosloff, S. Rice, P. Gaspard, S. Tersigni, D. Tannor,
Chem. Phys. \textbf{139} 201 (1989).

\bibitem{Zhu} H. Rabitz, W. Zhu, Accts. Chem. Res. \textbf{33} 572
(2000).

\bibitem{Rice} S. Rice, M. Zhao, \emph{Optical Control of Molecular
Dynamics}, Wiley (2000).

\bibitem{Broglie} L. de Broglie, in \emph{Rapport au V'ieme Congres de
Physique Solvay}, Gauthier-Villars, Paris (1930).

\bibitem{Nino} D. Durr, S. Goldstein, and N. Zanghi, ``Bohmian Mechanics as the Foundation of Quantum Mechanics,''
in \textit{Bohmian Mechanics and Quantum Theory: An Appraisal}, Boston Studies in the Philosophy of Science
\textbf{184} (1996), quant-ph/9511016.

\bibitem{Mitra1} A. Mitra and H. Rabitz, Phys. Rev. A \textbf{67} 043409 (2003).

\bibitem{Mitra2} Optimization results provided by A. Mitra.

\bibitem{Bohr} J. Gleick, \emph{Genius: the Life and Science of Richard
Feynman}, Pantheon (1992).

\bibitem{Guido} G. Bacciagaluppi, Found. Phys. Lett. \textbf{12} 1 (1999),
quant-ph/9811040.


\end{thebibliography}
\end{document}